\documentclass[aps,prd,floatfix,10pt,nofootinbib,showpacs,notitlepage,longbibliography]{revtex4-1} 
\usepackage{amsmath}
\usepackage{amsfonts}
\usepackage[colorlinks=true,linkcolor=blue,citecolor=blue,urlcolor=blue]{hyperref}
\usepackage{amssymb}
\usepackage{amsthm}    
\usepackage{graphicx}
\usepackage{enumerate}
\usepackage{csquotes}
\newtheorem*{conj}{Conjecture}

\newcommand{\nn}{\nonumber \\}
\newcommand{\eq}[1]{Eq.~(\ref{#1})}
\newcommand{\fig}[1]{Fig.~\ref{#1}}
\newcommand{\Fig}[1]{Fig.~\ref{#1}}

\newcommand{\bsub}{\begin{subequations}}
\newcommand{\esub}{\end{subequations}}
\newcommand{\be}{\begin{eqnarray}}
\newcommand{\ee}{\end{eqnarray}}
\newcommand{\om}{\ensuremath{\omega}}

\newcommand{\omc}{\ensuremath{\om_c}}

\usepackage{float}

\newcommand{\pd}{\ensuremath{\partial}}
\newcommand{\la}{\ensuremath{\lambda}}
\newcommand{\ep}{\ensuremath{\epsilon}}
\newcommand{\lp}{\ensuremath{\left(}}
\newcommand{\rp}{\ensuremath{\right)}}
\newcommand{\bi} {\begin{itemize}}
\newcommand{\ei} {\end{itemize}}
\newcommand{\ben}{\begin{enumerate}}
\newcommand{\een}{\end{enumerate}}
\newcommand{\bmat}{\begin{pmatrix}}
\newcommand{\emat}{\end{pmatrix}}
\newcommand{\eg}{ & \hspace*{-0.1 cm}=}
\newcommand{\abs}[1]{\left\lvert #1 \right\rvert}
\newcommand{\C}{\mathbb C}
\newcommand{\R}{\mathbb R}
\newcommand{\N}{\mathbb N}
\begin{document}

\title{No hair theorems for analogue black holes}


\author{Florent Michel}\email[]{florent.michel@th.u-psud.fr} 
\author{Renaud Parentani}\email[]{renaud.parentani@th.u-psud.fr}
\author{Robin Zegers}\email[]{robin.zegers@th.u-psud.fr}
\affiliation{Laboratoire de Physique Th\'eorique, CNRS, Univ. Paris-Sud, Universit\'e Paris-Saclay, 91405 Orsay, France.}

\begin{abstract}
We show that transonic one-dimensional flows which are analogous to black holes obey no-hair theorems both at the level of linear perturbations and in nonlinear regimes. Considering solutions of the Gross-Pitaevskii (or Korteweg-de Vries) equation, we show that stationary flows which are asymptotically uniform on both sides of the horizon are stable and act as attractors. Using Whitham's modulation theory, we analytically characterize the emitted waves when starting from uniform perturbations. Numerical simulations confirm the validity of this approximation and extend the results to more general perturbations and to the (non-integrable) cubic-quintic Gross-Pitaevskii equation. When considering time reversed flows that correspond to white holes, the asymptotically uniform flows are unstable to sufficiently large perturbations and emit either a macroscopic undulation in the supersonic side, or a nonlinear superposition of soliton trains. 
\end{abstract}

\maketitle

\section{introduction}

On a superficial level there are similarities between general relativity and hydrodynamics as they both 
rest on non linear partial differential equations depending on space and time. Interestingly, under some specific circumstances, this mere analogy is replaced by a precise correspondence. For instance, when working in the hydrodynamical approximation, i.e., for long wave lengths, the linearized wave equation in a moving fluid is identical to the d'Alembert equation of a scalar field propagating in a certain four dimensional curved space-time~\cite{Unruh:1980cg}. This remark has been the starting point of investigations aimed at understanding the validity domain of this correspondence. It is clear that the strict equivalence is lost when including dispersive effects which affect short wave length modes propagating in media~\cite{Jacobson:1991gr}. However an interesting and non trivial question is to identify the {\it consequences} of these dispersive effects: for instance, how would they affect the asymptotic properties of the Hawking radiation emitted by an analogue black hole flow~\cite{Unruh:1994je}? After a thorough analysis, it was found that they do not significantly affect the spectrum when certain inequalities involving the short dispersive length and the spatial gradient of the flow are met, see Refs.~\cite{Coutant:2011in,Robertson:2012ku} for recent updates.
 
The present work aims to address similar questions when including nonlinear effects. It is clear that the equations of general relativity and hydrodynamics differ at this level. However, as we explicitly show, both the linear and nonlinear stability of analogue black hole flows are in direct correspondence with those of black holes in general relativity~\cite{Misner1973}. Let us briefly remind the main properties of the latter. First, stationary black holes are characterized by very few parameters, namely their mass, angular momentum, and electric charge. Second, when perturbed during a finite time interval, the time evolution subsequently brings back the solution to one of those stationary solutions. These two properties are generally referred to as ``no hair theorems''. The precise validity limits of these theorems is still the subject of investigations~\cite{Chrusciel:2012jk}. Moreover, several hairy black hole solutions have been found when admitting exotic matter fields, see Refs. in~\cite{Chrusciel:2012jk}. 

When studying the behavior of one-dimensional transonic flows, solutions of the Gross-Pitaevskii (GP) or Korteweg-de Vries (KdV) equation, we observe similar properties. We work with time-independent external potentials (shape of the obstacle when using the KdV equation) which vary only in a finite domain of $x$. In this case, when considering the set of {stationary} flows which are asymptotically homogeneous (AH) as $x\to \pm \infty$ -- a condition analogous to asymptotic flatness for black holes -- a simple counting of integration constants reveals that there is at most a discrete set of solutions. Moreover, when the potential consists of a single step, we explicitly demonstrate that the solution is unique. In other words, one finds a single series of solutions which can be parameterized by the conserved current characterizing the flow. When the potential is smooth, we numerically found a series of solutions which smoothly connect to this series in the step like limit. In certain cases, we also found a disconnected series of solutions. These can be considered as hairy black holes as they contain a large fraction of a soliton attached to the sonic horizon. 

The analogy with general relativity is significantly reinforced when considering the stability of the AH black hole solutions. At the linear level, numerical and analytical results establish that local perturbations decay in time so as to reach an AH stationary solution. This conclusion applies both to the GP and the KdV equations. Hence, at the linear level, the AH black hole solutions of these equations are attractors. It has to be pointed out that the time reversed flows, analogous to white holes, are not attractors: in these flows, the scattering of linear perturbations generates a macroscopic undulation~\cite{Coutant:2012mf}. When adding non linearities, several behaviors are found at late time. Our findings are in agreement with, and add to, the results of Refs.~\cite{Mayoral:2010ck,Michel:2013wpa,Michel:2015pra,2015arXiv150900795D}. 

When taking into account the nonlinearities of the GP and KdV equations, by a combination of analytical and numerical methods, we show that there is a large domain of initial black hole configurations which evolve towards an AH stationary solution. In particular, using Whitham modulation theory applied to the GP equation~\cite{Kamchatnov}, we show that the ``hair'' present in the initial configuration is expelled away from the horizon by three nonlinear waves. At at small amplitudes, these three waves  give back the three types of linear waves emitted by an analogue black hole flow. 
As a result, this emission can be considered as the nonlinear version of the stimulated Hawking effect~\cite{Rousseaux:2007is,Macher:2009nz,Weinfurtner:2010nu}. The predictions of Whitham's theory are confirmed and extended by numerical analysis. Interestingly 
it seems that the integrability of the equations plays
an important role both in analogue models and in general relativity. In the first case, it provides the invariants needed to apply Whitham's theory and eventually characterize the time-evolution from a perturbed configuration to an AH, stable stationary solution. In the second case, an integrable nonlinear sigma model is used to constrain the set of asymptotically flat, stationary and axisymmetric solutions of Einstein-Maxwell theory, leading to the uniqueness theorem~\cite{Chrusciel:2012jk}. 

The paper is organized as follows. In Section~\ref{sec:2}, we first study the stationary solutions of the GP equation and show that the set of AH solutions is discrete. Then we study the linear stability of black hole flows and show that perturbations all decay polynomially in time. In Section~\ref{sec:NLstab}, we consider the nonlinear stability of black hole flows. We first use Whitham's theory in subsection~\ref{sec:analytical}, and then present numerical results for the full GP equation in subsection~\ref{sec:numres}. Some new results on white-hole flows are given in Section~\ref{Sec:WH}. We conclude in Section~\ref{sec:concl}. In appendixes we present various technical aspects concerning notions used in the main body of the text. Appendix~\ref{App:Whitham} reminds the derivation of Whitham's equations and derives the properties of the nonlinear waves used in the main text. Appendixes~\ref{App:KdV} and~\ref{App:linKdV} extend the analysis to KdV-like equations, while Appendix~\ref{App:CQNLS} treats the case of a cubic-quintic GP equation. 
\section{Uniqueness and linear stability of black-hole flows}
\label{sec:2}

\subsection{Asymptotically homogeneous transonic flows} 
\label{sub:Ahtf} 

We consider a one-dimensional flowing condensate whose wave-function $\psi(t,x)$ satisfies the GP equation. In a unit system where the atomic mass and reduced Planck constant are equal to $1$, this equation reads 
\be \label{eq:GPE}
i \pd_t \psi = -\frac{1}{2} \pd_x^2 \psi +V(x) \psi + g(x) \left\lvert \psi \right\rvert^2 \psi.
\ee
The static external potential is denoted by $V(x)$ and the effective one-dimensional coupling by $g(x)$. 
In order to have stable homogeneous configurations for constant $V$ and $g$, we shall consider only condensates with repulsive interactions between atoms, so that $g(x) > 0$. The stationary solutions can be written as
\be \label{eq:Madelung}
\psi(x,t) = \sqrt{\rho(x)} e^{i \int^x v(x') dx'}\, e^{-i \om t},
\ee
where $\omega$, $\rho(x)$, and $v(x)$ are real. $\rho(x)$ is the density of the condensate and $v(x)$ its local velocity. Plugging \eq{eq:Madelung} into \eq{eq:GPE} gives
\be \label{eq:eqf} 
\frac{\pd_x^2 \sqrt{\rho(x)}}{\sqrt{\rho(x)}} = 2 \lp V(x) - \om \rp + 2 g(x) \rho(x) + \frac{J^2}{\rho(x)^2}, 
\ee
where $J \equiv \rho(x) v(x)$ is the conserved current.

\subsubsection{Homogeneous potentials} \label{subsub:homo}

\begin{figure}
\begin{center}
\includegraphics[width=0.5 \linewidth]{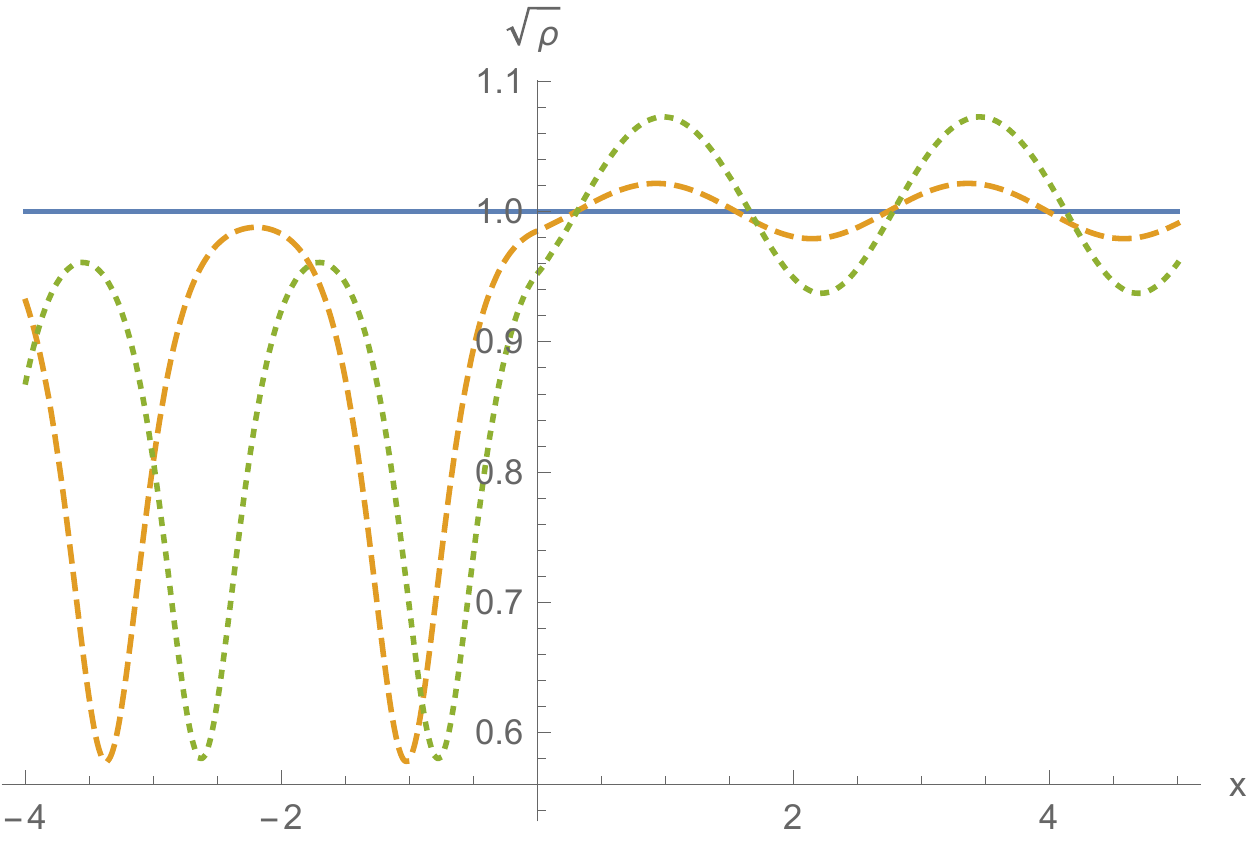}
\end{center}
\caption{We represent the profile of $\sqrt{\rho(x)}$ for three stationary transonic flows with $g_+ = 1$, $g_- = 8$, $\om - V_+ = 7/3$, $\om - V_- = 28/3$, and $J = \sqrt{8/3}$. 
These values are chosen as they admit a stationary, homogeneous, transonic solution with $\rho_0 = 1$ and a moderately large difference between the sound velocities on the two sides of the horizon. The sonic horizon is located on the discontinuity, at $x=0$. The continuous line corresponds to the unique homogeneous solution of Eqs.~(\ref{eq:rho0}, \ref{omc}). The two other flows (represented by a dotted and a dashed line) contain a soliton train on the subsonic (left) side, and an undulation on the supersonic (right) side. When the amplitude of the undulation goes to $0$, the distance between two consecutive solitons goes to infinity while their depth hardly varies.
} \label{fig:gensol}
\end{figure}

To prepare the analysis of transonic flows associated with space-dependent $g$ and $V$, it is useful to recall the main properties of the stationary solutions when $g$ and $V$ are independent of $x$. In this case, spatially bounded solutions of \eq{eq:eqf} exist if and only if
\be 
\label{Jmax}
J^2 \leq J_{\rm max}^2 \equiv \frac{8 \lp \om - V \rp^3}{27 g^2}.
\ee 
All bounded solutions are periodic and can be written using elliptic functions~\cite{Kamchatnov}. Two of them are homogeneous, and given by the two positive roots $\rho_p < \rho_b$ of the right-hand side of \eq{eq:eqf}. They merge when $|J| = J_{\rm max}$. The smallest one, $\rho_p$, describes a supersonic flow since the condensate velocity $ |J |/ \rho_p$ is larger than the speed of long-wavelength perturbations $c_p = \sqrt{g \rho_p}$. Instead, the flow described by $\rho_b$ is subsonic as $|J| / \rho_b < c_b = \sqrt{g \rho_b}$. Interestingly, the stationary perturbations on top of these two homogeneous solutions are radically different. Stationary flows with a density close to the supersonic value $\rho_p$ contain a zero frequency modulation of the density (hereafter called an undulation). Instead, flows with mean densities close to $\rho_b$ contain a train of solitons, see Fig.~\ref{fig:gensol}. 

Let us briefly explain how to get the main properties of these solutions. More details can be found in~\cite{Michel:2013wpa}. Multiplying \eq{eq:eqf} by $\sqrt{\rho} \pd_x \sqrt{\rho}$ and integrating the resulting equation gives 
\be \label{eq:GPErhpgen}
\frac{1}{4} \lp \pd_x \rho \rp^2 = g (\rho(x) - \rho_{1}) (\rho(x) -\rho_{2}) (\rho(x)-\rho_{3}),
\ee
where $\rho_{i}$, $i \in \left\lbrace 1,2,3 \right\rbrace$ are three constants. They are related to the current $J$ and frequency $\om$ through
\be 
\rho_{1} + \rho_{2} + \rho_{3} = 2 \frac{\om - V}{g}, \; \rho_{1} \rho_{2} \rho_{3} = \frac{J^2}{g}.
\ee
When there is no double root, the range of $\rho(x)$ is a connected component of the domain where the polynomial on the right-hand side of \eq{eq:GPErhpgen} is positive. The density $\rho$ is therefore asymptotically bounded only if that domain is itself bounded. This requires that the three constants $\rho_{i}$ are all real. We order them as $\rho_{1} \leq \rho_{2} \leq \rho_{3}$. Then, the two domains in which the right-hand side of \eq{eq:GPErhpgen} is positive are $\rho \in [\rho_1, \rho_2]$ and $\rho > \rho_3$. The second one corresponds to a divergent solution. The only bounded one thus corresponds to the first domain, and $\rho(x)$ oscillates between $\rho = \rho_1$ and $\rho = \rho_2$. 
The bounded solution is also characterized by a fourth parameter giving the phase of the oscillations. 

In the limit $\rho_2 \to \rho_1$, $\rho(x)$ becomes the homogeneous supersonic solution with density $\rho_p$ discussed after \eq{Jmax}. When $\rho_2$ is close to $\rho_1$, $\rho(x)$ describes a small-amplitude undulation on top of this homogeneous solution, see Fig.~\ref{fig:gensol}. In the limit $\rho_2 \to \rho_3$, one gets a homogeneous subsonic flow of density $\rho_b$, and when $\rho_2$ is close to $\rho_3$, $\rho(x)$ contains a train of widely-separated solitons. 

Interestingly, when $\rho_2 = \rho_3$ there exists another solution, called ``shadow soliton'', which diverges at a finite value of $x$ and also goes to the subsonic flow $\rho = \rho_3$ asymptotically. This last solution must be discarded when working with homogeneous functions $V$ and $g$ because of its diverging character. However, it must be included in the forthcoming analysis 
of the steplike case as the divergence can be erased by the change of $g$ and $V$. 

\subsubsection{Step-like potentials, unicity theorem}

To obtain stationary transonic flows, $g$ and/or $V$ must depend on $x$. To have simple solutions, we assume that $g$ and $V$ are given by 
\be \label{eq:step}
&& g(x) = \theta(-x) \, g_-  + \theta(x)\, g_+ , \;
\\ \nonumber
 && V(x) = \theta(-x)\,V_-  + \theta(x) \, V_+ ,
\ee
with $0 < g_+ < g_- $ and $V_+ > V_- $. Then, for any real value of $J$, there exists a globally homogeneous solution of \eq{eq:eqf} given by
\be \label{eq:rho0}
\rho_0 \equiv \frac{V_+ - V_-}{g_- - g_+}.
\ee
Its frequency $\om(J)$ increases with $J$ so that \eq{Jmax} is always satisfied 
on each side. 
Explicitly, one finds
\be 
\om(J)= V_- + g_- \rho_0 + \frac{J^2}{2 \rho_0^2} = V_+ + g_+ \rho_0 + \frac{J^2}{2 \rho_0^2}. 
\label{omc}
\ee
This solution is transonic iff 
\be \label{eq:BHcond} 
c_b \equiv \sqrt{g_- \rho_0} > \frac{|J|}{\rho_0} > \sqrt{g_+ \rho_0} \equiv c_p. \ee 
The flow is 
then subsonic for $x<0$ and supersonic for $x>0$. When $J > 0$, it corresponds to a black hole flow. This can be deduced from the large reduction of the wave vector experienced by counter-propagating waves, see next subsection and~\cite{Unruh:1980cg,Macher:2009nz,Barcelo:2005fc}. When $J$ and $v$ are negative, the transonic flow is analogous to a white hole. In that case, long wave length incoming modes are converted into short wave length modes. 

Interestingly, having fixed $g_+$, $g_-$, $V_+$ and $V_-$ of \eq{eq:step}, when $|J|$ obeys \eq{eq:BHcond}, the branch of solutions of \eq{omc} gives the {\it unique} AH transonic flow. Indeed, for this interval of $|J|$, all the other stationary bounded transonic solutions contain an undulation in the supersonic region and/or a train of solitons in the subsonic one, see Fig.~\ref{fig:gensol}. The uniqueness can be understood when considering the set of integration constants which characterize the general stationary solution. They are all fixed by the requirement that the solution be AH on both sides. 

\begin{figure}
\begin{center}
\includegraphics[width=0.49 \linewidth]{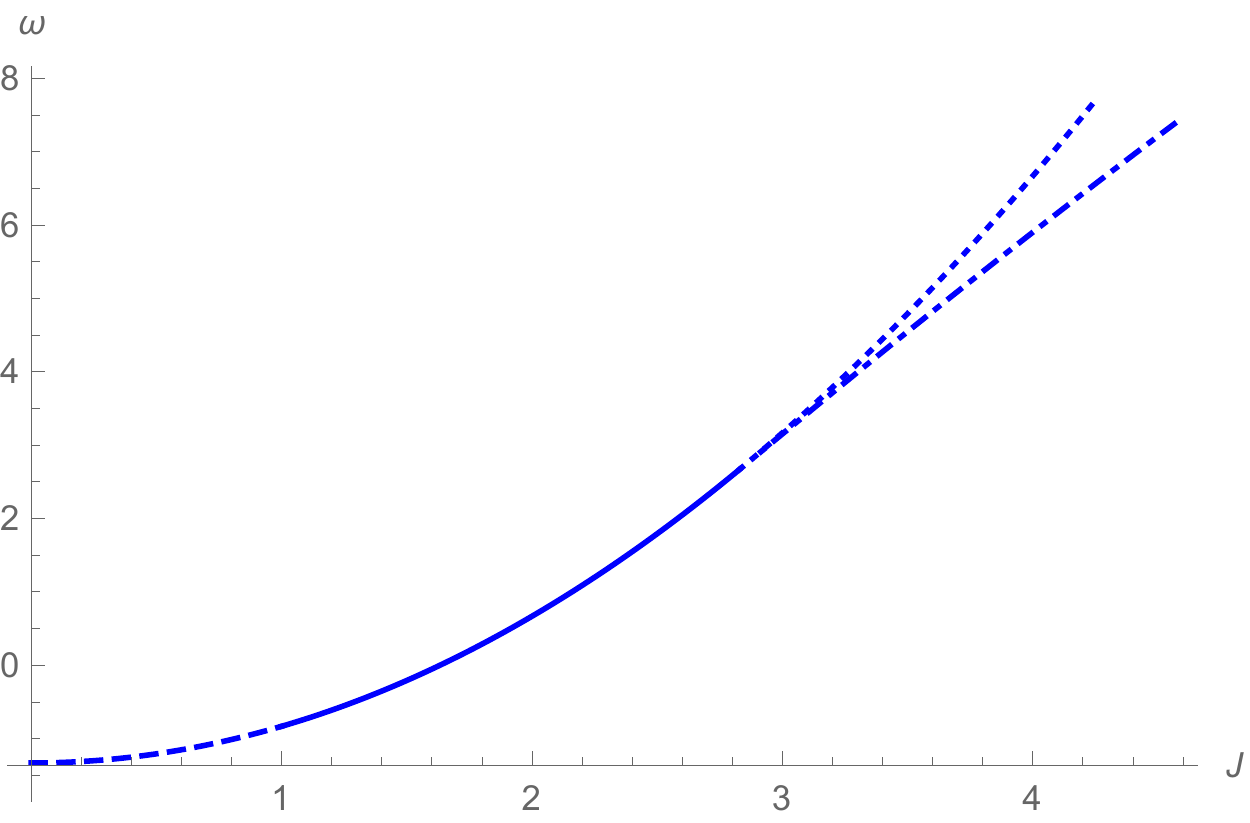}
\includegraphics[width=0.49 \linewidth]{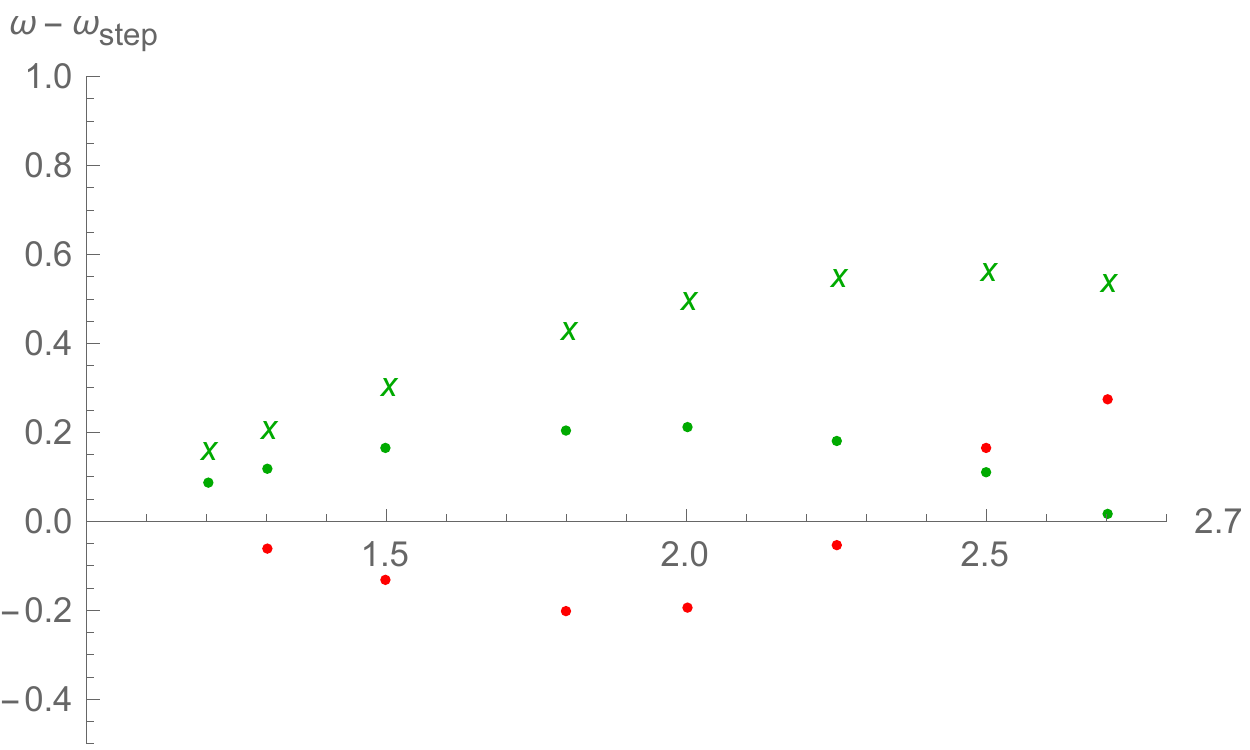} 
\end{center}
\caption{On the left panel, we show $\om(J)$ of the homogeneous solution and the ``waterfall'' configuration as functions of the current $J$ for the steplike potentials of \eq{eq:step}. The continuous line, which extends from $J = 1$ to $J = 2 \sqrt{2}$, corresponds to the homogeneous transonic flows. The dashed (respectively dotted) line corresponds to homogeneous solutions which are globally subsonic (respectively supersonic). The dot-dashed line corresponds to transonic waterfall solutions. $g_-$, $g_+$, $V_-$, and $V_+$ are the same as in \fig{fig:gensol}. The right panel shows the deviations of $\om(J)$ from its value in the steplike case, for some numerical solutions found with smooth $g(x)$ and $V(x)$ given by \eq{eq:vargV}. Red points correspond to $\sigma_V = 1.2, \,  \sigma_g = 0.5$, while green points and crosses correspond to $\sigma_g = 1.2,\,  \sigma_V = 0.5$. The eight green crosses show ``hairy'' solutions containing one soliton. Their density profiles are similar to that 
shown on the right panel of \fig{fig:flatsol}. 
} \label{fig:statsols}
\end{figure}

Yet, there is another series of stationary AH transonic flows. They contain a half-soliton for $x<0$, and correspond to the ``waterfall'' configuration studied in~\cite{PhysRevA.85.013621}. The corresponding value of $\om^{\rm wf}(J)$ is given by
\be
\om^{\rm wf}(J) = V_- + g_- \rho_-^{\rm wf} + \frac{J^2}{2(\rho_-^{\rm wf})^2} = g_+ \rho_+^{\rm wf} + V_+ + \frac{J^2}{2(\rho_+^{\rm wf})^2}, 
\label{omcwf}\ee
where $\rho_\mp^{\rm wf}$ is the density for $x \in \mathbb{R}^\mp$. Quite remarkably, these solutions are bounded only for values of $|J|$ larger than the upper bound of \eq{eq:BHcond}. When decreasing $J$ along the branch of waterfall solutions, the difference between the densities on the two sides of the horizon decreases. It vanishes precisely when $J = \sqrt{g_- \rho_0^3}$, at which point the homogeneous and the waterfall solutions coincide. Decreasing $J$ further, only the homogeneous solution of \eq{eq:rho0} remains bounded. The series of \eq{omc} and \eq{omcwf} are represented in \fig{fig:statsols}.
In brief, in the case of step like potentials, for all $|J|>  \sqrt{g_+ \rho_0^3}$, there is a unique AH transonic solution. Hence in this case, stationary transonic flows obey an analogous version of the black hole uniqueness theorem~\cite{Israel}.

These results may be understood as follows. The analysis of subsection~\ref{subsub:homo} applies on both sides of the horizon $x=0$. We thus have six integration constants $\rho_{i,\pm}$. Notice that these are not all independent when considering global solutions, since the current $J$ and frequency $\om$ take the same values on both sides. A global solution is in fact determined by 4 parameters, including $\om$ and $J$. Given these two quantities, the space of solutions is thus two-dimensional and parametrized by two integration constants. These fix the phase and amplitude of the undulation on the supersonic side, and the properties of the soliton train on the other side, as explained in \cite{Michel:2015pra}.

\subsubsection{Smooth potentials and hairy black holes}
 
When $V$ and $g$ smoothly vary with $x$, the above arguments are no longer sufficient to characterize the complete set of stationary bounded solutions of \eq{eq:GPE}. Yet, 
the set of AH stationary transonic flows remains discrete at fixed $J$. Indeed, imposing that the flow is AH and supersonic for $x \to \infty$ completely specifies a solution of \eq{eq:eqf} once $J$ and $\om$ are fixed. Then the condition that the solution be also AH for $x \to - \infty$ gives an extra condition, which constrains the frequency to belong to a 
(possibly empty) discrete set of branches $\om^i(J)$. (Notice that the homogeneous solution of \eq{eq:rho0} still exists if the variations of $g$ and $V$ are tuned in such a way that $\pd_x g / \pd_x V$ be a constant.) 

\begin{figure}
\begin{center}
\includegraphics[width=0.49\linewidth]{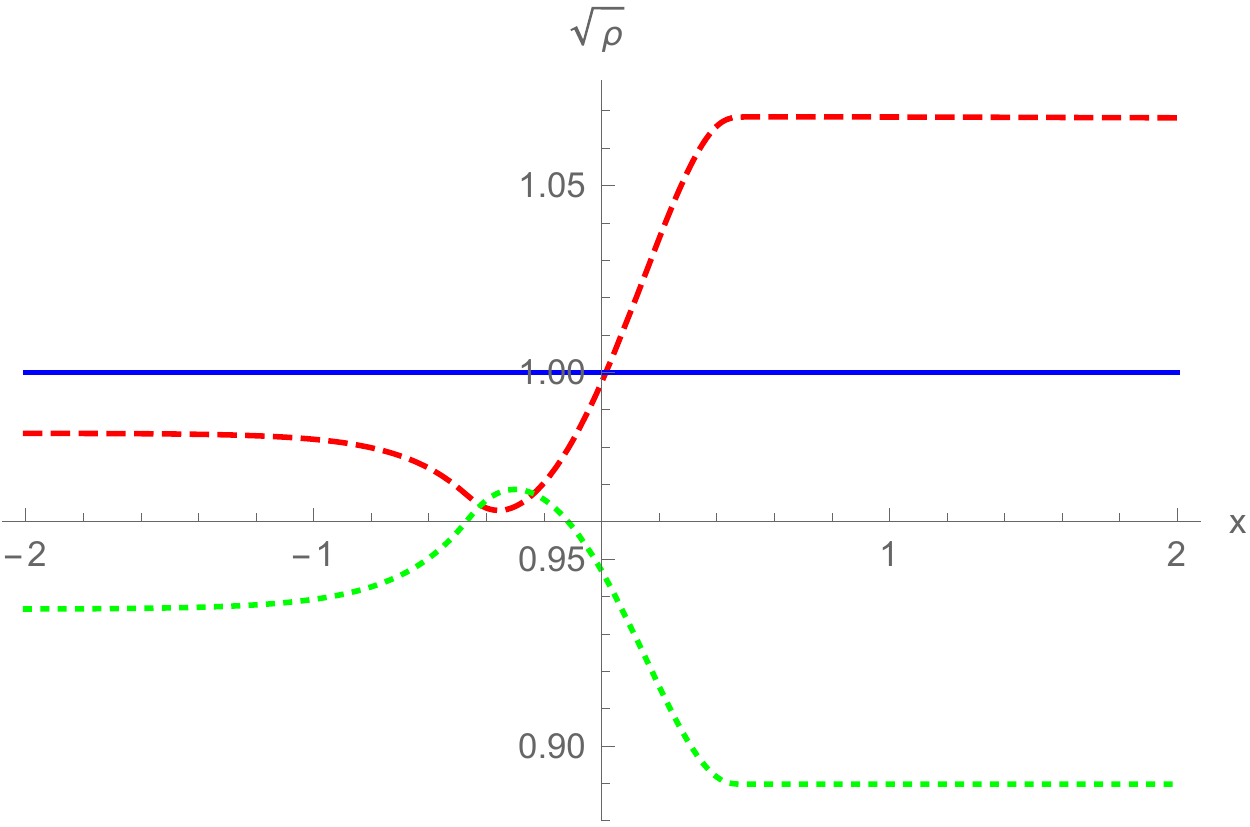}
\includegraphics[width=0.49\linewidth]{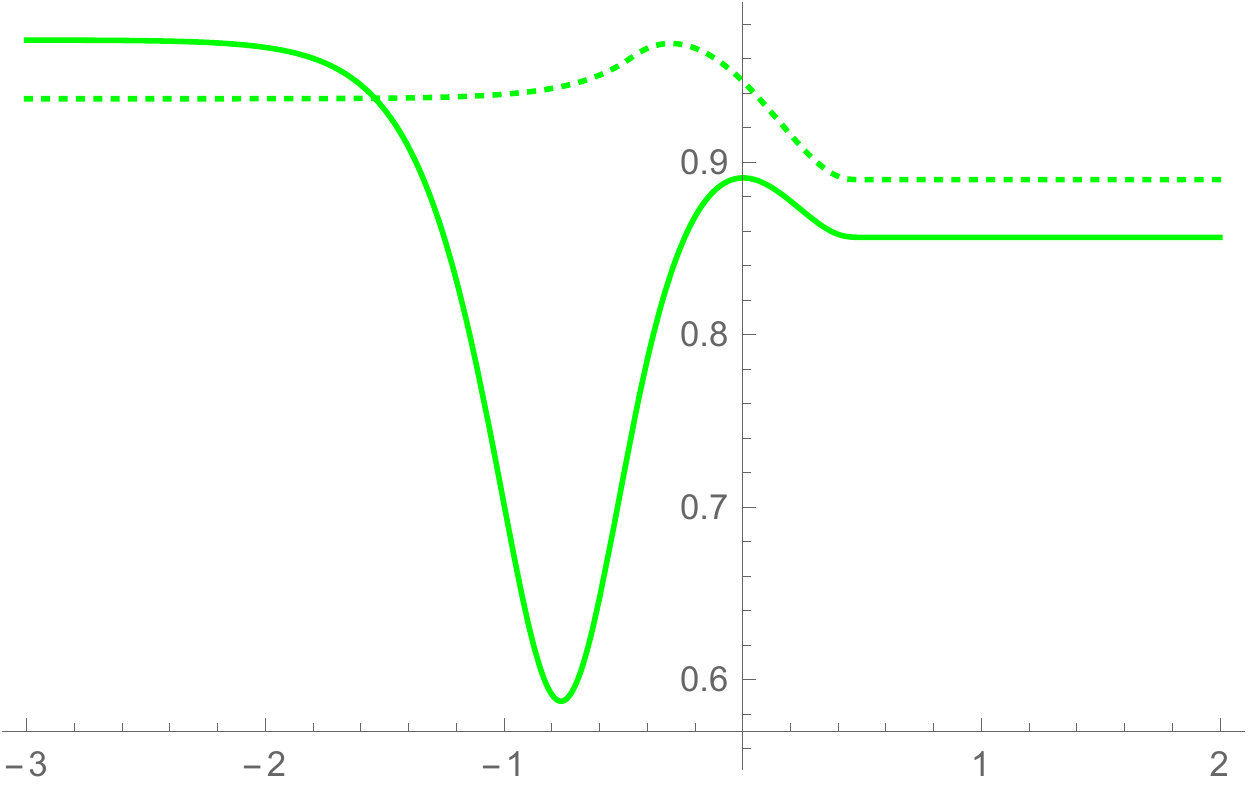} 
\end{center}
\caption{On the left panel we show the density profiles $\sqrt{\rho(x)}$ of stationary AH transonic flows obtained using the continuous profiles of \eq{eq:vargV} with different slopes $1 / \sigma_g$ and $1 / \sigma_V$. The values of $g_\pm$, $V_\pm$, 
and $J$ are the same as those used in Fig.~\ref{fig:gensol}. Hence the modifications are due to the finiteness of $\sigma_g$ and $\sigma_V$. The blue, continuous line corresponds to $\sigma_g = \sigma_V = 1$, the red, dashed one to $\sigma_V = 1.2, \,   \sigma_g = 1$, and the green, dotted one to $\sigma_V = 1, \,\sigma_g = 1.2$. On the right panel, we show the hairy solution (continuous) and the main solution (dotted) obtained with the last set of parameters. The hairy solution contains a nearly complete soliton attached to the horizon.} \label{fig:flatsol}
\end{figure}

To 
generalize the step like potentials of \eq{eq:step}, we considered the continuous profiles 
\be \label{eq:vargV}
&& g(x) = g_-  + (g_+ - g_-) \theta(x + \sigma_g / 2)\,[ \theta(\sigma_g/2 - x ) \lp \frac{x}{\sigma_g}  + \frac{1}{2}\rp +  \theta(x- \sigma_g/2)] , 
\\ \nonumber
 && V(x) = V_-  + (V_+ - V_-) \theta(x + \sigma_V /2)\,[ \theta(\sigma_V/2 - x ) \lp \frac{x}{\sigma_V} + \frac{1}{2}\rp+  \theta(x- \sigma_V/2)] , 
\ee
characterized by the domains $\sigma_g$ and $\sigma_V$ where $g$ and $V$ linearly interpolate 
between their asymptotic values. 
To obtain the stationary flows, we numerically solved \eq{eq:GPErhpgen}. As expected, we found AH solutions which are smoothly connected to those of the former subsection in the limit $\sigma_{g}, \sigma_{V} \to 0$. More interestingly, we also found a series of ``hairy'' AH solutions when $\sigma_V$ is smaller than $\sigma_g$. These solutions are disconnected from those of the main series because they contain an almost complete soliton which is attached to the sonic horizon, and which mainly lives in the subsonic region, see \fig{fig:flatsol}. Our numerical study indicates that its center is rejected at $x \to -\infty$ in the steplike limit, or in the limit where $\sigma_V = \sigma_g$. Our simulations also indicate that this solution is unstable under nonlinear perturbations, which may trigger the emission of the soliton at infinity. We hope to characterize this instability in a future work. On the right panel of \fig{fig:statsols},  we show the values of $\om(J)$ for the AH solutions obtained with \eq{eq:vargV}. The red dots describe solutions of the main series with $\sigma_V = 1.2, \,  \sigma_g = 0.5$. The green dots (crosses) correspond to hairy solutions obtained with $\sigma_g = 1.2, \,  \sigma_V = 0.5$. It is clear that these solutions belong to two different series of $\om(J)$. In \fig{fig:flatsol} we represent the density profile of three solutions of the main series and one hairy solution with $(\sigma_g,\sigma_V) \in \left\lbrace (1,1),(1,0.83),(0.83,1) \right\rbrace$.

In conclusion, when dealing with smooth profiles for $g$ and $V$, one always finds the main series of AH solutions characterized by a smooth density profile across the horizon. In addition, a discrete number of hairy solutions can exist. When we found such solutions, we observed that they are disconnected from the solutions of the main series because they contain some solitonic configuration attached to the region where the potential varies. 

\subsection{Linear stability of black hole flows} 
\label{sec:lin}

When considering the stability of transonic flows, it appears that those corresponding to black holes are much more stable than those describing white holes. This is non trivial because the S-matrix describing the scattering of linear perturbations on a white hole horizon is the inverse of that describing the scattering on the time reversed black hole flow~\cite{Macher:2009nz}. This relation implies that white hole flows 
possess the same degree of stability as black holes, namely, they are dynamically stable. (This means that the spectrum of linear perturbations contains no complex frequency modes.)  Yet, the late time evolution of perturbations scattered on a sonic horizon 
strongly depends on whether the modes experience a red-shift (black hole), or a blue-shift (white hole).
White hole flows are studied in Section~\ref{Sec:WH} where it is shown that they exhibit a linear infra-red instability. 
 In the present subsection we consider the linear stability properties of black-hole flows and show that they are stable under linear perturbations. 

\begin{figure}[h]
\begin{center}
\includegraphics[width=0.49 \linewidth]{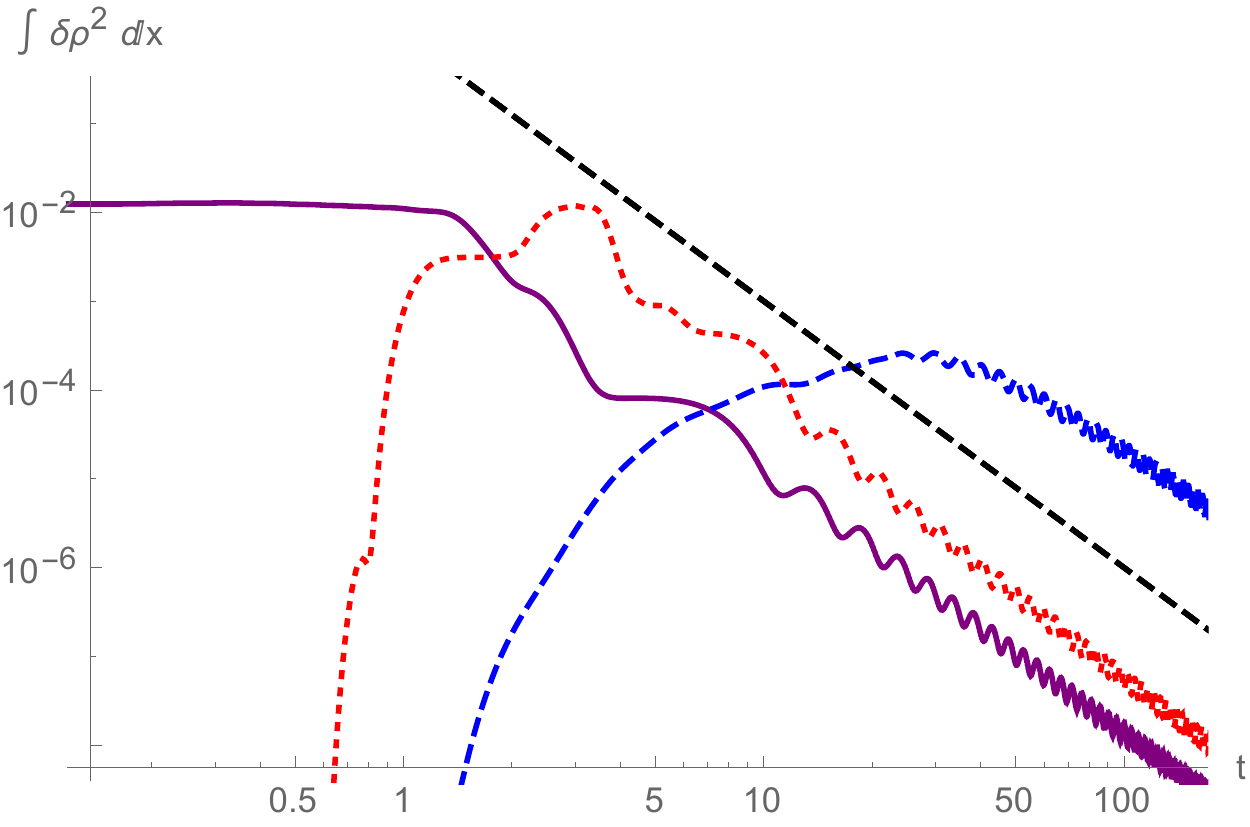} \includegraphics[width=0.49 \linewidth]{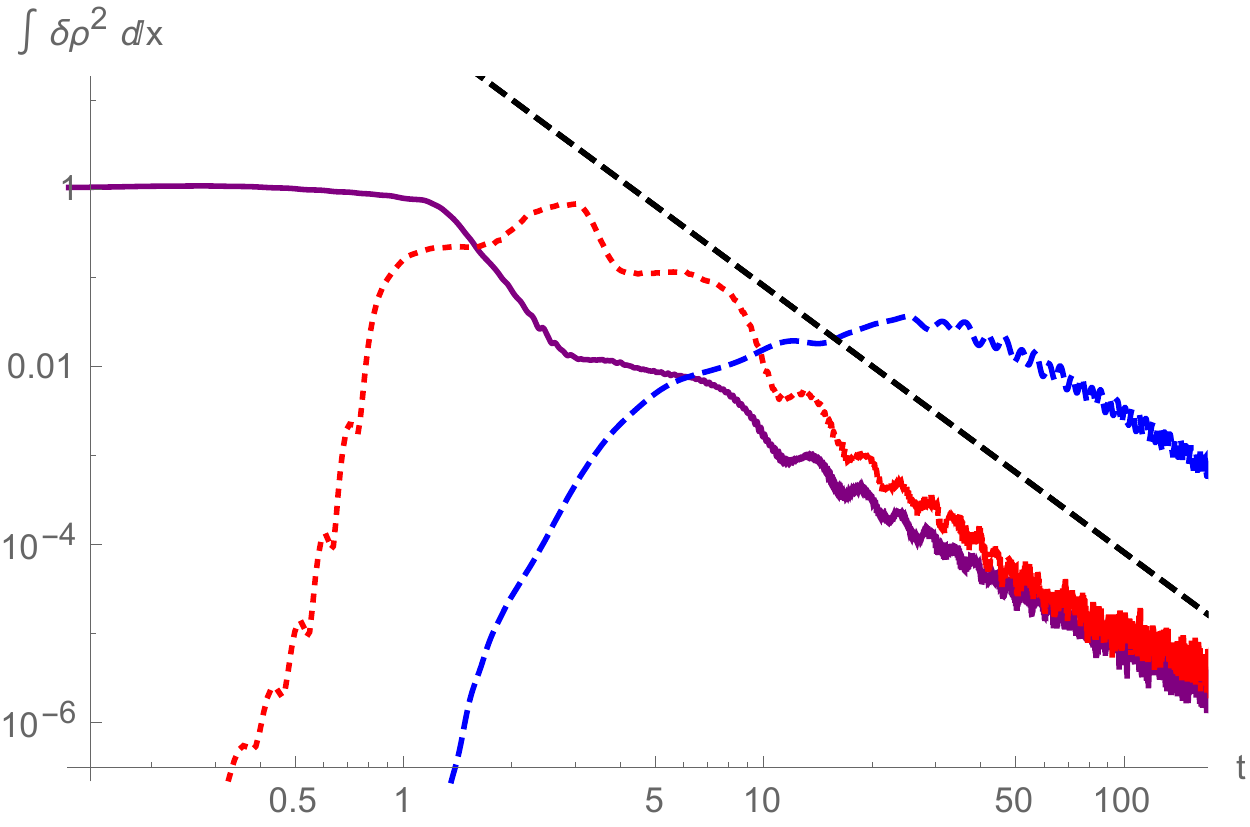} \end{center}
\caption{We show the evolution of the squared density perturbation, integrated between $x=-5$ and $x=5$, 
due to the scattering of three wave packets on a homogeneous black hole flow.
The perturbation is initially located in the supersonic region (red, dotted), in the subsonic one (blue, dashed), or centered on the horizon (continuous, purple). The initial value of their amplitude is $\delta \rho = 0.1$ on the left plot, and $\delta \rho = 0.9$ on the right plot. 
The oblique dashed line shows $t^{-3}$ 
(left) and $81 t^{-3}$ (right). When comparing the two plots, 
besides the factor of $9^2$ which relates the initial values of $\delta \rho^2$ one clearly sees that the behaviors are very similar.
The late-time oscillations have an angular frequency of $2 \om_{\rm max}$, 
and are due to the vanishing of the group velocity at $\om=\om_{\rm max}$, see \fig{fig:DR}. The background flow parameters 
are $g_+ = 0.86$, $g_- = 8.27$, $\om - V_+ = 2.19$, $\om - V_- = 9.61$, 
and $J = \sqrt{8/3}$. Their values have been chosen so that $\rho_{2,-} = \rho_{3,+} = 1$, $\rho_{2,+} = 2.25$, and $\rho_{3,-} = 0.49$.} \label{fig:lin3/2}
\end{figure}

To study the behavior of perturbations in qualitative terms, we present the results of some numerical simulations. We choose a spatial domain $I \subset \mathbb{R}$ containing the sonic horizon and look at the late-time evolution of the perturbations in $I$. Numerical simulations indicate that perturbations decay as $t^{-3/2}$ up to logarithmic factors. These observations hold for all AH solutions, both in a steplike potential of \eq{eq:rho0} and in smooth $V(x)$ and $g(x)$. To represent the various typical cases,  we consider the scattering of a gaussian perturbation $\delta \rho$ which is initially localized in the subsonic region, on the sonic horizon, or in the supersonic region. 
On the left panel of \fig{fig:lin3/2}, we work in the linear regime and study the evolution of a  perturbation of relative amplitude  $\delta \rho/\rho_0$ of order $0.1$ and of width equal to $1$ (in units of the healing length). 
On the right panel, to show that similar results hold for much larger perturbations, 
we work with $\delta \rho/\rho_0$ of order $1$ for the same spatial width. Then, we show the time dependence of the integral of $\delta \rho^2(t,x)$ over the segment $[-5,5]$, which contains the sonic horizon localized on $x=0$. At early times, the density fluctuation increases as the perturbation enters the integration region, or remains roughly constant when the perturbation is already in the integration region at $t=0$. At late times, in the three cases, 
one clearly sees that the average of $\delta \rho^2$ goes to zero as $t^{-3}$. 
The fact that 
the temporal behavior is very similar on the left panel (linear regime)
and on the right panel (nonlinear regime since $\delta \rho/\rho_0 \sim 1$),
reveals that the decay law  of $\delta \rho^2$ in $t^{-3}$ is extremely robust. It implies that all perturbations are diluted away so as to give at late time an AH stationary solution.

This behavior is similar to that of 
linear perturbations propagating in a black hole metric~\cite{Misner1973}. We briefly present at the end of the Section the nature of the correspondence between the present case and relativistic fields. We now give the analytical elements needed to understand this behavior. The full calculation is done in Appendix~\ref{App:linKdV} for the linearized KdV equation and a square perturbation.

\begin{figure}
\includegraphics[width=0.49 \linewidth]{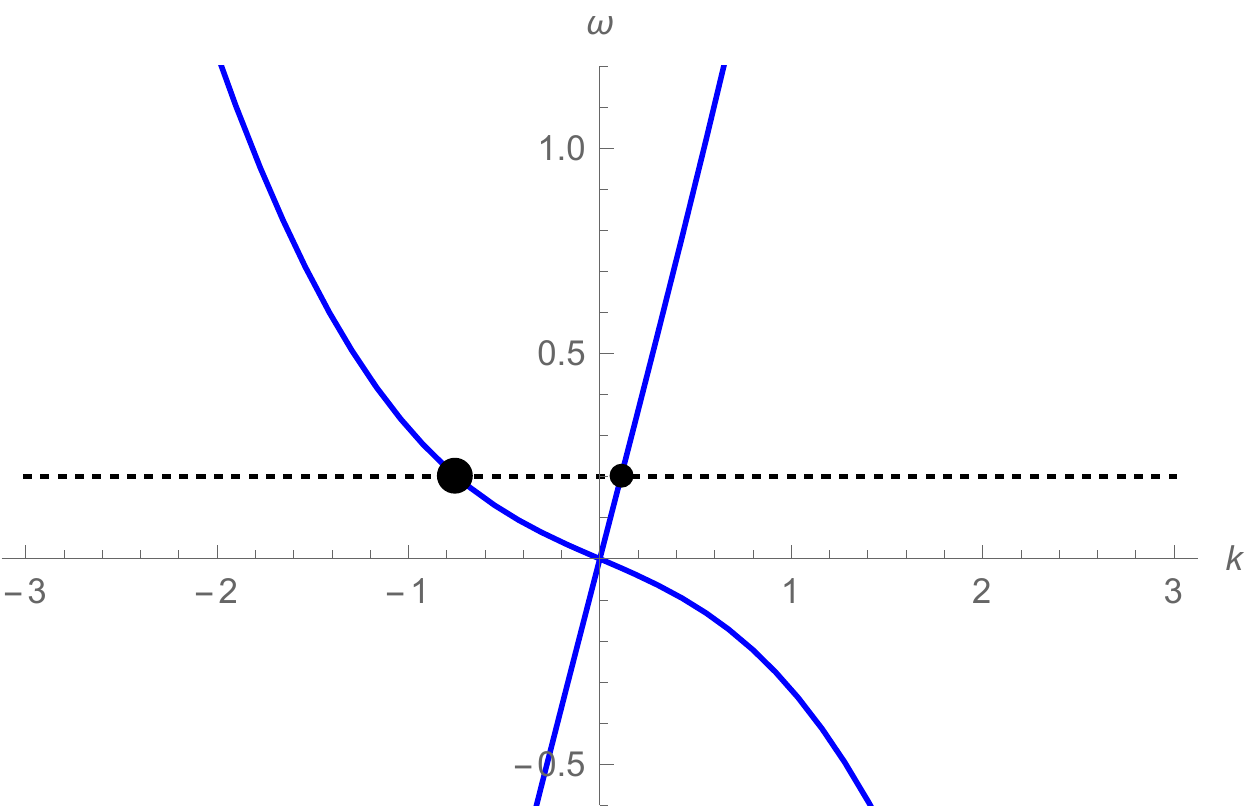}
\includegraphics[width=0.49 \linewidth]{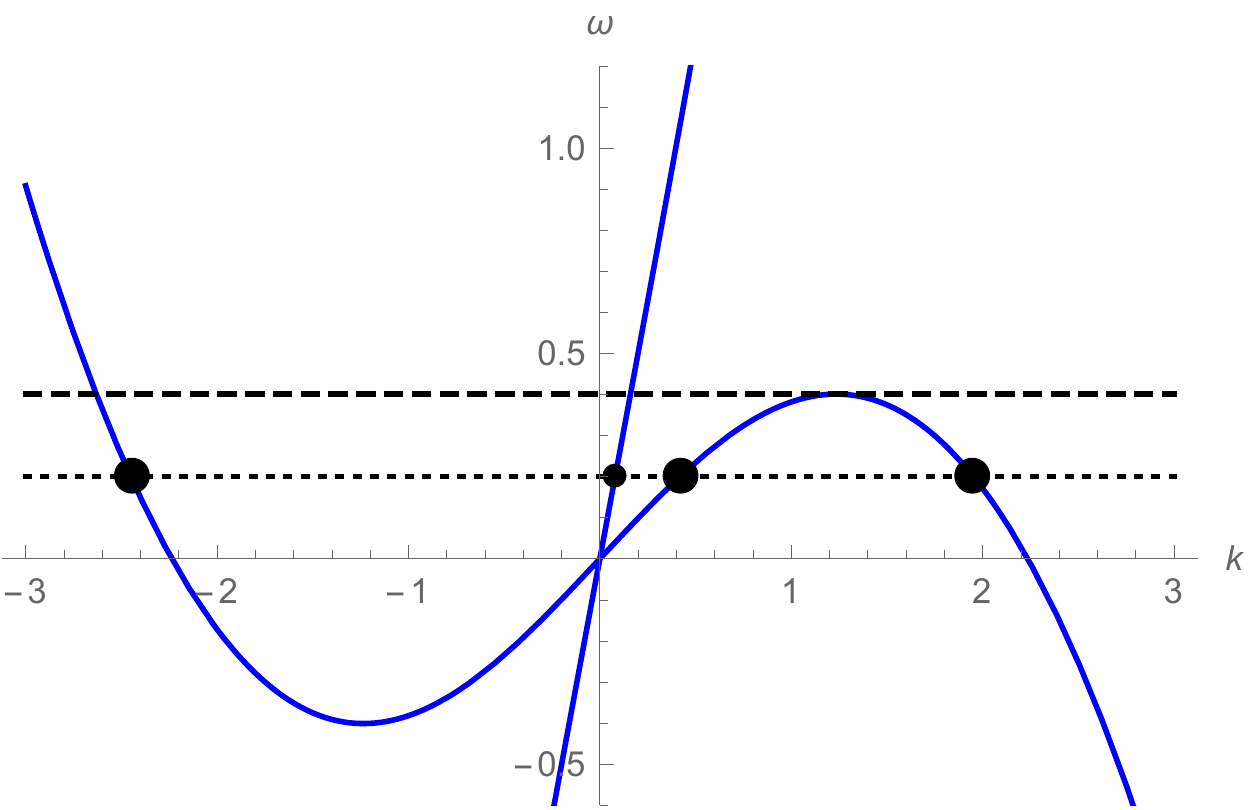}
\caption{Dispersion relation $\om$ versus $k$ of \eq{eq:disprelBdG} for $g \rho= 1$ and $v=0.8$ (left) and $1.5$ (right). The horizontal dashed line on the right plot shows the critical frequency $\om_{\rm max}$. The dotted line shows $\om = \om_{\rm max} / 2$. Large dots show the wave vectors of the counter-propagating modes (in the fluid frame), which give the dominant part of the scattering. The smaller dot on each panel shows the co-propagating mode, which only mildly affects the scattering~\cite{Macher:2009tw}. On the right panel, the two external roots with $k \sim -2.5$ and $k \sim 2$ describe dispersive (short wave length) waves. The 4 other roots describe phonic excitations, and propagate with a group velocity (in the lab frame) given by $v_{\rm gr} \approx v \pm c$. 
} \label{fig:DR}
\end{figure}

We look for perturbed solutions of the form
\be 
\psi(x,t) = \psi_0 (x) \, e^{- i\omega t} \lp 1 + \phi(x,t) \rp, 
\label{relpert}
\ee
where $\psi_0 = \sqrt{\rho_0(x)} e^{i \int^x v_0(x') dx'} $, see \eq{eq:Madelung}.
To first order in $\phi$, \eq{eq:GPE} gives
\be \label{eq:BdG} 
i \pd_t \phi(x,t)  = -\frac{1}{2} \pd_x^2 \phi(x,t) - \frac{\pd_x \psi_0(x)}{\psi_0(x)} \pd_x \phi(x,t) + g(x) \rho_0(x) \lp \phi(x,t) + \phi(x,t)^* \rp.
\ee 
Asymptotically, on either side, the background quantities $\rho_0$ and $v_0$ are constant. Hence we can look for solutions of the form 
\be \label{eq:formlin}
\phi_k(x,t) = \mathcal{U}_k \, e^{i (k x - \om t)} + \mathcal{V}_k^* \, e^{-i (k x - \om t)}. 
\ee
Plugging this form into \eq{eq:BdG}, we obtain a system of two linear equations on $\mathcal{U}_k$ and $\mathcal{V}_k$, which has non-trivial solutions if the dispersion relation
\be \label{eq:disprelBdG} 
(\om - v_0 k)^2 = g \rho_0 k^2 + \frac{k^4}{4}\, 
\ee 
(shown in \fig{fig:DR}) is satisfied. For $\om \in \mathbb{R}$, there are two roots in $k$ in the subsonic region. In the supersonic region, when $\abs{\om}$ is smaller than a critical frequency $\om_{\rm max}$, see Eq.~(25) in \cite{Macher:2009nz}, there are two additional roots. For $0 < \om < \om_{\rm max}$, there are thus six asymptotic plane waves (two in the subsonic region and four in the supersonic one), which give the asymptotic behavior of the globally defined bounded modes. These modes span a three-dimensional vector space. The scattering process is characterized by the relationship between the incoming and outgoing modes. The in (out) basis is composed of the three modes which asymptotically contain one wave with group velocity oriented towards (away from) the sonic horizon. These two bases are related by a $3\times 3$ Bogoliubov transformation. The behavior of its coefficients and the link with the Hawking effect can be found in~\cite{Macher:2009nz,Macher:2009tw}. 

We consider some initial perturbation $\phi_{\rm ini}(x) = \phi(x,t=0)$ and look for its late-time behavior. We assume that $\phi_{\rm ini}$ decays fast enough at infinity so that the coefficients of its expansion into incoming modes are finite. In each of the asymptotic regions, $\phi(x,t)$ has the form
\be 
\phi(x,t) = \int \sum_i \lp \mathcal{U}_{k_i} e^{i (k_i x - \om t)} + \mathcal{V}_{k_i}^* e^{-i (k_i x - \om t)}  \rp d \om,
\ee
where the subscript $i$ labels the branches of the dispersion relation. The coefficients $\mathcal{U}_{k_i}$ and $\mathcal{V}_{k_i}$ are determined by the overlap between $\phi_{\rm ini}$ and incoming modes, and by the expansion of the latter into asymptotic plane waves. Possible divergences in the expansion into outgoing modes come only from those of the 9 Bogoliubov coefficients. These were computed analytically in the steplike regime in~\cite{Finazzi:2012iu} and numerically for a smooth flow in~\cite{Macher:2009nz}. In both cases, it was found that the only divergence occurs for $\om \to 0$, and only affects the coefficients relating long-wavelength to short-wavelength modes. However, as explained in Appendix~\ref{App:linKdV}, the leading terms in these coefficients do not contribute to the amplitude of $\phi_0$  as they come in pairs with opposite signs which cancel each other. At late times, $\phi$ can be computed using a saddle-point approximation. The saddle-points are located where
\be 
\frac{d \varphi}{d \om} + \frac{d k_i}{d \om} x - t = 0.
\ee 
In this expression, $\varphi$ represents the phase of the prefactor ($\mathcal{U}_{k_i}$ or $\mathcal{V}_{k_i}$). In the limit $t \to \infty$ at fixed $x$, we must thus look for points where $\left\lvert d k_i / d \om \right\rvert \to \infty$, 
i.e., where the group velocity vanishes. (If the initial conditions are smooth, divergences in $\frac{d \varphi}{d \om}$ can arise only from terms in $k_j x_{i,j}(\om)$ in the expression of $\varphi$, where $x_{i,j}(\om)$ is a smooth function, coming from the integral giving the overlap, analogous to \eq{eq:A} for the KdV equation. Hence the divergences of  $\frac{d \varphi}{d \om}$ can only arise through divergences of $\frac{d k_i}{d \om}$.) Such divergence comes only at $\om = \pm \om_{\rm max}$, for the two roots which merge there. A straightforward calculation shows that close to this point $dk_i/d\om$ diverges as $\lp \om_{\rm max} - \abs{\om} \rp^{-1/2}$. Performing a Gaussian integration, we obtain that the dominant waves decrease as $t^{-3/2}$. 
This confirms that the dominant waves at late times have a frequency near $\om_{\rm max}$, in accordance with 
the numerical results of \fig{fig:lin3/2}. 
In principle, this linear analysis is valid provided ${\rm sup}_{x \in\mathbb{R}}\abs{\phi(x)} \ll 1$ at all times. However, numerical simulations indicate that the late-time behavior remains the same even for ``large'' initial perturbations, with ${\rm sup}_{x \in\mathbb{R}}\abs{\phi(x)}$ close to $1$ (see Fig.~\ref{fig:DR}, right panel). This is due to the dilution of the perturbation: as explained in the next section, homogeneous black hole solutions are stable, so that initial nonlinear perturbations will generally decay in time. Hence the perturbation close to the horizon effectively enters the linear domain at late times. 
In brief, numerical simulations and the above discussion establish that the AH black hole flows are 
linearly stable and that, in the vicinity of the sonic horizon, all linear perturbations decay as $t^{-3/2}$. This is very reminiscent to the no hair theorem of general relativity~\cite{Misner1973}.

To conclude this subsection, we briefly sketch the link between the decay we obtained and that exhibited by free scalar fields propagating in black hole metric. 
The comparison should be done at two different levels, 
namely when including or not the dispersive terms of \eq{eq:disprelBdG}.
When ignoring dispersion effects and the quantum pressure, the simplest way to proceed consists in separating $\phi$ into its real and imaginary parts~\cite{Barcelo:2005fc}. Straightforward algebra shows that the imaginary part of $\phi$ obeys
\be 
\partial_\mu \lp F^{\mu \nu} \partial_\nu \Im \phi \rp = 0,
\label{KGe}
\ee
where $\mu,\nu$ are $0,1$, and where Einstein's summation conventions are used. The matrix $F^{\mu \nu}$ is given by
\be 
F^{\mu \nu}(x) = 
\begin{pmatrix}
-1 & -v_0(x) \\ 
-v_0(x) & c_0^2(x) - v_0^2(x) , \end{pmatrix}
\ee
where we introduced the local value of the sound speed $c_0(x) = \sqrt{g(x) \rho_0(x)}$. Ignoring some subtleties related to the conformal part of the metric in 1+1 dimensions, \eq{KGe} is the d'Alembert equation in a stationary black hole with line element given by
\be
ds^2 = - c_0^2(x) dt^2 + (dx - v_0(x) dt)^2. 
\ee
When the flow is transonic, it describes a black hole metric and the event horizon is located where $|c_0(x)| = |v_0(x)|$. The analogy goes deeper 
as the ``analog'' surface gravity given by $\kappa = \partial_x (c_0 - v_0)$ (evaluated on the horizon) 
governs the red-shifting of the solutions of \eq{KGe}
which propagate against the flow. Namely the wave number $k$ of a localized wave packet leaving the near horizon region is red-shifted according to $k = k_0 e^{- \kappa t}$~\cite{Macher:2009nz}, exactly as found in General Relativity.
At the classical level, this guarantees that the decay of the solutions of \eq{KGe} will be similar to that 
found in general relativity~\cite{Misner1973}. Similarly, at the quantum level, 
the scattering of vacuum fluctuations on the sonic horizon should produce a thermal flux 
with a temperature given by the standard expression $k_B T_H =  \hbar \kappa/2 \pi$~\cite{Unruh:1980cg}.

\begin{figure}
\includegraphics[width=0.5 \linewidth]{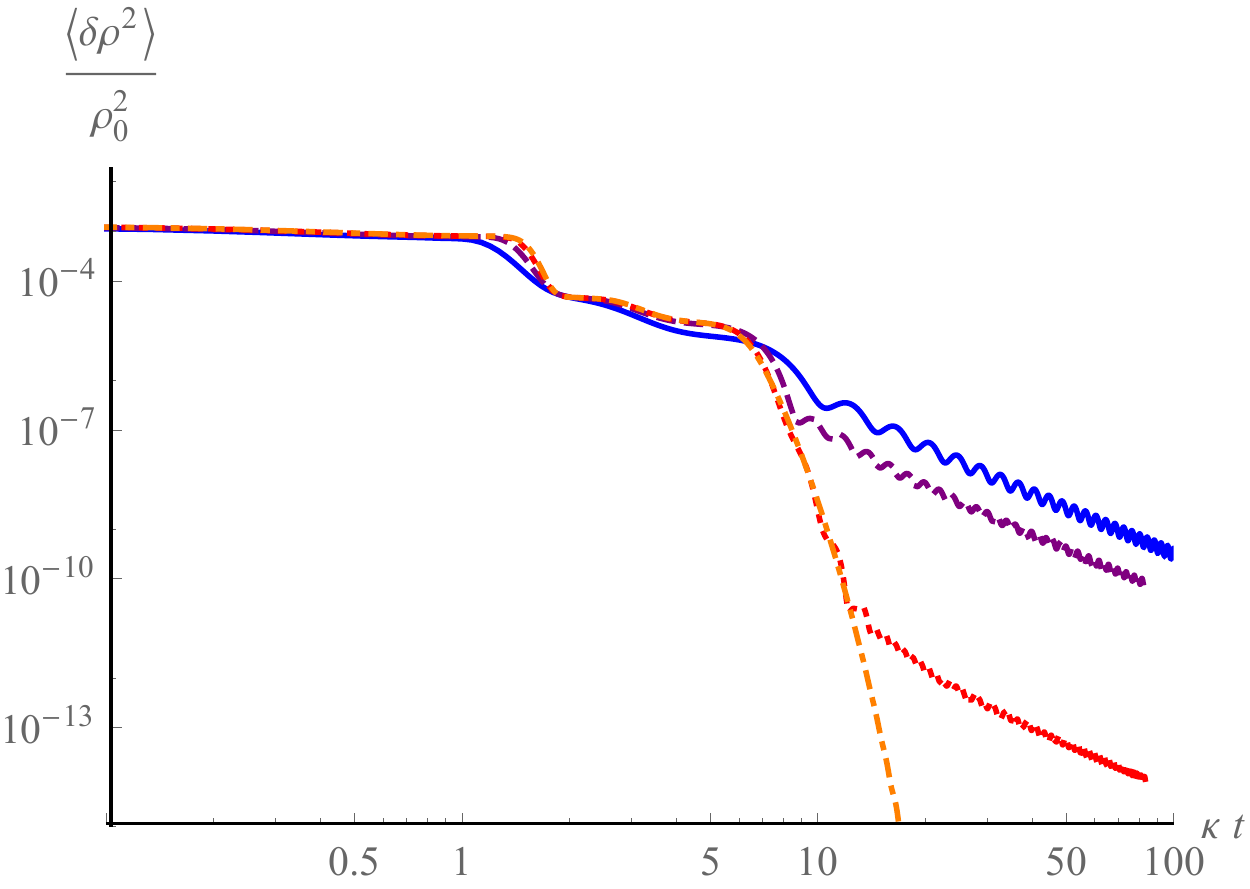}
\caption{We show the evolution of the squared relative density perturbation for an initially centered perturbation with the same parameters as in the left plot of \fig{fig:lin3/2}. We consider four different background flows
such that the quantities entering \eq{KGe} are held fixed. 
The four values of $\kappa/\omega_{\rm max}$ are $1.3$ 
(blue, continuous), $0.65$ (purple, dashed), $0.33$ (red, dotted), and $0.15$ (orange, dot-dashed). The functions $g$ and $\mu$ are given by \eq{eq:tanh}, where the asymptotic values are the same as in \fig{fig:lin3/2} and $\sigma = 0.56 c_0(x=0) / \kappa$.
}
\label{fig:varommax}
\end{figure}

Yet, to reproduce the late time decay law in $t^{-3/2}$, the dispersive effects of \eq{eq:disprelBdG}
must be taken into account, as they are governed by the behavior near the critical 
frequency $\omega_{\rm max}$.\footnote{When including them, the separation into real and imaginary parts of $\phi$ is no longer useful as they now couple to each other, see Eq.~(A6) in~\cite{Macher:2009nz}. In fact, this coupling complicates the relationship between the present settings and those characterizing dispersive fields in the context of alternative theories of gravity which break the local Lorentz invariance at short distances, see e.g.~\cite{Jacobson:2010mx}.
The interested reader might consult the appendixes A and B of \cite{Macher:2009nz} for further discussions.} 
To clarify the role of $\omega_{\rm max}$, it is interesting to consider the decay represented in Fig.~\ref{fig:lin3/2} when sending $\omega_{\rm max} \to \infty$ while keeping fixed all quantities appearing in \eq{KGe}. In \fig{fig:varommax} we represent the decay of the averaged squared density fluctuations for four different values of $\kappa/\omega_{\rm max}$, namely $1.3$ 
(blue, continuous), $0.65$ (purple, dashed), $0.33$ (red, dotted), and $0.15$ (orange, dot-dashed). To compare the four cases, the time is given in units of $\kappa$. For the same reason, the vertical axis represents the 
squared relative density perturbation $(\delta \rho/\rho_0)^2$, averaged over a domain which scales as $c_0/\kappa$ in the unit system used so far, see \eq{eq:GPE}. We clearly see that decreasing $\kappa/\omega_{\rm max}$ leaves the early evolution unchanged. This means that the decay 
of linear density perturbations at early times is governed by the relativistic equation \eqref{KGe}. 
At late-time we find the decay law in $t^{-3}$. We verified that the value of $(\delta \rho/\rho_0)^2$ when it starts to decay in $t^{-3}$ is  proportional to the squared of the Fourier component of the initial value of the relative
density perturbation with the critical wave vector $k_{\omega_{\rm max}}$. Hence, even when working at fixed $\kappa/\omega_{\rm max}$, the evolution of smooth perturbations (i.e., with no Fourier component for $k > k_{\omega_{\rm max}}$) is governed by \eq{KGe}. This is guaranteed by the fact that the propagation of the perturbation in the black hole flow redshifts the wave vectors $k$. In brief, \fig{fig:varommax} establishes that the dispersive effects of \eq{eq:disprelBdG} reduce the decay rate of perturbations near a black hole horizon when compared with the two-dimensional relativistic result. Importantly, this new behavior confirms that the hair
is lost at late time. 

\section{Nonlinear stability of black hole flows}
\label{sec:NLstab} 

\subsection{Analytical results} \label{sec:analytical} 

We now turn to the nonlinear evolution of initial perturbations on transonic black hole flows. In this subsection, we show analytical results obtained using Whitham's modulation theory~\cite{Whitham1}. Numerical results, which do not rely on the assumptions of this method, are shown afterwards. The reader interested in the derivation of Whitham's equations for the problem at hands may consult Appendix~\ref{App:Whitham} and the textbook~\cite{Kamchatnov}. 

Whitham's modulation theory offers a general scheme for finding approximate quasiperiodic solutions of (quasi) integrable nonlinear partial differential equations. The solutions we consider oscillate on a short scale while their amplitudes, mean values and wavelengths vary on much larger ones. The key idea is to average some of the conservation laws of the equation over the fast scale to obtain coupled equations for the slow evolution of a set of effective parameters, called \textit{Riemann invariants}, which describe the solution locally. This method is particularly useful when the equation is integrable by the inverse scattering method. As reminded in Appendix~\ref{App:Whitham}, the AKNS scheme then provides a general way to find the set of Riemann invariants. To the best of our knowledge, \eq{eq:GPE} with $x$-dependent functions $V$ and $g$ is not integrable. However, since Whitham's equations are local, when working with $V$ and $g$ of \eq{eq:step}, one can use the integrability for uniform $V,g$ to determine the solutions on each side of the discontinuity at $x=0$. The globally defined solutions are then obtained by using the matching conditions, i.e., continuity of $\psi$ and $\pd_x \psi$. 

We look for the time dependence of solutions characterized by asymptotic densities $\rho_\pm$ and velocities $v_\pm$ as $x \to \pm \infty$ different from those of the solution given by \eq{eq:rho0}. Our aim is to show that, for a wide range of values of $(\rho_\pm, v_\pm)$, the solution locally converges to that particular homogeneous solution. Doing so we shall first extend the stability analysis of the previous section by including nonlinear effects, as well as perturbations extending to infinity. Second, we shall obtain the main features of the emission process which progressively replaces the initial configuration by the homogeneous and stationary black-hole flow of \eq{eq:rho0}. We shall see that, in a large domain of parameter space, three macroscopic, nonlinear, scale-invariant waves are emitted. These can be conceived as the result of a nonlinear stimulated Hawking radiation. As shown below, these features are well reproduced by numerical simulations and, along with the above linear analysis, they provide a precise description of the late-time behavior of the solution when approaching the homogeneous black hole flow. They confirm that the solution given by \eq{eq:rho0} acts as an attractor. 

In this work we consider approximate solutions described by at most 4 Riemann invariants. Since the 
steplike $x$-dependence of $V$ and $g$ introduces no scale, one can further restrict our attention to scale-invariant solutions, for which the Riemann invariants depend only on $z \equiv x/t$. Our goal is to find the domain of parameter space in which the time-dependent solution interpolates between given asymptotic values of $\rho$ and $v$ as $x \to \pm \infty$, and a homogeneous black-hole solution in some spatial interval $I_t$ which contains $x=0$ and grows linearly in time. That is, we look for the solutions of the Whitham equations~(\ref{eq:Whz}) describing functions $\rho$ and $v$ such that
\begin{itemize}
\item $\rho(x,t) \to \rho_+$ and $v(x,t) \to v_+$ for $z \to +\infty$,
\item $\rho(x,t) \to \rho_-$ and $v(x,t) \to v_-$ for $z \to -\infty$,
\item $\rho(x,t) = \rho_0$ of \eq{eq:rho0} in an open interval of $z=x/t$ containing $0$.
\end{itemize} 
Such global solutions can be built using two types of nonlinear waves found when $g,V$ are constant: dispersive shock waves (DSW) and simple waves (SW), see Appendix~\ref{app:DSWNLS}. Two examples are shown in \fig{fig:solsNLS}. Along a SW, $\rho(t,x)$ and $v(t,x)$ are monotonic functions of the sole variable $z$. They are related through 
\be
z = v \pm \sqrt{g \rho},
\label{vgr}
\ee
where the sign $\pm$ is positive for a right-moving wave (in the fluid frame) and negative for a left-moving one. On the other hand, a DSW interpolates between small-amplitude oscillations (which vanish at the edge of the wave) and a soliton. 

When the variations of the density and velocity along a SW, or a DSW, are small, the wave has a small amplitude and propagates with a velocity (in the lab frame) close to the group velocity of long wavelength linear perturbations, i.e., with $v_{\rm gr} = v \pm \sqrt{g \rho}$, in agreement with \eq{vgr}. For a SW, this result remains true whatever the amplitude because it can be described locally as a superposition of non-dispersive waves. For a DSW instead, dispersion plays an important role when the amplitude becomes large, hence the more complicated expressions for the velocity, see Eqs.~(\ref{eq:NLSzb},\ref{eq:NLSzp}). As a result, in a black hole flow, provided their amplitudes are not too large, these waves propagate {\it away} from the horizon. They may thus be seen as a nonlinear version of outgoing wave-packets produced by the scattering of the initial configurations on the horizon. It has to be noticed that the three types of outgoing waves emitted by a black hole flow are governed by long wave length roots of the dispersion relation, see Fig.~\ref{fig:DR}. 
\begin{figure}[h]
\begin{center}
\includegraphics[width=0.49 \linewidth]{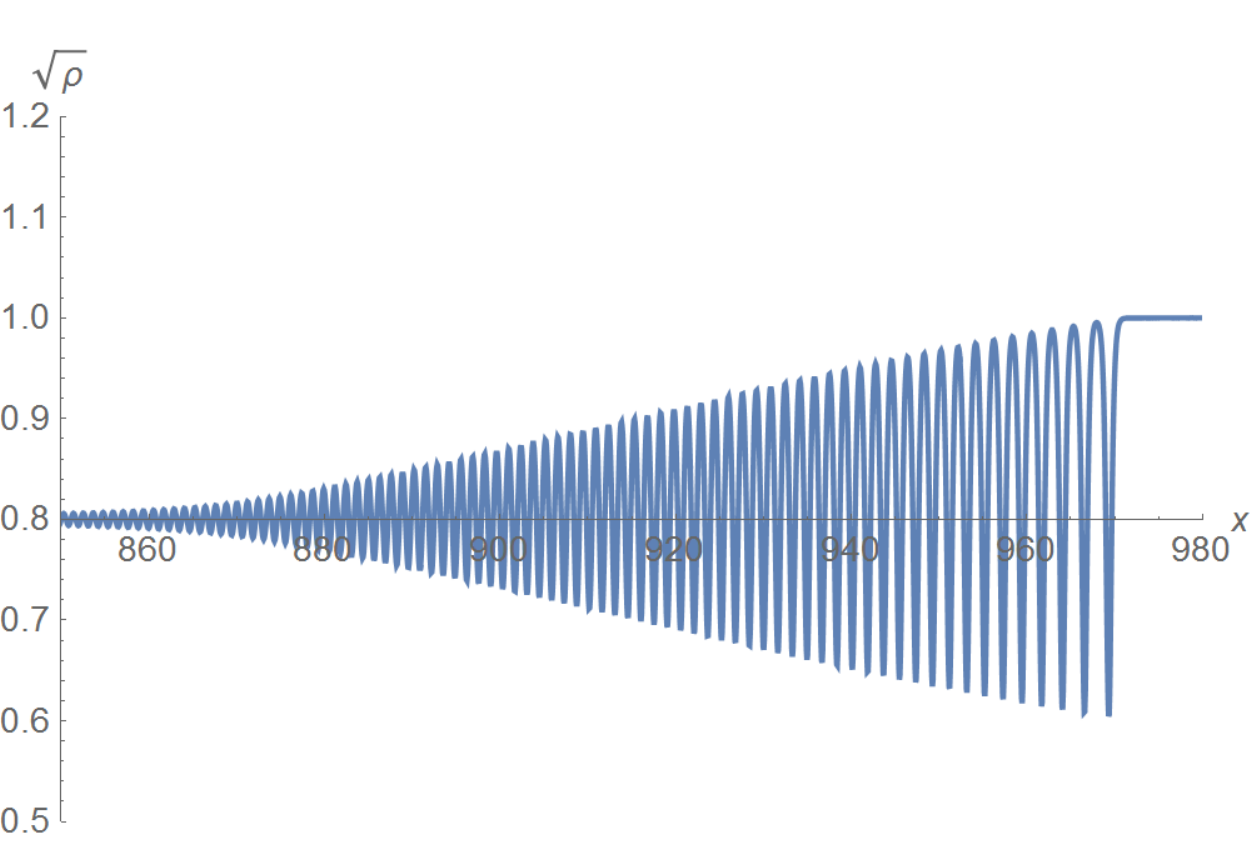}
\includegraphics[width=0.49 \linewidth]{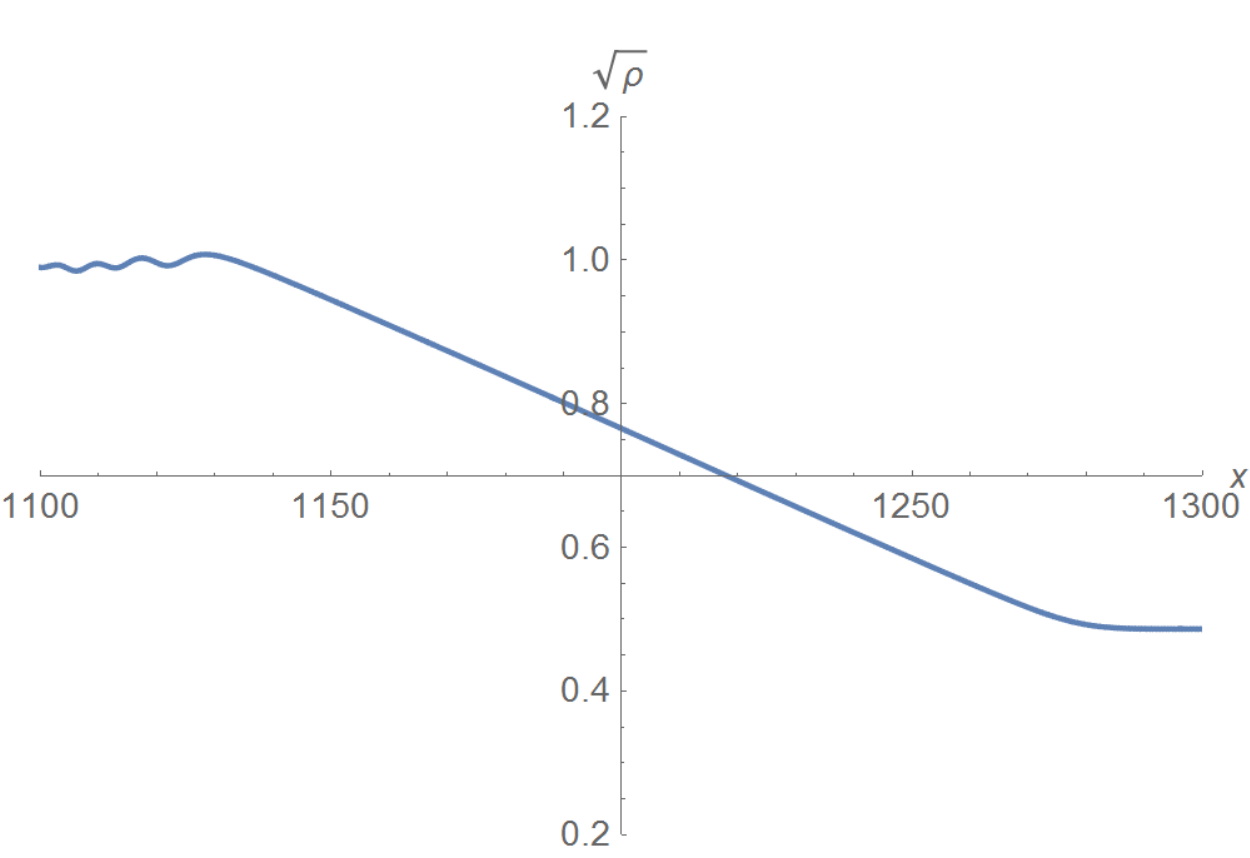}
\caption{Here we show the density profiles $\sqrt{\rho(x)}$ of a DSW (left) and a SW (right). Both are computed numerically in a domain where $g$ and $V$ are constant. The SW is strictly scale-invariant, as along this solution $\rho$ and $v$ depend only on $x/t$ up to corrections not captured by the Whitham equations. For the DSW, the envelope and wavelength of the oscillations depend only on $x/t$, while the rapid oscillations move with velocity $s_1/2$, where $s_1$ is defined in \eq{eq:Pla}. 
}\label{fig:solsNLS}
\end{center}
\end{figure} 

Global solutions can be obtained by matching exact solutions of the Whitham equation on each side of $x=0$, imposing continuity of $\psi$ and $\pd_x \psi$. Let us first study the case of symmetric asymptotic conditions, i.e., $\rho_+ = \rho_- \equiv \rho_i$ and $v_+ = v_- \equiv v_i$. We consider solutions with three waves (DSW or SW) in total, two propagating in $z > 0$ and one in $z < 0$. This is motivated by the behavior of linear modes emitted by a black hole. At the linear level, two waves are emitted in the supersonic region and one in the subsonic region~\cite{Macher:2009nz}. Since the DSW and SW we are looking for are nonlinear versions of outgoing wave-packets, it is natural to assume they follow the same behavior. The validity of this hypothesis will come a posteriori from the existence of solutions when $\rho_i$ is sufficiently close to the density $\rho_0$ of \eq{eq:rho0}. 
\begin{figure}[h]
\includegraphics[width=0.49 \linewidth]{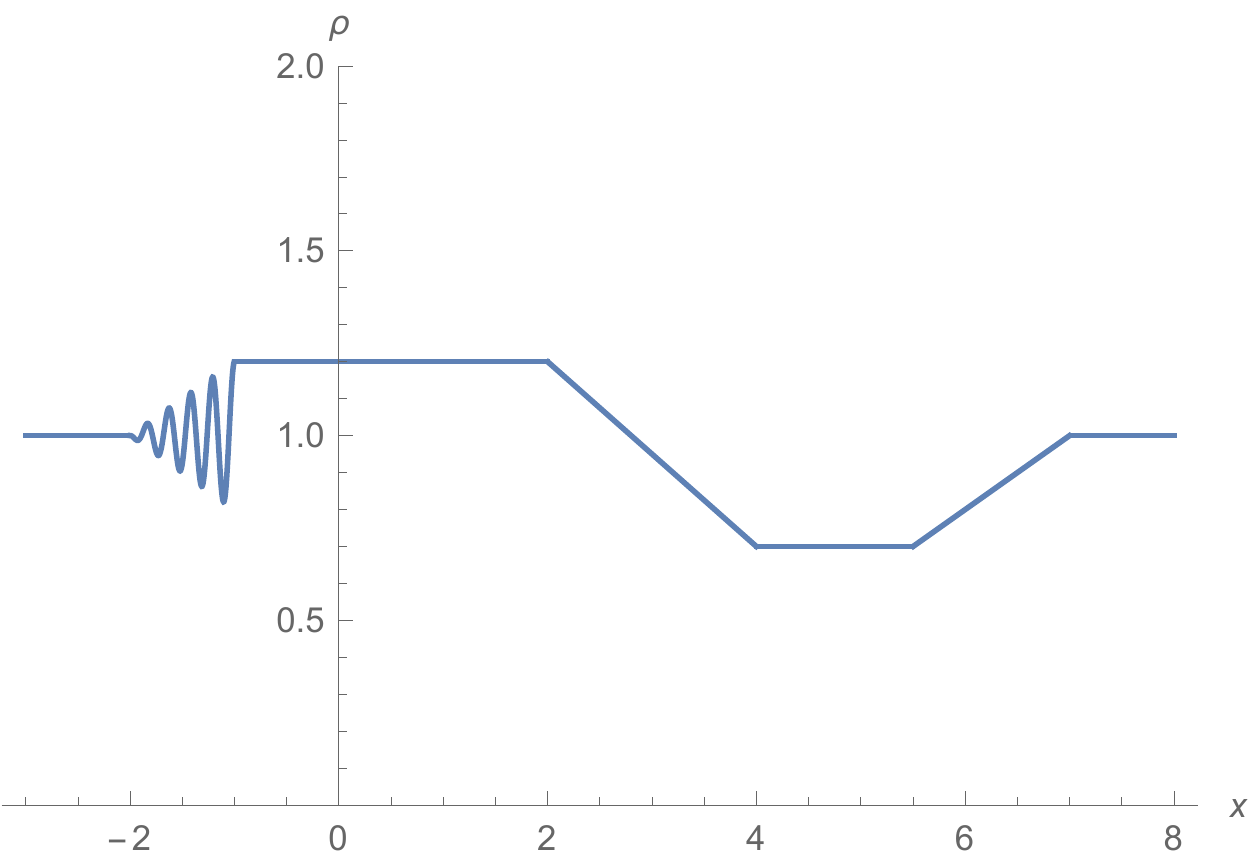}
\includegraphics[width=0.49 \linewidth]{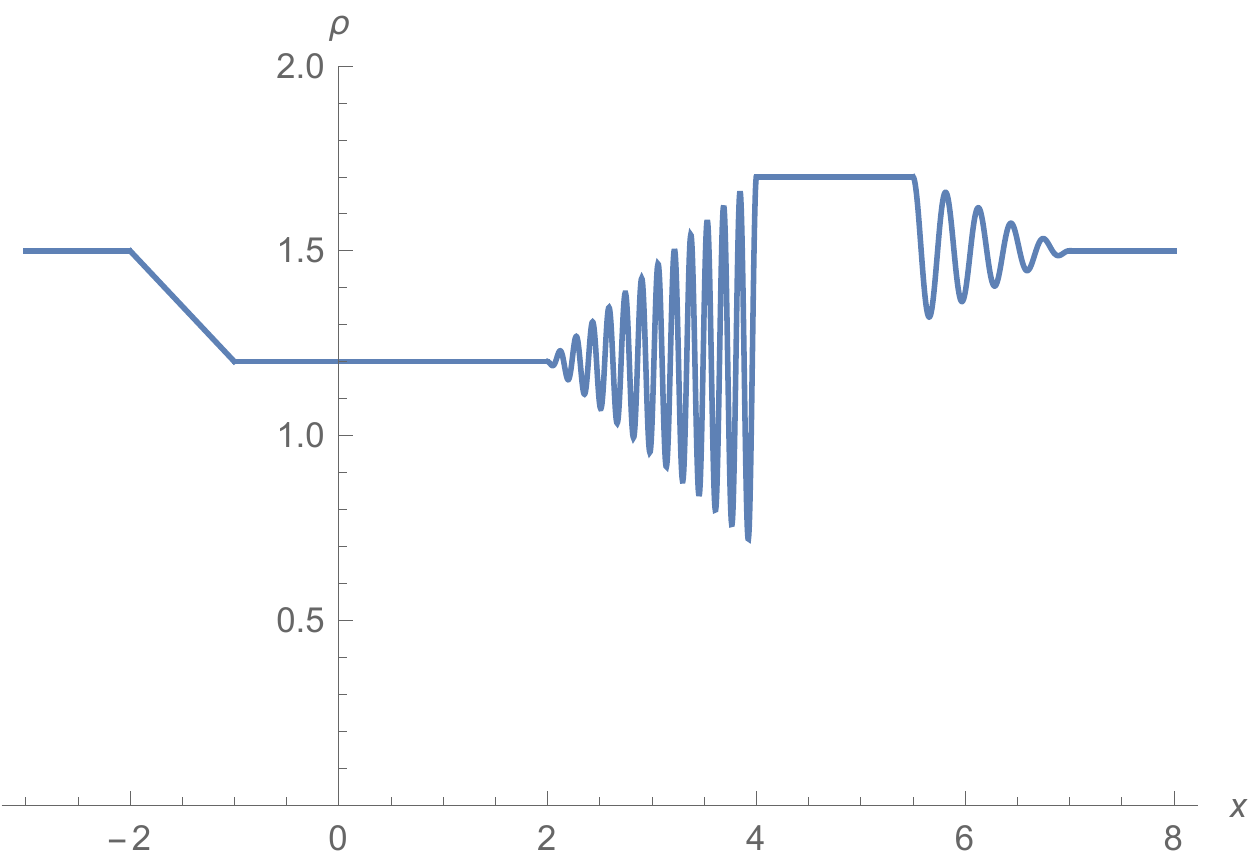}
\caption{Schematic drawing of two solutions of the Whitham equations corresponding to the emission of SW and DSW leaving behind them (in a region containing the horizon at $x=0$) the homogeneous black-hole solution of \eq{eq:rho0}. On the left panel, we show the case with one DSW and two SW which arises when $\rho_+ = \rho_- < \rho_0$. On the right panel, we show the other case with one SW and two DSW arising when $\rho_+ = \rho_- > \rho_0$.}\label{fig:schema}
\end{figure}

Using the properties of SW and DSW outlined in Appendix~\ref{app:DSWNLS}, we find two types of solutions depending on the sign of $\rho_i - \rho_0$. If $\rho_i - \rho_0 < 0$, the solution has one DSW for $z<0$ and two SW for $z>0$, separated by two homogeneous regions. A schematic plot is shown on the left panel of \fig{fig:schema}. On the left of the DSW and on the right of the two SW, the solution is (by construction) homogeneous with density $\rho_i$ and velocity $v_i$. Between the DSW and the leftmost SW, the density is {\it precisely} equal to $\rho_0$ of \eq{eq:rho0}. This equality can be understood from the fact that $\rho_0$ is the only value of $\rho$ allowing to match two homogeneous solutions at $x=0$. The conservation law \eq{eq:cons2} fixes the final value of the velocity around $x=0$ in terms of the asymptotic conditions on the left side: 
\be \label{eq:vf} 
v_f = v_- + 2 \sqrt{g_-} \lp \sqrt{\rho_-} - \sqrt{\rho_0} \rp.  
\ee 
In general, the late-time value of the current $J$ close to the horizon differs from the initial one. The difference is carried by the nonlinear waves.

\begin{figure}
\begin{center}
\includegraphics[scale=0.75]{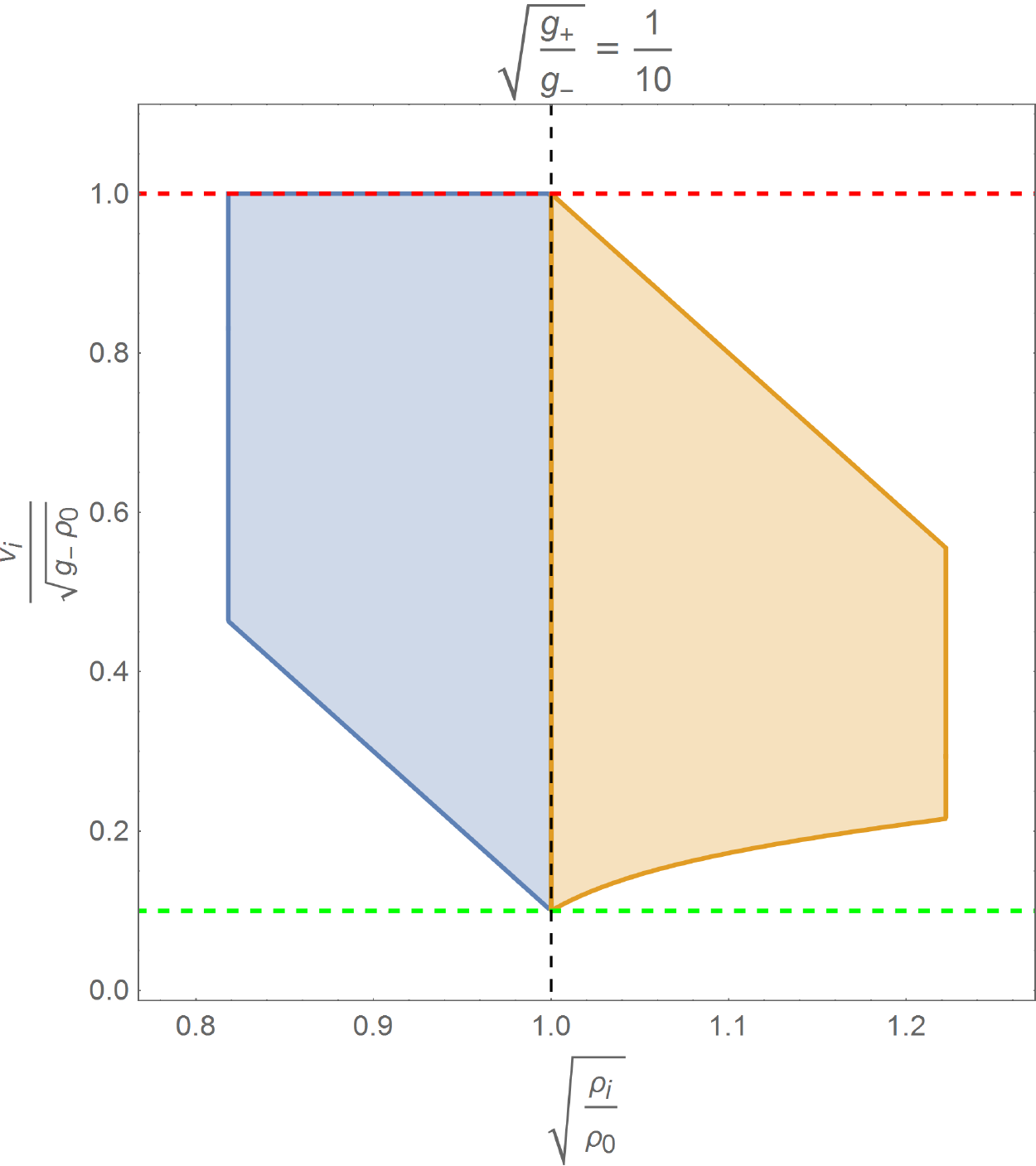}
\end{center}
\caption{We represent the domain of existence of the two solutions schematically shown in \fig{fig:schema}, for $g_- = 100 g_+$, in the plane $\sqrt{\rho_i}, v_i$ which give the asymptotic values of the perturbed solution. The large value of $g_- / g_+$ is chosen to show all the possible boundaries, the two vertical ones disappearing for $g_+ / g_- > 0.056$ (blue) and $g_+ / g_- > 0.02$ (orange). We work in adimensional units: $\rho_i$ is divided by the density of the homogeneous solution $\rho_0$, and $v_i$ is divided by the sound velocity in the left subsonic region: $c_b = \sqrt{g_-\, \rho_0}$. The blue region gives the domain of the solutions with one DSW and two SW, while the orange one gives that of the solutions with one SW and two DSW. The vertical dashed line shows the locus of $\rho_i = \rho_0$. The green horizontal dashed line gives $v_i = \sqrt{g_+\, \rho_0}$, while the red one gives $v_i = \sqrt{g_-\, \rho_0}$. They bound the domain where a homogeneous configuration $\rho = \rho_0$ has a black hole horizon at $x=0$. 
}\label{fig:dom}
\end{figure}

We now consider the validity domain of this solution. We find that it exists if and only if the two following conditions are satisfied:
\be \label{eq:cond1DSW2SW} 
\frac{\sqrt{g_-}-\sqrt{g_+}}{\sqrt{g_-}+\sqrt{g_+}} &< &\sqrt{\frac{\rho_i}{\rho_0}} < 1 , \nonumber 
\\
\label{eq:cond1DSW2SWbis} 
\sqrt{\frac{g_+}{g_-}} + 2 \lp 1- \sqrt{\frac{\rho_i}{\rho_0}} \rp &<& \frac{v_i}{\sqrt{g_- \rho_0}} < 1.
\ee
This domain is shown in blue in \fig{fig:dom} for $g_- = 100 g_+$. (We remind the reader that, for steplike potentials, $V_+$ and $V_-$ only intervene in fixing the value of $\rho_0$.) The important point is that for $\rho_i \to \rho_0$, these inequalities reduce to $\sqrt{g_+ \rho_0} < v_i < \sqrt{g_- \rho_0}$. This condition is equivalent to saying that the left asymptotic region is subcritical and the right one is supercritical, see \eq{eq:BHcond}. We also notice that the blue domain pinches off when $v_i$ saturates its lowest bound. For lower values of $v_i$, the flow is globally subsonic. In brief, if the asymptotic conditions are such that the flow is transonic and of the black hole type, then {\it one} solution with one DSW and two SW exists provided $\rho_0 - \rho_i > 0$ is not too large.

For $\rho_i >  \rho_0 $, one finds essentially the same results with one SW emitted to the left and two DSW emitted to the right. A schematic plot of the density profile is shown on the right panel of \fig{fig:schema}. The density and the velocity around $z=0$ are still given by $\rho = \rho_0$ and $v=v_f$ of \eq{eq:vf}. The conditions of existence of this solution are
\be \label{eq:cond1SW2DSW} 
1 < \sqrt{\frac{\rho_i}{\rho_0}} &<& \frac{\sqrt{g_-} + \sqrt{g_+}}{\sqrt{g_-} - \sqrt{g_+}} ,
\nonumber
\\
2 \sqrt{\frac{g_+ \rho_i}{g_- \rho_0}} - \frac{g_+/g_-}{\sqrt{\frac{\rho_i}{\rho_0}}+ \sqrt{\frac{g_+ \rho_i}{g_- \rho_0}}-1} &<& \frac{v_i}{\sqrt{g_- \rho_0}} < 3 - 2 \sqrt{\frac{\rho_i}{\rho_0}}.
\ee
The argument below \eq{eq:cond1DSW2SW} can also be applied here, 
giving that for $\sqrt{g_+ \rho_0} < v_i < \sqrt{g_- \rho_0}$ this solution exists provided $\rho_i - \rho_0 > 0$ is not too large. The corresponding domain is shown in orange in \fig{fig:dom}.

The same analysis can be carried out with different asymptotic conditions $\rho_+ \neq \rho_-$, $v_+ \neq v_-$. We must then consider 6 additional types of solutions. 4 of them are obtained from the above ones by replacing one of the two DSW by a SW or conversely. One solution has three SW. The last one has three DSW. The domains of existence of each of these solutions are given in Appendix.~\ref{App:domex}. There, it is also shown that one of these 8 solutions always exists in a neighbourhood of any asymptotic conditions compatible with a homogeneous black-hole flow, i.e., $\rho_+ = \rho_- = \rho_0$ and $v_+ = v_- \in ]\sqrt{g_+ \rho_0},\sqrt{g_- \rho_0}[$.

To summarize, when working with the steplike $V$ and $g$ of \eq{eq:step}, we have shown that sufficiently small initial perturbations are expelled at infinity, leaving at late time the homogeneous flow $\rho(x) = \rho_0$. The latter therefore acts as a \emph{local} attractor, in the sense that the solution $\rho(t,x)$ uniformly converges to $\rho_0$ over any bounded interval. In addition, the velocity profile $v(t,x)$ uniformly converges to $v_f$ of \eq{eq:vf}. To make this claim more precise, we propose the following 
\begin{conj}
	There exist two sets of positive real numbers $\{R_{n}\in\R_+^*:n=0,\dots, p\}$ and $\{V_{n}\in\R_+^*:n=0,\dots, p\}$ such that for every initial data $\rho_i,v_i \in C^p(\R)$ satisfying the three conditions:
\begin{enumerate}
	\item $\rho_i$ and $v_i$ are homogeneous outside of some bounded interval, with the asymptotic values $v_\pm=\lim_{x\to\pm\infty}v_i(x)$ such that $\sqrt{g_+ \rho_0} > v_\pm > \sqrt{g_- \rho_0}$;
	\item $\|\rho_i-\rho_0\|_{\R,\infty} < R_0$ and $\|v_i - v_f\|_{\R,\infty}< V_0$;
	\item $\|\pd_x^n \rho_i\|_{\R,\infty} < R_{n}$ and $\|\pd_x^n v_i\|_{\R,\infty}< V_{n}$	for every $n\in\{1,2,\dots, p\}$;
\end{enumerate}
we have, for every bounded interval $I\subset\R$, $$\lim_{t\to+\infty}\|\rho(t)-\rho_0\|_{I,\infty}=0 \quad \text{and} \quad\lim_{t\to+\infty}\|v(t)-v_f\|_{I,\infty}=0.$$ 	\end{conj}
\noindent 
We have used here the standard notation $\|f\|_{I,\infty}=\sup_I|f|$ for any function $f:\R\to\C$ and any interval $I \subset \mathbb{R}$. Although a natural analogue of the asymptotic flatness condition for the initial data, condition 1. may turn out to be too restrictive. We expect that an exponential convergence, or even maybe polynomial ones, should be enough. Similarly, the sufficient level of regularity of the initial data -- i.e. the actual value of $p\in\N\cup\{\infty\}$ -- is not entirely clear to us at this stage, with $p=2$ and $p=\infty$ as natural candidates. Conditions 2. and 3. are, despite their dependence on the choice of $p$, the important point, providing a sense in which the initial perturbation is small enough. We hope to be able to sort these questions out in a future work. Let us nonetheless emphasize that all the numerical simulations presented in the next section support the above conjecture. They also indicate that the conjecture should hold when replacing the steplike functions of \eq{eq:step} by smooth ones, such as those of \eq{eq:tanh}. 
Finally, under the above three conditions, we conjecture that the late-time properties of the three nonlinear waves moving away from the horizon should only depend on the asymptotic initial conditions $v_\pm$ and $\rho_\pm$. In other words, the Fourier components of the smooth profiles $v_i(x)$ and $\rho_i(x)$ are diluted away at very late time, as was rigorously shown in the linear treatment in the former section, and as was also found in our 
simulations, see below.  

To conclude this section, we 
notice that the present analysis does not apply to white-hole flows. Indeed, a crucial point in our calculations is that the three non linear waves move {\it away} from the discontinuity at $x=0$. For black hole flows, this is realized for the domain of initial conditions represented in Fig.~\ref{fig:dom}. In white-hole flows instead, the nonlinear waves move towards $x=0$ in the supersonic region, see Section~\ref{Sec:WH}. When reaching this point, the Whitham theory breaks down. We then expect that white-hole flows will show a more complex behavior than black-hole ones, in accordance with~\cite{Michel:2015pra}. New results concerning white-hole flows are presented in Section~\ref{Sec:WH}. 

\subsection{Numerical results}
\label{sec:numres} 

We numerically solved the GP equation for several reasons. First, by solving the GP equation directly, the results of the previous section can be checked without relying on the approximations of Whitham's modulation theory. Second, the differences in the emission process induced by smooth functions $V$ and $g$ can be studied. Finally, we wanted to study what happens outside the domain of existence of solutions with three waves. In all simulations, we used the code of~\cite{Michel:2015pra}. 

\begin{figure}[h]
\begin{center}
\includegraphics[width=0.49 \linewidth]{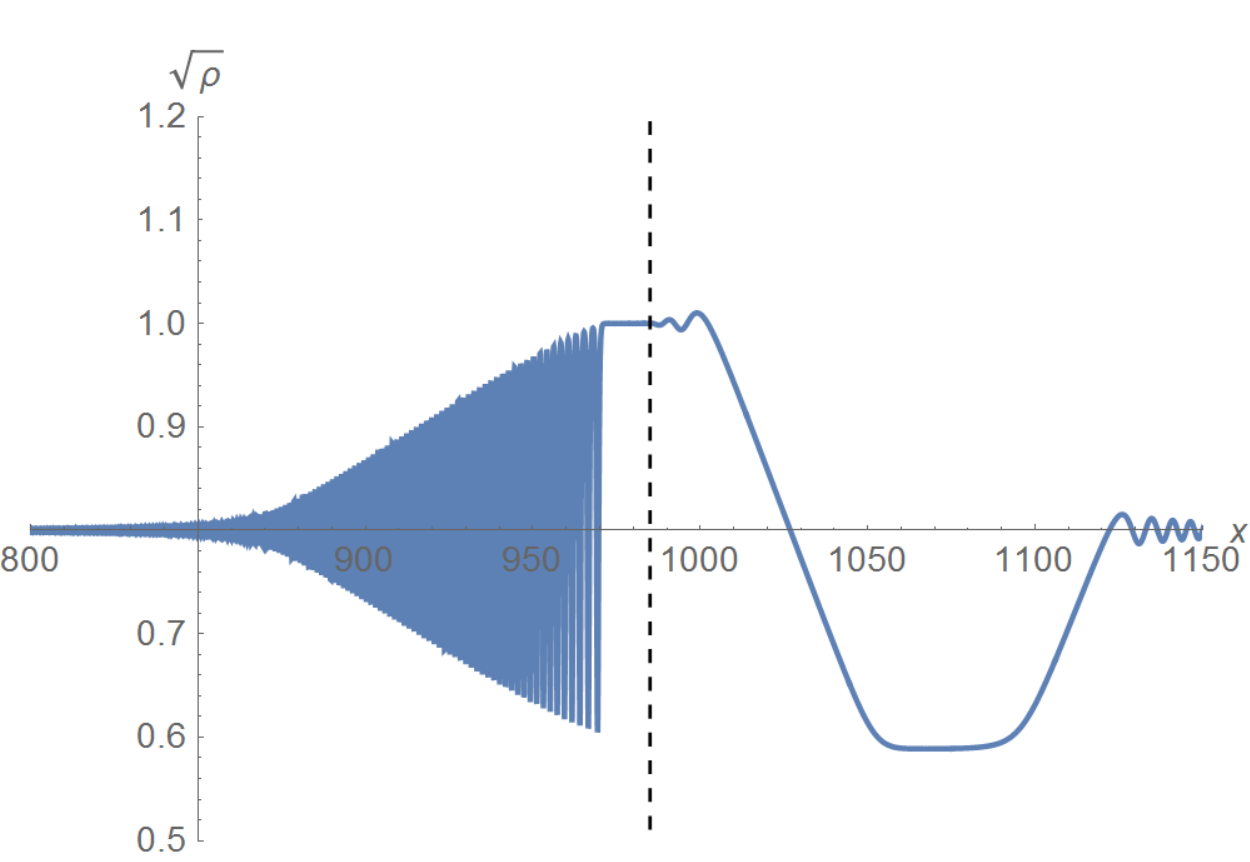}
\includegraphics[width=0.49 \linewidth]{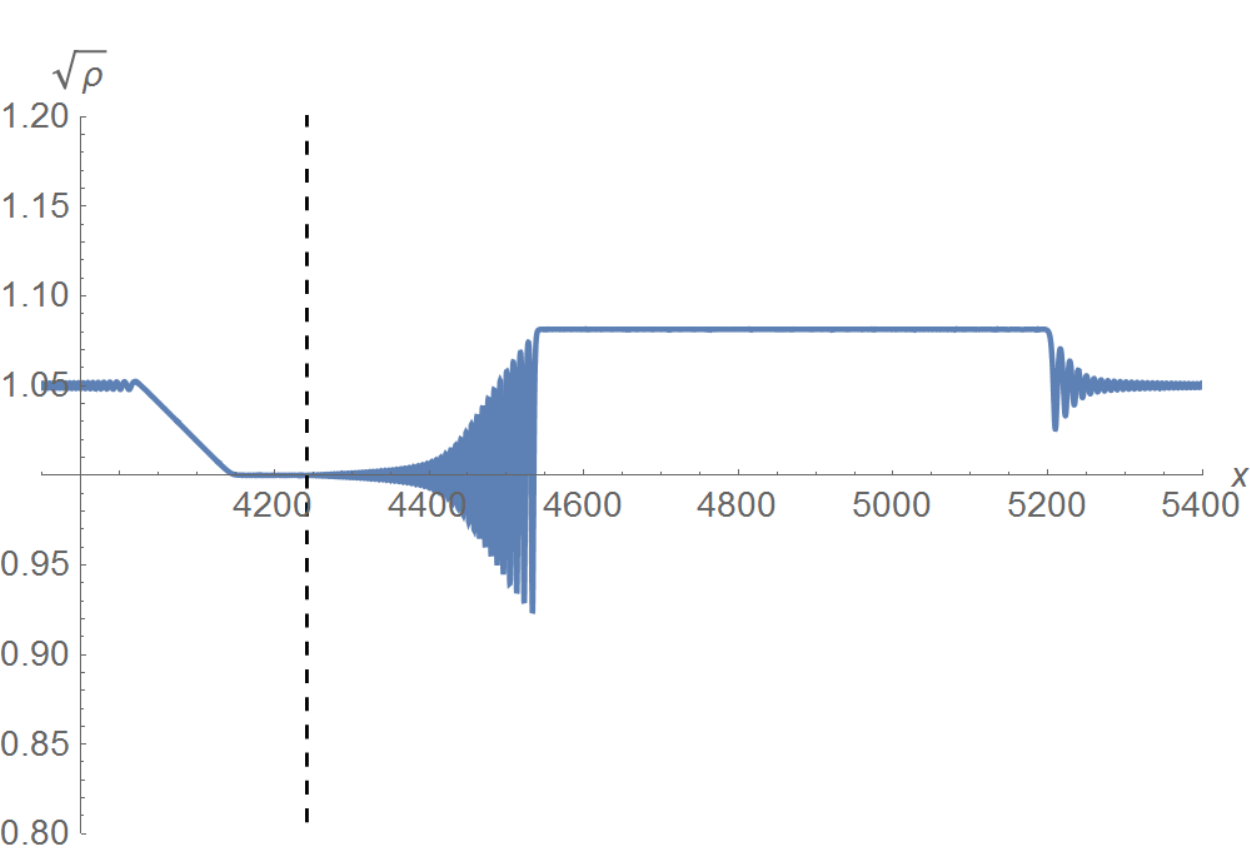}
\caption{Numerical solutions of the GP equation with one DSW and two simple waves (left) and one simple wave and two DSW (right). The vertical dashed line indicates the position of the horizon. In both cases, the initial configuration is homogeneous and $J=\sqrt{8/3}$. For the left plot, the initial density is $\rho_i=0.81$, $t=136$, and $g_-$, $g_+$, $\om - V_-$, and $\om - V_+$ are the same as in \fig{fig:lin3/2}. For the right plot, the initial density is $\rho_i =1.1$, the solution is shown at $t=400$, and the parameters are $g_- = 3.7$, $g_+ = 0.72$, $\om - V_- = 5$, and $\om - V_+ = 2.1$. They are chosen so that $\sqrt{\rho_{2,-}} = \sqrt{\rho_{3,+}} = 1$, $\sqrt{\rho_{2,+}} = 1.6$, and $\sqrt{\rho_{3,-}} = 0.9$.} \label{fig:solsNLSb} 
\end{center}
\end{figure}

In \fig{fig:solsNLSb}, we show two solutions obtained numerically when starting at $t=0$ from a configuration with homogeneous velocity $v_i$ and density $\rho_i \neq \rho_0 = 1$. On the left panel, $\rho_i$ and $v_i$ satisfy \eq{eq:cond1DSW2SW}, while on the right panel they satisfy \eq{eq:cond1SW2DSW}.
At late times, we observe that these solutions are superpositions of the three waves of Section~\ref{sec:analytical} plus perturbations whose amplitude decays in time as $t^{-{3/2}}$. We checked that the properties of the SW and DSW, in particular their amplitudes, the position of their edges, and their domains of existence, agree with those given by Whitham's equations in the limit $t \to \infty$. 
We performed additional simulations by replacing the homogeneous initial conditions $v_i, \, \rho_i$ by smooth varying ones in a bounded domain, and we observed that this agreement was preserved. It would be interesting to identify sufficient conditions on local variations of $v_i$ and $\rho_i$ ensuring that the late time properties of the three nonlinear waves only depend on the asymptotic values of $v_i$ and $\rho_i$. 
Moreover, the fact that perturbations still decay as $t^{-3/2}$ shows that the three macroscopic waves do not change the late-time behavior of perturbations obtained in Section~\ref{sec:lin} close to the horizon. 

To complete our analysis, we replaced steplike $g$ and $V$ by smooth functions. We first chose their variations so that a solution with homogeneous density $\rho_0$ still exists, i.e., such that $V(x) + g(x) \rho_0$ is a constant. We worked with functions of the form
\be \label{eq:tanh}
&&g(x) = \frac{g_+ + g_-}{2} + \frac{g_+ - g_-}{2} \tanh (x /\sigma), \nonumber 
\\
&& V(x) = \frac{V_+ + V_-}{2} + \frac{V_+ - V_-}{2} \tanh (x /\sigma),
\ee
where $\sigma > 0$. When the asymptotic conditions are inside the domain described by Eqs.~(\ref{eq:cond1DSW2SW},\ref{eq:cond1SW2DSW}), we observed that the late-time properties of the solution are the same as in the steplike case: the homogeneous black-hole solution is reached for $t \to \infty$ through the emission of three nonlinear waves with the same properties. However, the typical formation time of the SW and DSW now depends on $\sigma$. It is linear in $\sigma$ for $\sigma \approx 0$. We verified that these results qualitatively extend to the case where $\sigma$ takes different values for $g$ and $V$. In that case, there exists no stationary solution with a homogeneous density. However, as discussed in subsection~\ref{sub:Ahtf}, there is a one-parameter family of solutions 
with angular frequencies $\omega(J)$  and asymptotically constant values of $\rho$. Our numerical results indicate that, if the initial conditions are sufficiently close to this series, one of its solutions is reached at $t \to \infty$ through the emission of three waves. In brief, all simulations confirm that initial configurations within the domain specified by Eqs.~(\ref{eq:cond1DSW2SW},\ref{eq:cond1SW2DSW}) all evolve, at late times, towards an AH solution.

We finally performed numerical simulations starting from initial conditions with asymptotic behaviors {\it outside} the domains described by Eqs.~(\ref{eq:cond1DSW2SW},\ref{eq:cond1SW2DSW}). These conditions can be violated in two different ways. In the first case, corresponding to crossing one of the two vertical boundaries in \fig{fig:dom}, the two waves in the supersonic region $x>0$ overlap each other, producing a complicate interference pattern. Yet, our simulations indicate that the solution still converges to a homogeneous black-hole flow at late times, as the overlapping waves still escape to infinity. In the second case, one of the waves in the region $x>0$ (respectively $x<0$) has its left (respectively, right) edge moving to the left (respectively, right). [The wave with one edge moving towards the horizon lies in the region $x>0$ when crossing the lower boundary, or in the region $x<0$ when crossing the upper one. The sign of $\rho_i - \rho_0$ then gives its type (DSW or SW).] When the wave is a SW, the late-time solution contains part of a soliton or shadow soliton~\cite{Michel:2013wpa}, and asymptotes to an asymptotically homogeneous density different from $\rho_0$ on the corresponding side. 

When the corresponding wave is a DSW, the situation is more complicated. In the simplest case, the solution still becomes stationary at late times over any finite interval of $x$. It then contains a stationary density modulation attached to $x=0$. We also found cases where the solution apparently never reaches a stationary profile around $x=0$, instead emitting a density modulation with a non-vanishing phase velocity, which at the nonlinear level corresponds to a soliton train. 
Our numerical investigation was not extensive enough to determine with confidence the conditions in which one or the other behavior occurs, although it seems that a stationary solution is reached when the DSW with an edge moving towards the horizon is on the left $x<0$, while the emission of soliton trains occurs when it is on the right $x>0$. 

\section{White-hole flows}
\label{Sec:WH} 

In this section we briefly study the time-evolution of perturbations on white-hole flows. As discussed at the end of Section~\ref{sec:analytical}, there is an important difference between the evolutions of black- and white-hole flows: while the former expel perturbations at infinity, the latter have a tendency to accumulate them close to the horizon, 
as can be understood from the fact that white-holes behave as the time-reversed of black holes. One can thus expect that white-hole flows will act as ``repellents'' rather than attractors. We here show that it is indeed the case. When starting from the homogeneous solution (which is an attractor for black hole flows), depending on whether the perturbation gives rise to a (sufficiently large) decrease or an increase of the near horizon density, the flow is destabilized by nonlinear effects and develops either only a macroscopic undulation, or sends a train of solitons accompanied by a macroscopic undulation. When working to linear order, one finds that
small perturbations generally leave at late times a stationary undulation with a large amplitude, 
thereby signaling an infra-red instability of the background flow. 
\subsection{Linear perturbations}

\begin{figure}
	\begin{center}
		\includegraphics[width=0.49 \linewidth]{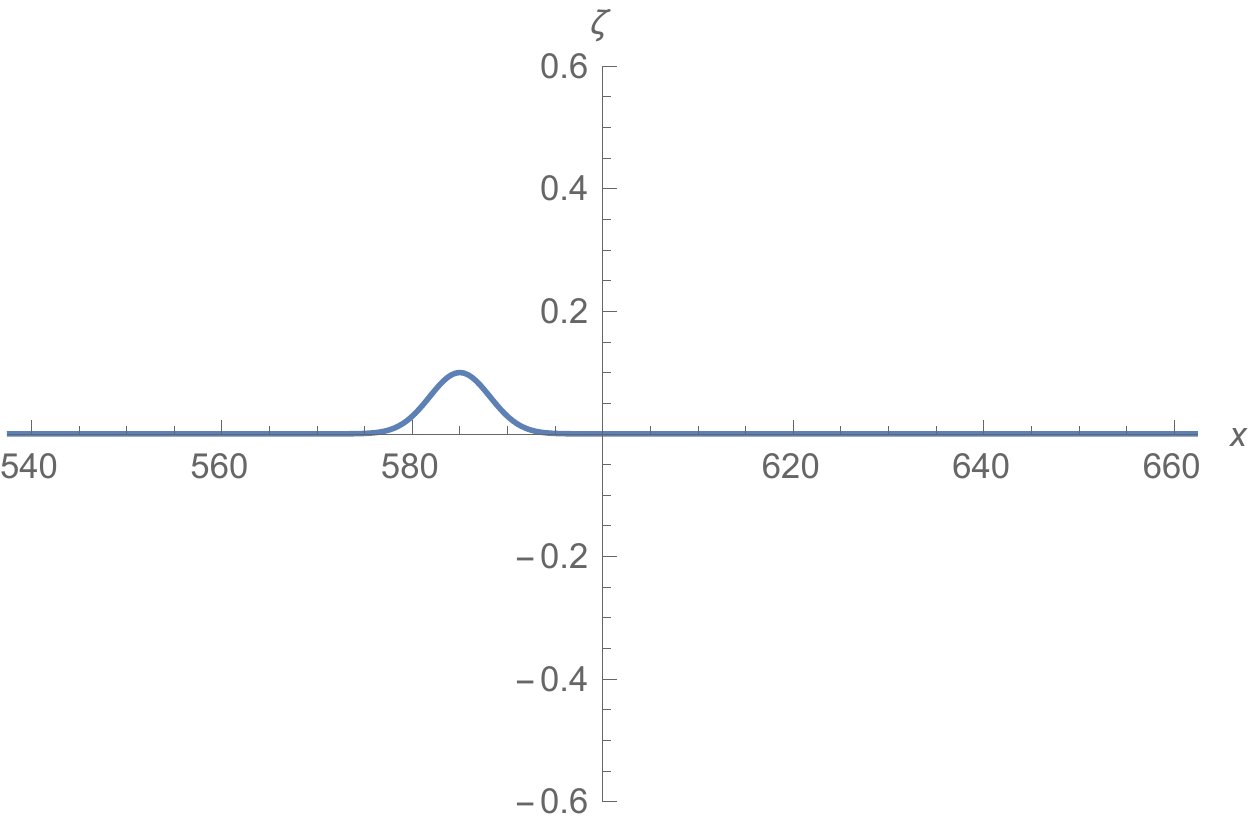}
		\includegraphics[width=0.49 \linewidth]{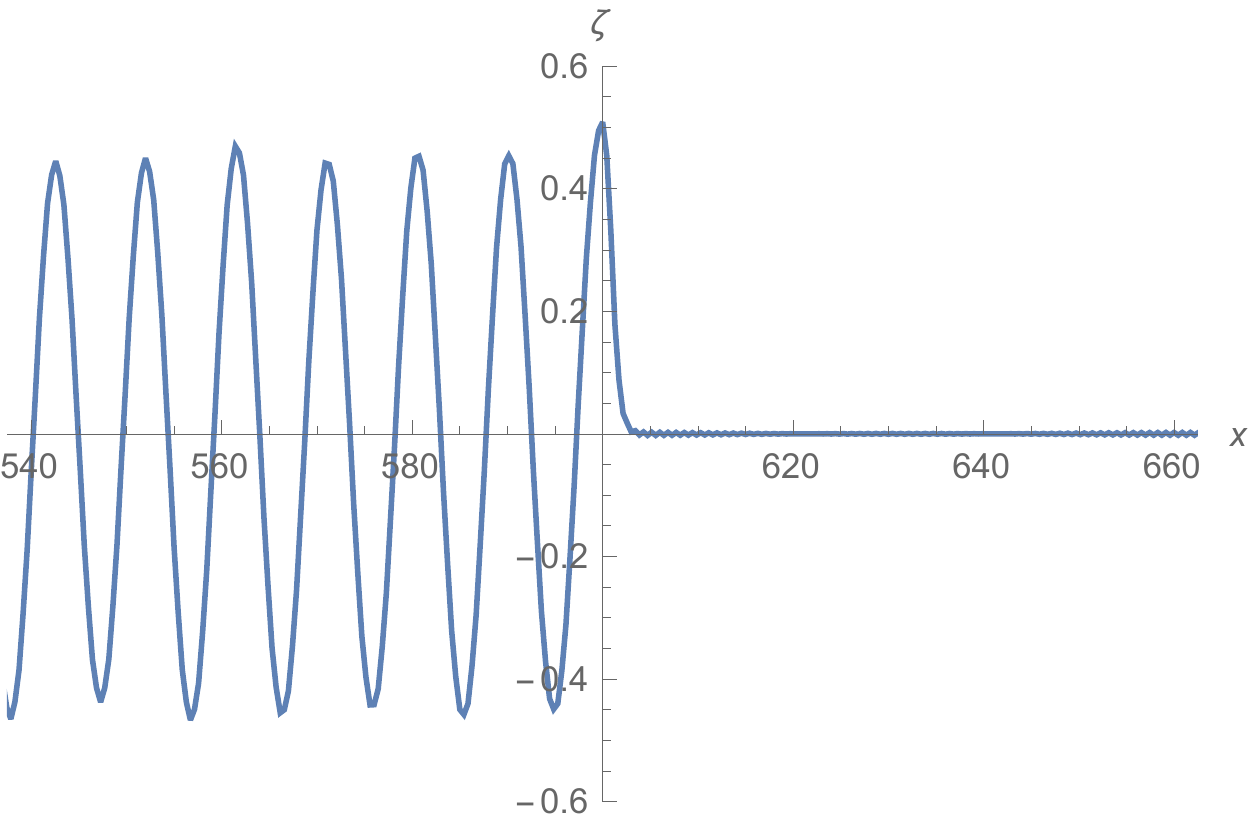}
	\end{center}
	\caption{Scattering of a localized perturbation on a white hole configuration for the linearized KdV equation. We work with $v = J / h$, where $J = -1$ (i.e., a flow to the left) and $h(x) = 2+\tanh(600 - x)$. The white hole horizon is located at $x = 600$. The initial perturbation is a gaussian with amplitude $0.1$ and width $10$. The left panel shows the initial perturbation of the water height and the right one shows the late-time undulation, which is here localized in the subsonic region because the dispersion relation is subluminal. Time-dependent effects are still visible through the small variations of amplitudes between two adjacent oscillations. }\label{fig:linKdVWH}
\end{figure}

Let us first consider white-hole solutions of the KdV equation,
as they are technically simpler to characterize. We work with functions $v$ and $h$ given by \eq{eq:ansatzvh}, with $v_-,v_+ < 0$, $v_- + \sqrt{g h_-} < 0$, and $v_+ + \sqrt{g h_+} > 0$. The trivial solution $\zeta = 0$ then corresponds to a white hole flow. We consider some initial perturbation $\zeta(x,t=0)$. The corresponding initial conditions on $\psi$ are 
\be
\psi(x,t=0) = \int_0^x \zeta(y,0) d y .
\ee
The time-evolution of $\psi$ can be determined by expanding it into out-modes. Due to the symmetry between in-modes of black-hole flows and out-modes of white-hole flows~\cite{Macher:2009tw}, the calculation is similar to the one sketched in Appendix~\ref{App:linKdV}. In particular, the structure of divergences is the same, except that those which multiplied incoming waves now multiply outgoing ones, and conversely. This introduces an important difference, as the divergence in $1/\om$ of the coefficients of dispersive waves for an extended perturbation, which did not contribute for a black-hole as it multiplied incoming waves, now multiplies outgoing waves~\cite{Mayoral:2010ck}. As explained in \cite{Coutant:2012mf}, it adds a saddle-point contribution for $\om \approx 0$, which generates a stationary undulation with a large amplitude.\footnote{In~\cite{Coutant:2012mf} only perturbations localized in $\psi$ were considered. In that case, corresponding to perturbations in $\zeta$ with a vanishing integral, the amplitude of the undulation vanishes as $t \to \infty$. When the perturbation satisfies $\int_\mathbb{R} \zeta dx \neq 0$ instead, we verified that its amplitude goes to a finite, non-vanishing constant for $t \to \infty$.} 
Numerical simulations using the linearized KdV equation confirm  that this extends to smooth white hole configurations, see \fig{fig:linKdVWH}. Since the scattering coefficients on a white-hole flow described by the GP equation have the same divergences for $\om \to 0$ as those obtained using the KdV equation, see~\cite{Macher:2009tw}, the same argument tells us that a stationary undulation shall also be produced in a condensate. 
In all cases, the macroscopic character of the undulation amplitude indicates that nonlinear effects will play a crucial role, which implies that linear equations are unable to predict the late time evolution.  

\subsection{Nonlinear evolution} 

The nonlinear evolution of perturbations on a white-hole flow was studied numerically in~\cite{Mayoral:2010ck,Michel:2015pra}. In the present appendix we report some new numerical results. To make contact with the nonlinear analysis of black-hole flows of the previous section, we focus on the case of the GP equation. We obtained qualitatively similar results for the KdV equation, with the signs of the perturbations to the boundary conditions reversed. A rule of thumb, which works for the GP and KdV equations as well as for a superluminal KdV equation obtained by changing the sign of the dispersive term, is that analogue white hole flows are most unstable to perturbations with the sign of the difference between the stationary soliton and the corresponding homogeneous solution. For the GP equation, the soliton is a local underdensity, so the strongest instabilities come from perturbations of the density with a negative sign. For the KdV equation instead, the soliton is a local surelevation of the free surface. Correspondingly, a positive perturbation on $\zeta$ leads to generally wilder behaviors than a negative one. 

\begin{figure}[h]
	\begin{center}
		\includegraphics[width=0.49 \linewidth]{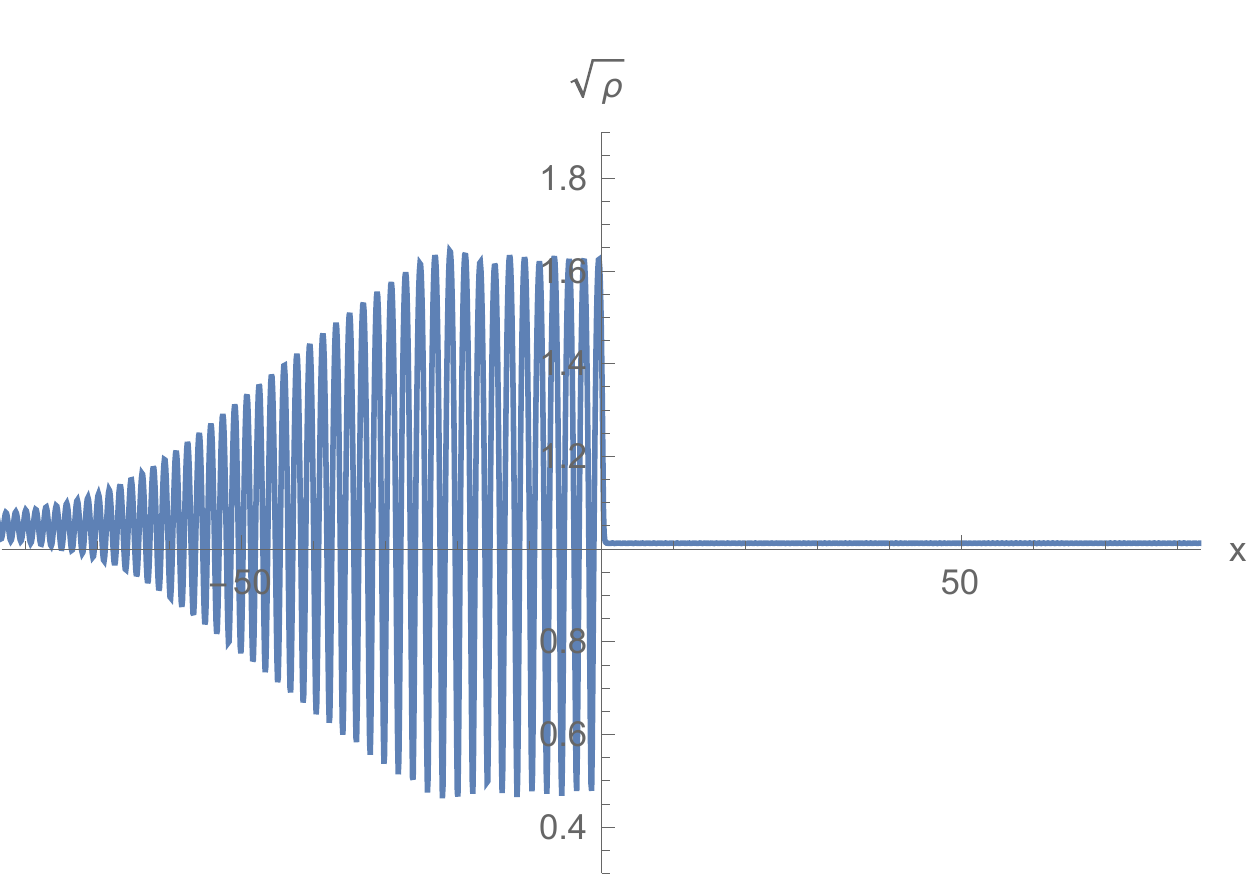} 
		\includegraphics[width=0.49 \linewidth]{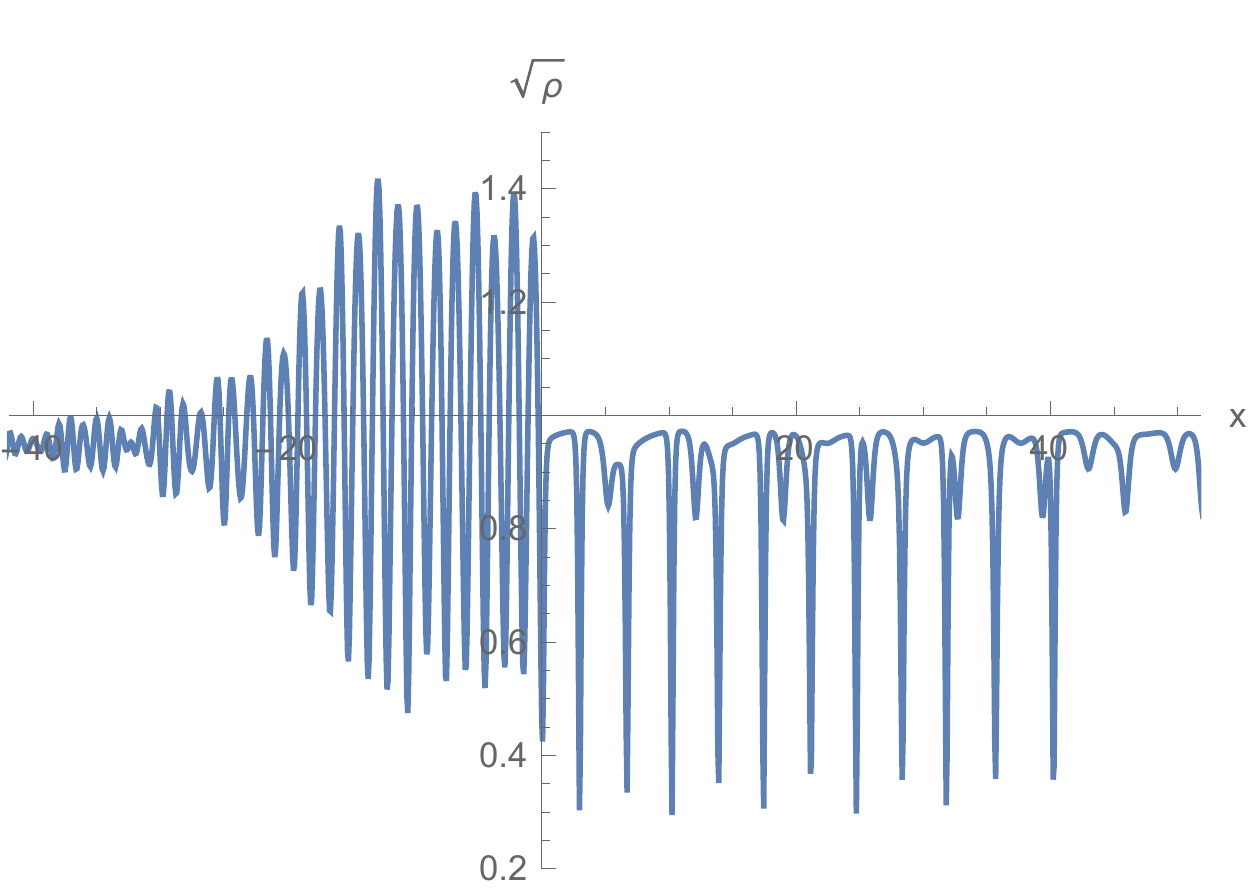}
		\caption{We show the undulation (left) and soliton trains (right) produced when the initial 
density differs from that of the homogeneous white hole flow with $\rho_0 = 1$. The vertical axis corresponds to the horizon. On the left (supersonic) side, the parameters are 
$g= 0.625$ and $V=-2.625$, while on the right side they are $g= 40$ and $V=-42$.
The initial configuration has a homogeneous density equal to $1.1$ (left) and $0.9$ (right), and a current $J=2$. The solution is shown at $t=30$ (left) and $t=10$ (right).}\label{fig:solsNLSb2}
	\end{center} 
\end{figure}

To start the analysis, we work with steplike functions $g$ and $V$, given by \eq{eq:step} with $g_+ > g_-$ and $V_+ < V_-$. We first consider an initially {\it homogeneous} configuration with $\rho = \rho_i$ close to $\rho_0$ and $v \in ]\sqrt{g_- \rho_0}, \sqrt{g_+ \rho_0}[$. Two typical solutions at intermediate times are shown in \fig{fig:solsNLSb2}.
When $\rho_i > \rho_0$, a finite-amplitude undulation develops in the supersonic region. Its amplitude is linear in $\sqrt{\rho_i - \rho_0}$ and is constant at late time. When $\rho_i < \rho_0$ instead, the horizon emits superposed soliton trains in the subsonic region (three of them can be seen in the figure), and a perturbed undulation in the supersonic one. A large-amplitude perturbation is produced periodically close to the horizon, with a frequency linear in $\sqrt{\rho_0 - \rho_i}$, which then separates into several solitons with different velocities in the subsonic region, plus a perturbation propagating on top of the undulation in the supersonic one. It should be emphasized that for homogeneous initial configurations there is no threshold on the value of $\rho_i - \rho_0$. When it is positive (negative), one obtains a stationary undulation 
(non-stationary soliton train). For both signs of $\rho_i - \rho_0$, the nonlinear solution is characterized by the non-analytic parameter $|\rho_i - \rho_0|^{1/2}$. In other words, irrespectively of the perturbation amplitude, the Bogoliubov-de Gennes equation \eqref{eq:BdG} cannot be used to determine the late time configuration. 

We now study the evolution of localized perturbations. To this end, we choose initial conditions of the form
\be 
\psi(x,t=0) = \sqrt{\rho_0 + A \, e^{- \lp x-x_0 \rp^2 / \lambda^2}} e^{i v_i x}.
\ee
For $A > 0$, we observe only the emission of an undulation in the supersonic region. For $A < 0$, we observe the emission of a finite number of solitons in the subsonic region, as well as a perturbed undulation in the supersonic one. In both cases, the amplitude of the undulation is roughly linear in $A$ at early times provided $A \ll \rho_0$. Interestingly, we observe that it slowly decreases in time due to nonlinear effects, apparently going to zero for $t \to \infty$. As in the case of homogeneous initial configurations, irrespectively of value of $|A|$, the linear equation \eqref{eq:BdG} cannot determine the late time solution. This is due to the accumulation of the low frequency configurations on the sonic horizon~\cite{Coutant:2012mf}. 

Finally, we performed numerical simulations with functions $g$ and $V$ of the form \eq{eq:tanh} with $V_+ + g_+ \rho_0 = V_- + g_- \rho_0$, so that a solution with a homogeneous density $\rho_0$ exists. The main difference with the above results is that, when working with localized perturbations, solitons are emitted only if $A$ is below a negative threshold value $A_s < 0$. For $A$ not too close to $A_s$, we observed that the time needed to produce the first soliton scales as $( |A| \lambda /\sigma)^{-1/2}$. This can be understood as the condition for obtaining a sufficiently large underdensity so as to allow for the emission of a soliton. The simulations we performed were not precise enough to accurately determine the scaling of $A_s$ in $\sigma$ and $\lambda$, although we found that $A_s < 0$ increases with $\lambda$ and $1/\sigma$, going to 0 for $\lambda \to \infty$ or $1/\sigma \to \infty$. It would be interesting to further investigate these questions along with the validity of the linear equation \eqref{eq:BdG} for $|A| < |A_s|$.

\section{Conclusions}
\label{sec:concl} 

In this paper we studied one-dimensional transonic solutions of the GP and KdV equations. We showed that they exhibit behaviors which are in close analogy with those of black hole solutions of general relativity. At the level of stationary solutions, we showed that the set of solutions which are asymptotically homogeneous (AH) on both sides is discrete at fixed value of the current $J$. When considering steplike potentials, we demonstrated that the series of solutions parameterized by $J$ is unique. For smooth potentials, we numerically found a series of AH flows which is smoothly connected to the above series, as the solutions coincide in the high gradient limit. However, under some conditions, we also found a second series of AH solutions which is disconnected from the first one. These hairy solutions possess a large fraction of a soliton attached to the sonic horizon. Our preliminary investigations indicate that they are less stable than those of the first series because the soliton can be sent away from the horizon by a sufficiently large perturbation. In the rest of the paper, we focused on the stability properties of the first series of AH solutions. 

At the level of linear perturbations, we found, both analytically and numerically, that the near horizon amplitude of all localized perturbations decays in time as a power law. This establishes that AH black hole flows are linearly stable. It should be noticed that the $S$-matrix which governs this linear scattering is the same as that encoding the Hawking effect in the present settings. One clearly sees here the close link between the (stimulated) Hawking process, i.e., the wave amplification upon scattering on the horizon, and the expulsion of all incident perturbations away from the horizon. 
Moreover, in the limit where the dispersive momentum scale is sent to infinity this expulsion follows the relativistic prediction. 

To study the stability when including nonlinearities of the GP equation, we proceeded along two different approaches. We first used the approximate scheme of Whitham's modulation theory to characterize in analytic terms the late time evolution of the solutions for steplike potentials. We showed that some field configurations are expelled from the horizon region to infinity by three nonlinear waves, known in the literature as dispersive shock waves and simple waves. Importantly, in the vicinity of the sonic horizon, the solution tends to one of the AH that we formerly characterized. It should be pointed out that Whitham's theory also provides a characterization of the domain of (homogeneous) initial conditions which evolve at late time to an AH transonic flow. Finally, when taking the limit of small amplitude, it can be verified that these results generalize those of the linear analysis.

We then performed numerical simulations. We first showed that the late time behavior of a much wider class of initial configurations agrees with that predicted by Whitham's theory, namely, an AH transonic flow is obtained by the emission 
of three nonlinear waves plus perturbations that decay in time. We verified that the properties of the three waves are in agreement with those obtained using Whitham's theory. We also showed that the time dependent perturbations, which are not accounted for in our nonlinear analytical approach, decay in time with the same power law as that found in the linear analysis. All these results indicate that AH transonic flows are local attractors for neighboring flows. In a future work, using the integrability of the GP equation, we hope to be able to demonstrate this property. Since AH black hole solutions are local attractors, they can be produced without fine-tuning the initial conditions nor the potential $V$. This should help observing both the spontaneous and stimulated analogue Hawking emission. 

We also numerically studied the behavior of solutions when the initial conditions are outside the validity domain of the solutions obtained with Whitham's theory. In this regime of large deviations with respect to the attractor solutions, several behaviors have been observed. In some cases, we found that the emitted nonlinear waves can leave behind them an undulation which propagates backwards towards the black hole horizon. In other cases, we observed the emission of soliton trains. This variety of behaviors is similar to that observed in Section~\ref{Sec:WH} when studying the time behavior of white hole flows. For these flows, the late time properties are rather complicated even when the perturbations have a small amplitude. Yet, the observed behaviors can be separated into two types. In this respect, our analysis indicates that the AH solutions are a kind of ``separators'', rather than attractors, as the type of the solution is chosen according to the sign of the density fluctuation in the near horizon region. When there is a sufficiently large increase of the density, the solution displays a macroscopic undulation in the supersonic domain, whereas it gives rise to an emission of soliton trains when there is a sufficiently large density decrease. These two types of behaviors have been already found in the context of the dynamical instability (called the black hole laser) which is found when a stationary flow crosses twice the sound speed~\cite{2015arXiv150900795D,Michel:2015pra}. Our study clearly shows that it is the white hole horizon which is responsible for the wide variety of temporal evolutions that was observed. 

The present work leaves several open questions which deserve further study. First, it would be interesting to prove rigorously our conjecture using inverse scattering techniques. In the same vein, investigating deformed GP or KdV equations including non-integrable terms could shed light on the relations between the mathematical properties of the equation and the ``no-hair'' results. As a first example, we study numerically the case of the cubic-quintic GP equation in Appendix~\ref{App:CQNLS}. Another possible extension concerns higher-dimensional systems. In general relativity, no-hair and uniqueness results crucially depend on the dimensionality of space~\cite{Chrusciel:2012jk}. It would certainly be enlightening to see how the dimensionality affects black hole stability in systems described by the GP or hydrodynamic equations. 

\section*{Acknowledgement}
We thank A. M. Kamchatnov, N. Pavloff, and G. Shlyapnikov for interesting discussions. This work was supported by the French National Research Agency under the Program Investing in the Future Grant No. ANR-11-IDEX-0003-02 associated with the project QEAGE (Quantum Effects in Analogue Gravity Experiments).

\appendix

\section{Whitham equations}
\label{App:Whitham}

In this appendix we give the main steps in obtaining the Whitham modulation equations for \eq{eq:GPE} and their solutions used in the main text. The interested reader will find in~\cite{Kamchatnov} and references therein a full derivation. When possible we use the same notations and conventions as in this reference. 

The Whitham modulation theory~\cite{Whitham1} was developed to study oscillating solutions of partial differential equations with slowly varying parameters. It rests on the two following ideas. First, if there is a clear separation between the fast scale of oscillations and the slow scale of variation of the parameters, an averaging procedure can decouple them. Second, averaging conservation laws generically gives the most accurate and best-controlled results. This theory is then particularly useful when one has enough conservation laws to characterize the space of solutions one is interested in. As such, it is no surprise that deep links exist with integrability. A generic way to obtain the Whitham equations for an integrable system is to use the AKNS scheme, developped in~\cite{Ablowitz:1974ry} and applied to the Whitham theory in~\cite{Kamchatnov1994387}. Here we briefly review this procedure, following the presentation of~\cite{Kamchatnov}. 

\subsection{The AKNS scheme}

The basic idea of the AKNS scheme is to reformulate a partial differential equation one wishes to solve as the compatibility condition of a linear system of the form
\be \label{eq:sys} 
\left\lbrace
\begin{array}{cc}
	\pd_x \psi_\lambda(x,t) = U(x,t;\lambda) \psi(x,t), \\
	\pd_t \psi_\lambda(x,t) = V(x,t;\lambda) \psi(x,t),
\end{array}
\right.
\ee
where $\psi_\lambda(x,t)$ is a two-component complex vector and $U(x,t;\lambda)$, $V(x,t;\lambda)$ are two matrices of the form
\be \label{eq:formAKNS}
U(x,t;\lambda) = 
\begin{pmatrix}
	F(x,t;\lambda) & G(x,t;\lambda) \\ 
	H(x,t;\lambda) & -F(x,t;\lambda)
\end{pmatrix}, \; 
V(x,t;\lambda) = 
\begin{pmatrix}
	A(x,t;\lambda) & B(x,t;\lambda) \\ 
	C(x,t;\lambda) & -A(x,t;\lambda)
\end{pmatrix},
\ee
where $A$, $B$, $C$, $F$, $G$, and $H$ are differential operators, analytic in $\la$. Here $\lambda$ is a complex spectral parameter, independent on $t$ and $x$. To simplify the notations, from now on the dependence in $\lambda$ will not be written explicitly when no confusion is possible. The compatibility condition of \eq{eq:sys} is $\pd_t U - \pd_x V + [U,V] = 0$, i.e.,
\be \label{eq:comp1}
\left\lbrace
\begin{array}{cc}
	\pd_t F - \pd_x A + C G - B H = 0, \\
	\pd_t G - \pd_x B + 2 \lp B F - A G \rp = 0, \\
	\pd_t H - \pd_x C + 2 \lp A H -C F \rp = 0.
\end{array}
\right.
\ee
Importantly, this system must be equivalent to our original partial differential equation for all values of $\lambda$. Then $\lambda$ will generate an infinite number of conserved quantities which can be used to solve the problem, either exactly using the inverse scattering method when it applies, or approximately using the Whitham equations. 

To see this, it is convenient to define two linearly independent solutions $\psi$ and $\phi$ of the linear problem and the three scalar quantities
\be \label{eq:deffgh}
f \equiv -\frac{i}{2} \lp \psi_1 \phi_2 + \psi_2 \phi_1 \rp, \; g \equiv \psi_1 \phi_1, \; \text{and} \; h \equiv -\psi_2 \phi_2.
\ee
Partial derivatives of $f$, $g$, and $h$ can be computed straightforwardly. We find
\be \label{eq:comp2}
\begin{array}{cc}
	\pd_x f = i G h-i H g, \\
	\pd_t f = i B h - i C g, \\
	\pd_x g = 2 F g + 2 i G f, \\
	\pd_t g = 2 A g + 2 i B f, \\
	\pd_x h = -2 F h-2 i H f, \\
	\pd_t h = -2 A h-2 i C f.
\end{array}
\ee 
Conversely, the compatibility conditions of the system \eq{eq:comp2} give back \eq{eq:comp1} provided $g f h \neq 0$. \eq{eq:comp2} can thus be seen as a rewriting of the original problem. This formulation is particularly useful for deriving conserved or slowly-varying quantities, as we now explain. 

We first notice that $f^2 - g h$ is directly related to the generalized wronskian of $(\psi,\phi)$:
\be 
f^2 - g h = -\frac{1}{4} W^2, \; W \equiv \psi_1 \phi_2 - \psi_2 \phi_1.
\ee 
Since $U$ and $V$ are traceless, $\pd_t W = \pd_x W = 0$. So, $f^2 - g h$ depends only on the spectral parameter $\lambda$. As we shall see, for the solutions we are interested in this quantity is a polynomial in $\lambda$, which we denote as $P(\lambda)$. The crux of the construction is that $g$ satisfies the following conservation law: 
\be \label{eq:consrv}
\pd_t \lp \frac{G}{g} \rp - \pd_x \lp \frac{B}{g} \rp = 0.
\ee
\eq{eq:consrv} can be used to generate an infinite number of conservation laws after expanding $g$ in powers of $\lambda$. For our purposes, it is enough to retain only a finite number of terms, giving as many Whitham equations. 

Let us assume that we can find solutions of \eq{eq:comp2} where $g$ reads 
\be 
g(x,t;\lambda) = g_0(x,t;\lambda) \lp \lambda - \mu(x,t) \rp,
\ee
where $g_0$ is a smooth function which does not vanish at $\lambda = \mu(x,t)$. Evaluating the partial derivatives of $g$ gives
\be 
\pd_t \mu = -2 i \frac{B}{g_0} \sqrt{P(\mu)}, \; \pd_x \mu = -2 i \frac{G}{g_0} \sqrt{P(\mu)}.
\ee

So far we worked with exact solutions of the problem. Let us now consider a solution with two well-separated length-scales: a ``fast'' scale on which it oscillates periodically and a ``slow'' one on which the parameters describing the local oscillations, such as their amplitude, mean value and wave vector, vary. We can then apply the above procedure in two different ways:
\begin{itemize}
	\item One can define global functions $f_g, g_g, h_g$ which describe the exact solutions but have in general no analytical closed form.
	\item One can also define local functions $f_l, g_l, h_l$ by neglecting all variations on the ``slow'' scale.
\end{itemize}
Locally, these global and local solutions have the same form by definition, except for the normalization of the vectors $\psi, \phi$ used to define them. In defining the global functions, this normalization must be independent of $x$ and $t$, say $P_g(\lambda) = 1$. However, in general $f_l, g_l, h_l$ will take a simpler form when using a normalization such that $P_l$ depends on $x$ and $t$. Then,
\be 
(f_l(x,t), g_l(x,t), h_l(x,t)) = \sqrt{P_l(x,t;\lambda)} (f_g(x,t), g_g(x,t), h_g(x,t)).
\ee
To avoid unnecessarily cumbersome notations, in the following we shall not write the index $l$ explicitly, as we shall work only with the ``local'' functions. The exact conservation law \eq{eq:consrv} applied to $g_g$ gives
\be 
\pd_t \lp \sqrt{P(x,t,\lambda)} \frac{G(x,t)}{g(x,t)} \rp - \pd_x \lp \sqrt{P(x,t,\lambda)} \frac{B(x,t)}{g(x,t)} \rp = 0.
\ee
We now assume that $G/g_0$ is a constant, which is the case for the solutions we will consider. Then, at fixed $t$, $d\mu \propto \sqrt{P(\mu)} dx$. We can then average the conservation law by integrating over a few wavelengths and replacing the integral over $x$ by one over $\mu$, taken over a contour encircling its locus in the complex plane:
\be \label{eq:intcons} 
\pd_t \lp \oint \sqrt{P(x,t;\lambda)} \frac{G(x,t;\lambda)}{\sqrt{P(x,t;\mu)} g_0(x,t;\lambda) (\lambda - \mu)} d \mu \rp - \pd_x \lp \oint \sqrt{P(x,t;\lambda)} \frac{B(x,t;\lambda)}{\sqrt{P(x,t;\mu)} g_0(x,t;\lambda) (\lambda - \mu)} d \mu \rp \approx 0. \nn
\ee
The last step is to extract the singularities in $\lambda$ from \eq{eq:intcons}. These come from the simple roots $\lambda_i$ of $P(x,t;\lambda)$, which give after differentiation terms in $1/\sqrt{\lambda - \lambda_i}$. Cancelling them gives the general form of the Whitham modulation equations:
\be 
\lp \oint \frac{G(x,t;\lambda_i)}{\sqrt{P(x,t;\mu)}g_0(x,t;\lambda_i) (\lambda_i - \mu)} d\mu \rp \pd_t \lambda_i - \lp \oint \frac{B(x,t;\lambda_i)}{\sqrt{P(x,t;\mu)}g_0(x,t;\lambda_i) (\lambda_i - \mu)} d\mu \rp \pd_x \lambda_i \approx 0.
\ee
Notice that when neglecting the slow evolution, each of the two integrals depends only on the local parameters of the solution. If the ansatz chosen for $f$, $g$, and $h$ and leading to the polynomial form of $W$ is sufficiently general, these parameters can all be expressed in terms of the roots $\la_i$ of $P$. Then, for each root $\lambda_i$, the Whitham equations may be written in the more transparent form
\be 
\lp \pd_t + v_i(\left\lbrace \lambda_j \right\rbrace) \pd_x \rp \lambda_i = 0,
\ee
where
\be \label{eq:vi}
v_i \equiv -\frac{\oint \frac{B(x,t;\lambda_i)}{\sqrt{P(x,t;\mu)}g_0(x,t;\lambda_i) (\lambda_i - \mu)} d\mu}{\oint \frac{G(x,t;\lambda_i)}{\sqrt{P(x,t;\mu)}g_0(x,t;\lambda_i) (\lambda_i - \mu)} d\mu}.
\ee
The roots $\la_i$ are called the Riemann invariants. 

\subsection{Applications and characteristic velocities}

Let us apply the above formalism to the GP equation. Direct calculation using \eq{eq:comp1} shows that the compatibility condition obtained when choosing
\be \label{eq:UVNLS}
U = \begin{pmatrix}
	-i \lambda & i a \psi \\
	-i a \psi^* & i \lambda
\end{pmatrix}, \; 
V = \begin{pmatrix}
	-i \lambda^2 - i \frac{a^2}{2} \left\lvert \psi^2 \right\rvert & i \lambda a \psi - \frac{a}{2} \pd_x \psi \\
	-i a \lambda \psi^* & i \lambda^2 + i \frac{a^2}{2} \left\lvert \psi^2 \right\rvert
\end{pmatrix}
\ee
(which are obviously of the form \eq{eq:formAKNS}) is
\be 
i \pd_t \psi + \frac{1}{2} \pd_x^2 \psi -a^2 \left\lvert \psi^2 \right\rvert \psi = 0.
\ee
This is exactly \eq{eq:GPE} with a uniform two-body coupling\footnote{To avoid conflicts of notations, in this appendix we denote the two-body coupling as $a^2$, and use the symbol ``$g$'' only for the function defined in \eq{eq:deffgh}.} 
$a^2$  and a vanishing potential. Notice that a uniform potential $V$ can be absorbed in a redefinition of $\psi$ through $\psi \to e^{-i V t} \psi$. So, the choice \eq{eq:UVNLS} allows us to find solutions in the presence of uniform $V$ and $a^2$. We look for solutions where $f$ is real-valued and $h = g^*$. The system \eq{eq:comp2} then becomes
\be \label{eq:compNLS1}
\left\lbrace 
\begin{array}{ll}
	\pd_x f \eg -a \lp \psi g^*+\psi^* g \rp \\
	\pd_t f \eg -a \lp \lp \lambda \psi + \frac{i}{2} \pd_x \psi \rp g^*+ \lp \lambda \psi^* -\frac{i}{2} \pd_x \psi^* \rp g \rp \\
	\pd_x g \eg -2 i \lambda g - 2 a \psi f \\
	\pd_t g \eg -2 i \lp \lambda^2 + \frac{a^2}{2} \abs{\psi^2} \rp - 2 a \lp \la \psi + \frac{i}{2} \pd_x \psi \rp f
\end{array}
\right.  .
\ee 
We now further restrict to solutions of the form
\be \label{eq:ansatz}
\begin{array}{ll}
	f(x,t;\la) \eg \la^2-f_1(x,t) \lambda + f_2, \nn
	g(x,t;\la) \eg i a \psi(x,t) \lp \la - \mu(x,t) \rp.
\end{array}
\ee
The justification of this ansatz will come {\it a posteriori} by obtaining all the periodic, stationary solutions. Expanding \eq{eq:compNLS1} in powers of $\la$ and eliminating the trivial equalities gives a system of 8 differential equations on $f_1$, $f_2$, and $\mu$. It is useful to parametrize the solution using the coefficients of $P(\lambda)$. In our case, it is a fourth-order polynomial of the form
\be \label{eq:Pla}
P(\la) = \la^4 - s_1 \la^3 + s_2 \la^2 - s_3 \la + s_4.
\ee
$f_1$ and $f_2$ are related to $s_1$ and $s_2$ through
\be 
f_1 = \frac{s_1}{2}, \; f_2 = \frac{s_2}{2} - \frac{s_1^2}{6}+\frac{a^2}{2} \abs{\psi^2}.
\ee
Variations of $\mu$ are given by
\be \label{eq:mu}
\left\lbrace
\begin{array}{rl}
	\pd_x \mu(x,t) \eg -2 i \sqrt{P(\mu(x,t)} \\
	\pd_t \mu(x,t) \eg \frac{s_1}{2} \pd_x \mu(x,t)
\end{array}
\right. .
\ee
The corresponding equations on $\psi$ are more conveniently written in terms of
\be 
\tilde{\psi}(x,t) \equiv e^{i \lp \frac{s_1^2}{4} - s_2 \rp t} \psi(x,t). 
\ee
We obtain
\be 
\left\lbrace
\begin{array}{rl}
	i \pd_x \tilde{\psi}(x,t) \eg 2 \lp \frac{s_1}{2} - \mu(x,t) \rp \tilde{\psi}(x,t) \\
	i \pd_t \tilde{\psi}(x,t) \eg \frac{s_1}{2} i \pd_x \tilde{\psi}(x,t)
\end{array}
\right. .
\ee
Notice that $\mu$ and $\tilde{\psi}$ depend on $x$ and $t$ only through $\xi \equiv x + \frac{s_1}{2} t$. The density perturbations thus move with the velocity $-s_1 / 2$ without changing their shape. Using the two other conserved quantities $s_3$ and $s_4$ to determine the relationship between $\mu$ and $\psi$,  one obtains a closed equation on $\abs{\psi}$ of the form 
\be \label{eq:psiW}
\frac{d \abs{\psi^2}}{d \xi} = \pm \frac{2}{a^2} \sqrt{-\mathcal{R} \lp \abs{\psi^2} \rp},
\ee
where $\mathcal{R}$ is a third-order polynomial with roots at $(\lambda_4 + \lambda_3 - \lambda_1 - \lambda_2 )^2/(4 a^2)$, $(\lambda_4 - \lambda_3 - \lambda_1 + \lambda_2 )^2/(4 a^2)$, and $(\lambda_4 - \lambda_3 + \lambda_1 - \lambda_2 )^2/(4 a^2)$, and $\la_i$, $i =1..4$ are the four roots of $P(\la)$, i.e., the Riemann invariants. They are ordered so that $\la_1 \leq \la_2 \leq \la_3 \leq \la_4$. Importantly, one can check that all the periodic solutions of the GP equation can be recovered from \eq{eq:psiW}, which justifies the ansatz \eq{eq:ansatz}.

Let us now turn to slowly-modulated solutions, for which the parameters $\la_i$ slowly vary in space and time. Using Eqs.~(\ref{eq:vi},\ref{eq:mu}), we find that the characteristic velocities are
\be 
v_i = \frac{1}{2} \lp \frac{L}{\pd_{\lambda_i}L} - s_1 \rp,
\ee
where $L$ is the wavelength of the periodic solution, given by
\be 
L = \frac{a^2}{\sqrt{(\la_4 - \la_2) (\la_3 - \la_1)}} K \lp \frac{(\la_4 - \la_3) (\la_2 - \la_1)}{(\la_4 - \la_2) (\la_3 - \la_1)} \rp,
\ee
where $K$ is the complete elliptic integral of the first kind~\cite{Abramowitz}.

\subsection{Dispersive shock waves and simple waves} 
\label{app:DSWNLS}

We now look for scale-invariant solutions depending only on $z \equiv x/t$. The Whitham equations then take the simple form 
\be \label{eq:Whz}
\lp v_i - z \rp \frac{d \la_i}{dz} = 0.
\ee
That is, each Riemann invariant is constant except in a domain where the associated velocity is equal to $z$. We further restrict to solutions for which $\abs{\psi}$ is asymptotically homogeneous at $z \to \pm \infty$. This means that $\mathcal{R}$ must have a double root in each asymptotic region, i.e., two Riemann invariants must be equal there. We thus have two possibilities to build a non-trivial solution:
\begin{itemize}
	\item If two Riemann invariants are equal and strictly homogeneous while a third one is varying, the solution is a simple wave (SW);
	\item If one Riemann invariant $\lambda_{i_0}$ varies between $\lambda_{i_0-1}$ at $z=z_-$ and $\lambda_{i_0+1}$ at $z=z_+$, the solution is a dispersive shock wave (DSW).
\end{itemize}
In the first case, the solution shows no oscillations as two Riemann invariants remain equal all along. In the second case, it shows oscillations which start in the linear regime on one side and become widely spaced solitons when approaching the other side. A SW and a DSW are shown in \fig{fig:solsNLS}. 

Let us first focus on SW. A careful analysis shows that there are only two possibilities, corresponding to SW propagating to the left or to the right in the fluid frame. They are characterized by
\be \label{eq:cons1} 
\left\lbrace
\begin{array}{rl}
	\pd_z \lp \pm a \sqrt{\rho} -\frac{v}{2} \rp \eg 0 \\
	v \pm a \sqrt{\rho} \eg z
\end{array}
\right. ,
\ee
where $\rho \equiv \left\lvert \psi \right\rvert^2$ is the density and $v \equiv - i (\pd_x \psi) / \psi$ the velocity of the condensate. 

There are also two different DSW: one along which $\la_2$ varies between $\la_1$ and $\la_3$ and one along which $\la_3$ varies between $\la_2$ and $\la_4$. Each of them interpolates between a subsonic homogeneous solution at $z=z_b$ and a supersonic one at $z=z_p$. In the first case, these extremal values of $z$ are given by
\be \label{eq:NLSzb} 
z_b = v_p + a \sqrt{\rho_b}
\ee
and 
\be \label{eq:NLSzp}
z_p = \frac{8 a \sqrt{\rho_b} \lp v_p - v_b \rp + v_p^2 - \lp v_b + 2 a \sqrt{\rho_b} \rp^2}{2 \lp v_p-v_b-2 a \sqrt{\rho_b} \rp},
\ee
where an index $b$ (respectively, $p$) indicates a quantity evaluated at $z=z_b$ (respectively $z=z_p$). The constraints on this solution, coming from the conservation of $\lambda_4$ and the assumption $\lambda_1 < \lambda_3$, are 
\be \label{eq:cons2}
a \sqrt{\rho_p} - \frac{v_p}{2} = a \sqrt{\rho_b} - \frac{v_b}{2}, \; \rho_b > \rho_p.
\ee
Importantly for our purposes, a direct calculation gives $z_p - z_b>0$. The second DWS is obtained by flipping the sign in front of $a$. The constraints on this solution are
\be 
a \sqrt{\rho_p} + \frac{v_p}{2} = a \sqrt{\rho_b} + \frac{v_b}{2}, \; \rho_b > \rho_p.
\ee
A direct calculation gives $z_p - z_b<0$. 

\subsection{Three-waves solutions of the Whitham equations in the single step configuration}
\label{App:domex}

We now give the domain of existence (in the space of asymptotic conditions) of the 8 solutions of the Whitham equations mentioned in Section~\ref{sec:analytical}. They are given by the above conditions and those on the signs of the velocities of the nonlinear waves. The solution with 3 SW exists for
\be 
\left\lbrace
\begin{array}{l}
	\rho_- \leq \rho_0 \\
	\sqrt{g_- \rho_0} - \sqrt{g_- \rho_-} - \sqrt{g_+ \rho_0} - \sqrt{g_+ \rho_+} \leq \frac{v_- - v_+}{2} \leq \sqrt{g_- \rho_0} - \sqrt{g_- \rho_-} - \left\lvert \sqrt{g_+ \rho_0} - \sqrt{g_+ \rho_+} \right\rvert \\
	\sqrt{g_+ \rho_0} + 2 \sqrt{g_- \rho_0} -2 \sqrt{g_- \rho_-} \leq v_- \leq 3 \sqrt{g_- \rho_0} - 2 \sqrt{g_- \rho_-}
\end{array}
\right. .
\ee
The solution with one DSW on the left exist if and only if
\be 
\left\lbrace
\begin{array}{l}
	\rho_- \leq \rho_0 \\
	\sqrt{g_- \rho_0} - \sqrt{g_- \rho_-} - \sqrt{g_+ \rho_0} - \sqrt{g_+ \rho_+} \leq \frac{v_- - v_+}{2} \leq \sqrt{g_- \rho_0} - \sqrt{g_- \rho_-} - \left\lvert \sqrt{g_+ \rho_0} - \sqrt{g_+ \rho_+} \right\rvert \\
	2 \sqrt{g_- \rho_1} + \sqrt{g_+ \rho_1} -2 \sqrt{g_- \rho_-} \leq u_- \leq \sqrt{g_- \rho_1}
\end{array}
\right. .
\ee
The solution with one DSW between two SW exists if and only if
\be 
\left\lbrace
\begin{array}{l}
	\rho_- \geq \rho_0 \\
	\sqrt{g_- \rho_0} - \sqrt{g_- \rho_-} + \sqrt{g_+ \rho_0} - \sqrt{g_+ \rho_+} \leq \frac{v_- - v_+}{2} \leq \sqrt{g_- \rho_0} - \sqrt{g_- \rho_-} - \sqrt{g_+ \rho_0} + \sqrt{g_+ \rho_+} \\
	v_- + 2 \sqrt{g_- \rho_-} - 3 \sqrt{g_- \rho_1} \leq 0
\end{array}
\right. .
\ee
The solution with one DSW on the right of two SW exists if and only if
\be 
\left\lbrace
\begin{array}{l}
	\rho_- \geq \rho_0 \\
	\sqrt{g_- \rho_0} - \sqrt{g_- \rho_-} - \sqrt{g_+ \rho_0} + \sqrt{g_+ \rho_+} \leq \frac{v_- - v_+}{2} \leq \sqrt{g_- \rho_0} - \sqrt{g_- \rho_-} + \sqrt{g_+ \rho_0} - \sqrt{g_+ \rho_+} \\
	\sqrt{g_+ \rho_0} +2 \sqrt{g_- \rho_1} -2 \sqrt{g_- \rho_-} \leq u_- \leq 3 \sqrt{g_- \rho_1} -2 \sqrt{g_- \rho_-}
\end{array}
\right. .
\ee
The solution with one SW on the left of two DSW exists if and only if
\be 
\left\lbrace
\begin{array}{l}
	\rho_- \geq \rho_0 \\
	\sqrt{g_- \rho_-} - \sqrt{g_- \rho_0} - \left\lvert \sqrt{g_+ \rho_+} - \sqrt{g_+ \rho_0} \right\rvert \leq \frac{v_- - v_+}{2} \leq \sqrt{g_- \rho_0} - \sqrt{g_- \rho_-} + \sqrt{g_+ \rho_+} + \sqrt{g_+ \rho_0} \\
	v_+^2 - v_+ v_- + 2 v_- \sqrt{g_+ \rho_+} - 2 v_+ \lp \sqrt{g_- \rho_-} - \sqrt{g_- \rho_1} + 2\sqrt{g_+ \rho_+} \rp + 4 \lp \sqrt{g_- \rho_-} - \sqrt{g_- \rho_0} + \sqrt{g_+ \rho_+} \rp \sqrt{g_+ \rho_+} -2 g_+ \rho_0 \leq 0\\
	v_- + 2 \sqrt{g_- \rho_-} - 3 \sqrt{g_- \rho_0} \leq 0 \\
\end{array}
\right. .
\ee
The solution with one SW between two DSW exists if and only if
\be 
\left\lbrace
\begin{array}{l}
	\rho_- \leq \rho_0 \\
	\sqrt{g_+ \rho_+} - \sqrt{g_- \rho_-} - \sqrt{g_+ \rho_0} + \sqrt{g_- \rho_0} \leq \frac{v_- - v_+}{2} \leq \sqrt{g_+ \rho_0} + \sqrt{g_- \rho_0} - \sqrt{g_+ \rho_+} - \sqrt{g_- \rho_-} \\
	v_+ \geq -4 \sqrt{g_+ \rho_+} \\
	v_- + 2 \lp \sqrt{g_- \rho_-} - \sqrt{g_- \rho_0} \rp - \sqrt{g_+ \rho_0} \geq 0 \\
	v_- \geq \sqrt{g_- \rho_0}
\end{array}
\right. .
\ee
The solution with one SW on the right of two DSW exists if and only if
\be 
\left\lbrace
\begin{array}{l}
	\rho_- \leq \rho_0 \\
	v_- \leq \sqrt{g_- \rho_0} \\
	\sqrt{g_- \rho_0} - \sqrt{g_- \rho_-} - \sqrt{g_+ \rho_+} + \sqrt{g_+ \rho_0}  \leq \frac{v_- - v_+}{2} \leq \sqrt{g_- \rho_0} - \sqrt{g_- \rho_-} + \sqrt{g_+ \rho_+} - \sqrt{g_+ \rho_0} \\
	v_+^2 - v_+ v_- + 2 v_- \sqrt{g_+ \rho_+} - 2 v_+ \lp \sqrt{g_- \rho_-} - \sqrt{g_- \rho_1} + 2\sqrt{g_+ \rho_+} \rp + 4 \lp \sqrt{g_- \rho_-} - \sqrt{g_- \rho_0} + \sqrt{g_+ \rho_+} \rp \sqrt{g_+ \rho_+} -2 g_+ \rho_0 \leq 0\\
\end{array}
\right. .
\ee
Finally, the solution with 3 DSW exists if and only if
\be 
\left\lbrace
\begin{array}{l}
	\rho_- \leq \rho_0 \\
	v_- \leq \sqrt{g_- \rho_0} \\
	\sqrt{g_- \rho_0} - \sqrt{g_- \rho_-} + \left\lvert \sqrt{g_+ \rho_+} - \sqrt{g_+ \rho_0} \right\rvert \leq \frac{v_- - v_+}{2} \leq \sqrt{g_- \rho_0} - \sqrt{g_- \rho_-} + \sqrt{g_+ \rho_+} + \sqrt{g_+ \rho_0} \\
	v_+^2 - v_+ v_- + 2 v_- \sqrt{g_+ \rho_+} - 2 v_+ \lp \sqrt{g_- \rho_-} - \sqrt{g_- \rho_1} + 2\sqrt{g_+ \rho_+} \rp + 4 \lp \sqrt{g_- \rho_-} - \sqrt{g_- \rho_0} + \sqrt{g_+ \rho_+} \rp \sqrt{g_+ \rho_+} -2 g_+ \rho_0 \leq 0\\
\end{array}
\right. .
\ee

\begin{figure}[h!]
	\begin{center}
		\includegraphics[width=0.49 \linewidth]{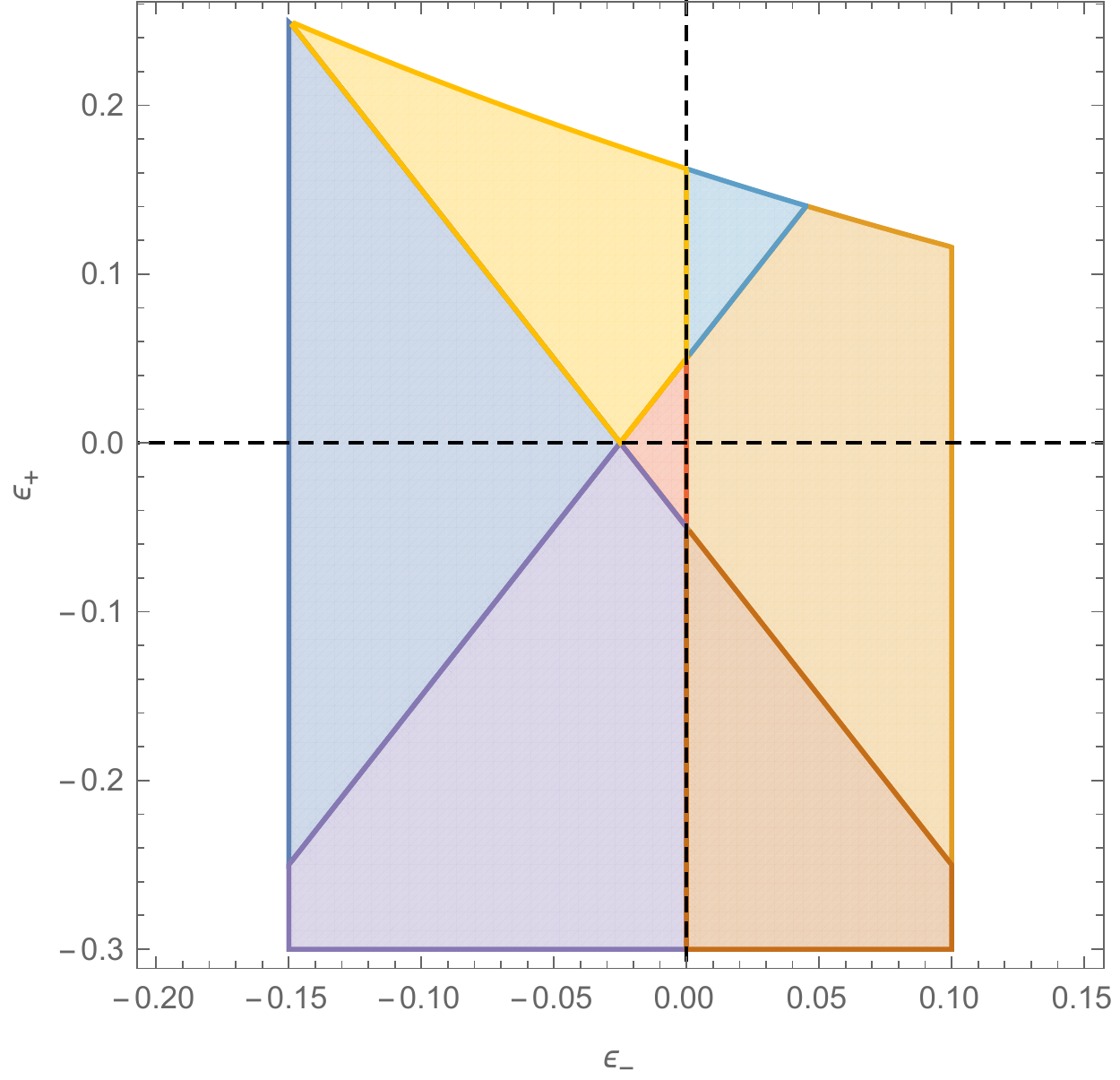}
		\includegraphics[width=0.49 \linewidth]{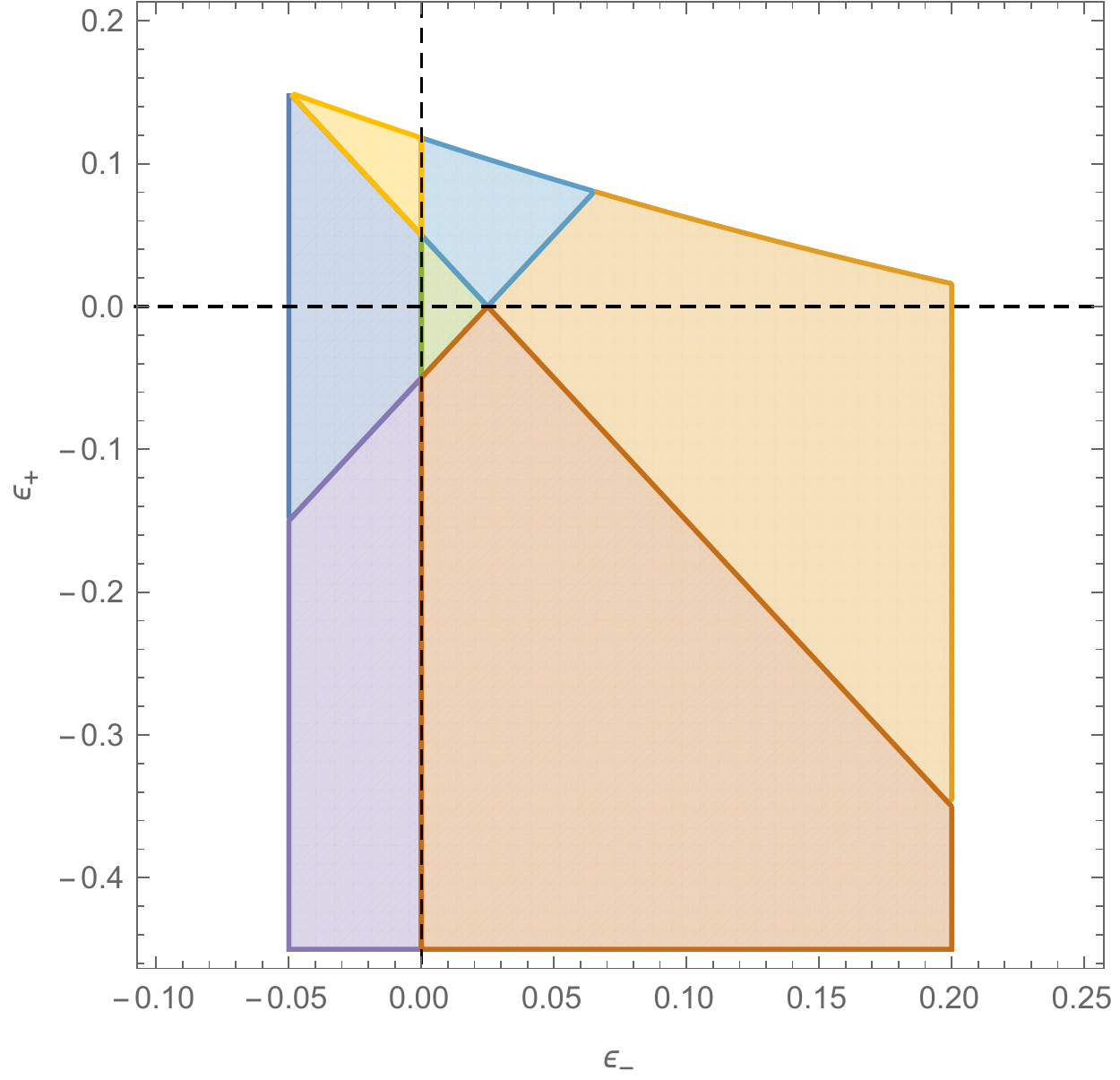}
	\end{center}
	\caption{Domain of existence of the 8 solutions with 3 waves in the $(\ep_-, \ep_+)$ plane, where $\ep_i \equiv \sqrt{\rho_i / \rho_0} - 1$, for $g_- = 4 g_+$. The asymptotic velocities are given by $v_- = 0.8 \sqrt{g_- \rho_0}, \; v_+ = 3 \sqrt{g_+ \rho_0}$ (left) and $v_- = 0.6 \sqrt{g_- \rho_0}, \; v_+ = 2.6 \sqrt{g_+ \rho_0}$ (right). The horizontal dashed line shows the locus $\ep_+ = 0$, while the vertical one shows $\ep_- = 0$. The domains of the solutions described in the text are shown, in the same order, in green, blue, cyan, brown, orange, purple, yellow, and red.}\label{fig:8sols}
\end{figure}

The domains of existence of these solutions are shown in \Fig{fig:8sols} for fixed asymptotic velocities. 
Interestingly, one can show that one of them always exists provided the asymptotic conditions are sufficiently close to those obtained from a homogeneous black-hole flow. 
To see this, let us assume that $\sqrt{g_+ \rho_0} < v_- < \sqrt{g_- \rho_0}$ 
(this condition is always satisfied for a flow close to a homogeneous black-hole one). Then, 
\begin{itemize}
	\item if $-4 \sqrt{g_+ \rho_0} < v_- - v_+ < 0$ and $\rho_0 - \rho_- \geq 0$ is sufficiently small, the solution with three SW exists in an open interval of $\rho_+$ containing $\rho_0$;
	\item if $-2 \sqrt{g_+ \rho_0} < v_- - v_+ < 0$ and $\rho_- - \rho_0 \geq 0$ is sufficiently small, the solution with one DSW on the left of two SW exists in an open interval of $\rho_+$ containing $\rho_0$;
	\item if $\max \lp -4 \sqrt{g_+ \rho_0}, \frac{1}{2} \lp 4 \sqrt{g_+ \rho_0} - v_- - \sqrt{g_- \rho_0} \sqrt{8+ \frac{v_-^2}{g_+ \rho_0}} \rp \rp < v_+ - v_- < 0$ and $\rho_- - \rho_0 \geq 0$ is sufficiently small, then the solution with a SW on the left of two DSW exists in an open interval of $\rho_+$ containing $\rho_0$;
	\item if $\max \lp -4 \sqrt{g_+ \rho_0}, \frac{1}{2} \lp 4 \sqrt{g_+ \rho_0} - v_- - \sqrt{g_- \rho_0} \sqrt{8+ \frac{v_-^2}{g_+ \rho_0}} \rp \rp < v_+ - v_- < 0$ and $\rho_0 - \rho_- \geq 0$ is sufficiently small, then the solution with three DSW exists in an open interval of $\rho_+$ containing $\rho_0$.
\end{itemize}
To be complete, one must also consider the case $v_- = v_+$, $\rho_+ \neq \rho_-$. There are then six possibilities. Assuming $\rho_+$ and $\rho_-$ are both close to $\rho_0$,
\begin{itemize}
	\item If $\rho_- > \rho_0$,
	\begin{itemize}
		\item If $\sqrt{\frac{\rho_+}{\rho_0}} - 1 > \sqrt{\frac{g_-}{g_+}} \lp \sqrt{\frac{\rho_-}{\rho_0}} - 1 \rp$, we obtain a solution with one DSW between two SW; 
		\item If $ \left\lvert \sqrt{\frac{\rho_+}{\rho_0}} - 1 \right\rvert < \sqrt{\frac{g_-}{g_+}} \lp \sqrt{\frac{\rho_-}{\rho_0}} - 1 \rp$, we have one SW on the left of two DSW;
		\item If $\sqrt{\frac{\rho_+}{\rho_0}} - 1 <- \sqrt{\frac{g_-}{g_+}} \lp \sqrt{\frac{\rho_-}{\rho_0}} - 1 \rp$, we have one DSW on the right of two SW;
	\end{itemize}
	\item If $\rho_- < \rho_0$,
	\begin{itemize}
		\item If $\sqrt{\frac{\rho_+}{\rho_0}} - 1 > - \sqrt{\frac{g_-}{g_+}} \lp \sqrt{\frac{\rho_-}{\rho_0}} - 1 \rp$, we have one SW on the right of two DSW;
		\item If $ \left\lvert \sqrt{\frac{\rho_+}{\rho_0}} - 1 \right\rvert < -\sqrt{\frac{g_-}{g_+}} \lp \sqrt{\frac{\rho_-}{\rho_0}} - 1 \rp$, we have one DSW on the left of two SW;
		\item If $\sqrt{\frac{\rho_+}{\rho_0}} - 1 < \sqrt{\frac{g_-}{g_+}} \lp \sqrt{\frac{\rho_-}{\rho_0}} - 1 \rp$, we have one SW between two DSW.
	\end{itemize}
\end{itemize}
So, if the asymptotic conditions are sufficiently close to those from a black-hole flow with a homogeneous density, there always exists a solution of the Whitham equations with three waves moving away from $x=0$ and leaving such a flow behind them. 

\section{Nonlinear evolution for KdV equations}
\label{App:KdV}

To complete the analysis of Sections~\ref{sub:Ahtf}, \ref{sec:analytical} and appendix~\ref{App:domex}, we here extend it to the case of the KdV equation with variable coefficients. Our first goal is to show explicitly that the main results are not restricted to the specific case of the GP equation. We also aim at unveiling the qualitative differences which arise when dealing with a subluminal dispersion relation. To disentangle these two aspects, we also consider a superluminal version of the KdV equation. 

\subsection{KdV equation}

\subsubsection{KdV equation with variable coefficients}

To make a parallel with results obtained from the GP equation at both the linear and nonlinear levels, it is convenient to use a KdV equation which derives from an action principle giving a canonical Hamiltonian structure. Let us consider a real scalar field $\psi$ with Lagrangian density
\be \label{eq:LKdV}
\mathcal{L} = \lp \pd_x \psi \rp \lp \pd_t \psi \rp + \lp \sqrt{g h(x)} + v(x) \rp \lp \pd_x \psi \rp^2 + \frac{1}{2} \sqrt{\frac{g}{h(x)}} \lp \pd_x \psi \rp^3 - \frac{h(x)^2}{6} \sqrt{g h(x)} \lp \pd_x^2 \psi \rp^2,
\ee
where $g$ is the gravitational acceleration, $h(x)$ the background water depth, and $v(x)$ the background flow velocity. The momentum conjugate to $\psi$ is $\zeta \equiv \pd_x \psi$. The Euler-Lagrange equation from \eq{eq:LKdV} is
\be \label{eq:KdV}
\pd_t \zeta + \pd_x \lp \lp \sqrt{g h(x)} + v(x) \rp \zeta + \frac{3}{4} \sqrt{\frac{g}{h(x)}} \zeta^2 + \pd_x \lp \frac{h(x)^2}{3} \sqrt{g h(x)} \pd_x \zeta \rp \rp = 0.
\ee
The dispersion relation for a linear perturbation $\delta \zeta \propto e^{-i \om t + i k x}$ in a homogeneous region is
\be 
\om - \lp \sqrt{g h} + v + \frac{3}{2} \sqrt{\frac{g}{h}} \zeta \rp k + \frac{h^2}{6} \sqrt{g h}k^3 = 0.
\ee
There is an analogue horizon where $\sqrt{g h} + v + \frac{3}{2} \sqrt{\frac{g}{h}} \zeta = 0$. It corresponds to a black hole horizon if this quantity passes from negative to positive when increasing $x$. 

\subsubsection{Stationary black-hole solutions in the steep regime}

When considering stationary solutions, integrating \eq{eq:KdV} over $x$ gives
\be 
\lp \sqrt{g h} + v \rp \zeta + \frac{3}{4} \sqrt{\frac{g}{h}} \zeta^2 + \pd_x \lp \frac{h^2}{6} \sqrt{g h} \pd_x \zeta \rp = C,
\ee 
where $C$ is a real integration constant. One can show that if $h$ and $v$ are homogeneous, then at most two asymptotically homogeneous solutions exist for each value of $C$. These are the homogeneous, subcritical solution and the soliton, which is asymptotically supercritical. 

Let us assume that $v(x)$ and $h(x)$ are piecewise constant functions with only one discontinuity at $x=0$: 
\be \label{eq:ansatzvh}
v(x) = \left\lbrace 
\begin{array}{rl}
	v_+ & x>0 \\
	v_- & x<0
\end{array}
\right.
, \;
h(x) = \left\lbrace 
\begin{array}{rl}
	h_+ & x>0 \\
	h_- & x<0
\end{array}
\right. .
\ee
We look for transcritical AH solutions for $v<0$. That is, we impose that $\zeta$ becomes homogeneous for $x \to \pm \infty$, and that the solution is supercritical for $x \to -\infty$ and subcritical for $x \to \infty$. Using the matching conditions at $x=0$, namely continuity of $\zeta$ and $\pd_x \zeta$, we find that the solution must then either be strictly homogeneous (in that case, the soliton on the left is sent to $x \to -\infty$) or contain a half soliton. 

A straightforward calculation shows that there exists two homogeneous solutions given by
\be 
\zeta=0
\ee
and
\be
\zeta = \frac{4}{3} \frac{\sqrt{g h_-}-\sqrt{g h_+}+v_- - v_+}{\sqrt{\frac{g}{h_+}}- \sqrt{\frac{g}{h_-}}}.
\ee
We choose parameters such that the trivial solution $\zeta = 0$ is a black-hole solution: $\sqrt{g h_-} + v_- < 0$ and $\sqrt{g h_+} + v_+ > 0$. Then, the second homogeneous solution is not transcritical. A long but straightforward calculation shows that a black-hole solution with a half-soliton exists if and only if $h_+ > 16 h_-$. So, for $h_+ < 16 h_-$ the trivial solution is the only AH transcritical stationary solution. The uniqueness result demonstrated in Section~\ref{sub:Ahtf} for the GP equation thus also applies to the KdV equation.

\subsection{Whitham modulation equations} 
\label{WKdV}

To apply the general formalism presented in Appendix~\ref{App:Whitham}, it is convenient to rewrite the KdV equation \eq{eq:KdV} in a canonical form. We assume the background flow is homogeneous and define the non-dimensional variables $Y$, $T$, and $u$ through
\be 
Y \equiv \frac{1}{h} \lp x- \lp \sqrt{g h} + v \rp t \rp, \; 
T \equiv \sqrt{\frac{g}{h}} \frac{t}{6}, \; \text{and} \;
u (Y,T) \equiv \frac{3}{2 h} \zeta (Y,T).
\ee
\eq{eq:KdV} then becomes
\be 
\pd_T u + 6 u \pd_Y u + \pd_Y^3 u = 0.
\ee
Direct calculation shows that it is the compatibility condition of the system \eq{eq:sys} with the choice
\be \label{eq:UVKdV}
U = \begin{pmatrix}
	-i \la & -1 \\
	u & i \la
\end{pmatrix}, \; 
V = \begin{pmatrix}
	-4 i \la^3 + 2 i u \la + \pd_Y u & -4 \la^2 + 2 u \\
	4 u \la^2 - 2 i \la \pd_Y u-2 u^2 - \pd_Y^2 u & 4 i \la^3 - 2 i \la u - \pd_Y u
\end{pmatrix} . 
\ee
The procedure is similar to the one we used for the GP equation, except that we can not restrict to solutions where $h = g^*$. Writing down the system \eq{eq:comp2} explicitly gives, after a few lines of algebra
\be 
\left\lbrace 
\begin{array}{ll}
	f + \la g - \frac{i}{2}  \pd_Y g  = 0  \\
	h + \frac{1}{2} \pd_Y^2 g + i \la \pd_Y g + u g = 0 \\
	\pd_Y f^2 - g h = 0 \\
	\pd_T f^2 - g h = 0 \\
	\pd_T g - 2 \lp g \pd_Y u - \lp 2 \la^2 + u \rp \pd_Y g \rp = 0
\end{array}
\right. .
\ee
We now look for solutions where $g$ has the form
\be 
g(Y,T; \la) = \la^2 - \mu(Y,T).
\ee
Notice that $P(\la) = f^2 - g h$ is now a third-order polynomial in $\la^2$. A straightforward calculation gives 
\be 
\left\lbrace 
\begin{array}{ll}
	u+s-2 \mu = 0 \\
	\lp \pd_T - 2 s \pd_Y \rp \mu = 0 \\
	\lp \pd_Y \mu \rp^2 + P(\la^2 = \mu) = 0
\end{array}
\right.,
\ee
where $s = \mu_1 + \mu_2 + \mu_3$ and $\mu_1 \leq \mu_2 \leq \mu_3$ are the three roots of $P(\la)$ in $\la^2$. One can check that this system describes all the periodic solutions which are stationary in a Galilean frame.

Turning to modulated solutions, the parameters $\mu_i$ become slowly-varying functions of $Y$ and $T$ obeying the Whitham equations:
\be \label{eq:WhithamKdV}
\lp \pd_T + v_i \pd_Y \rp \mu_i = 0,
\ee
where
\be 
v_i \equiv 2 \lp \frac{L}{\pd_{\mu_i} L} - s \rp 
\ee
and $L$ is the local wavelength, given by
\be 
L = \frac{1}{\sqrt{\mu_3 - \mu_1}} K \lp \frac{\mu_3 - \mu_2}{\mu_3 - \mu_1} \rp.
\ee
We now look for scale-invariant solutions depending only on $Z \equiv Y/T$. Then \eq{eq:WhithamKdV} becomes
\be 
\lp v_i - Z \rp \frac{d \mu_i}{d Z} = 0. 
\ee
There are two non-trivial solutions of this system giving asymptotically homogeneous values of $u$. The (unique) simple wave, where one Riemann invariant varies over a finite interval while the two other ones are equal and constant, corresponds to $u = Z/6$. In the original variables of \eq{eq:KdV}, it becomes 
\be \label{eq:SWKdV}
\zeta(x,t) = \frac{2}{3} \sqrt{\frac{h}{g}} \lp \frac{x}{t} - \lp \sqrt{g h} + v \rp \rp
\ee
over the domain of variation of $\zeta$. 

There is also one dispersive shock wave, along which $\mu_1$ and $\mu_3$ are constant while $\mu_2$ varies from $\mu_1$ to $\mu_3$. Its two edges are located at \be 
\frac{x}{t} = z_1 = -\frac{\sqrt{g h}}{3} \lp 2 \mu_1 + \mu_3 \rp + \sqrt{g h} +v
\ee
and
\be 
\frac{x}{t} = z_3 = -\sqrt{g h} \lp 2 \mu_3 - \mu_1 \rp + \sqrt{g h} +v.
\ee
Since $\mu_3 > \mu_1$, we have $z_1 > z_3$. $\mu_1$ and $\mu_3$ are related to the values of $\zeta$ outside the shock wave through $\zeta(x/t < z_3) = -2 h \mu_1 / 3$ and $\zeta(x/t > z_1) = -2 h \mu_3 / 3$. Existence of this solution thus requires $\zeta(x/t < z_3) > \zeta(x/t > z_1)$. These two solutions will be used to build global ones in the next subsection. 

\subsection{Solving the Whitham equations}

\begin{figure}
	\begin{center}
		\includegraphics[width=0.49 \linewidth]{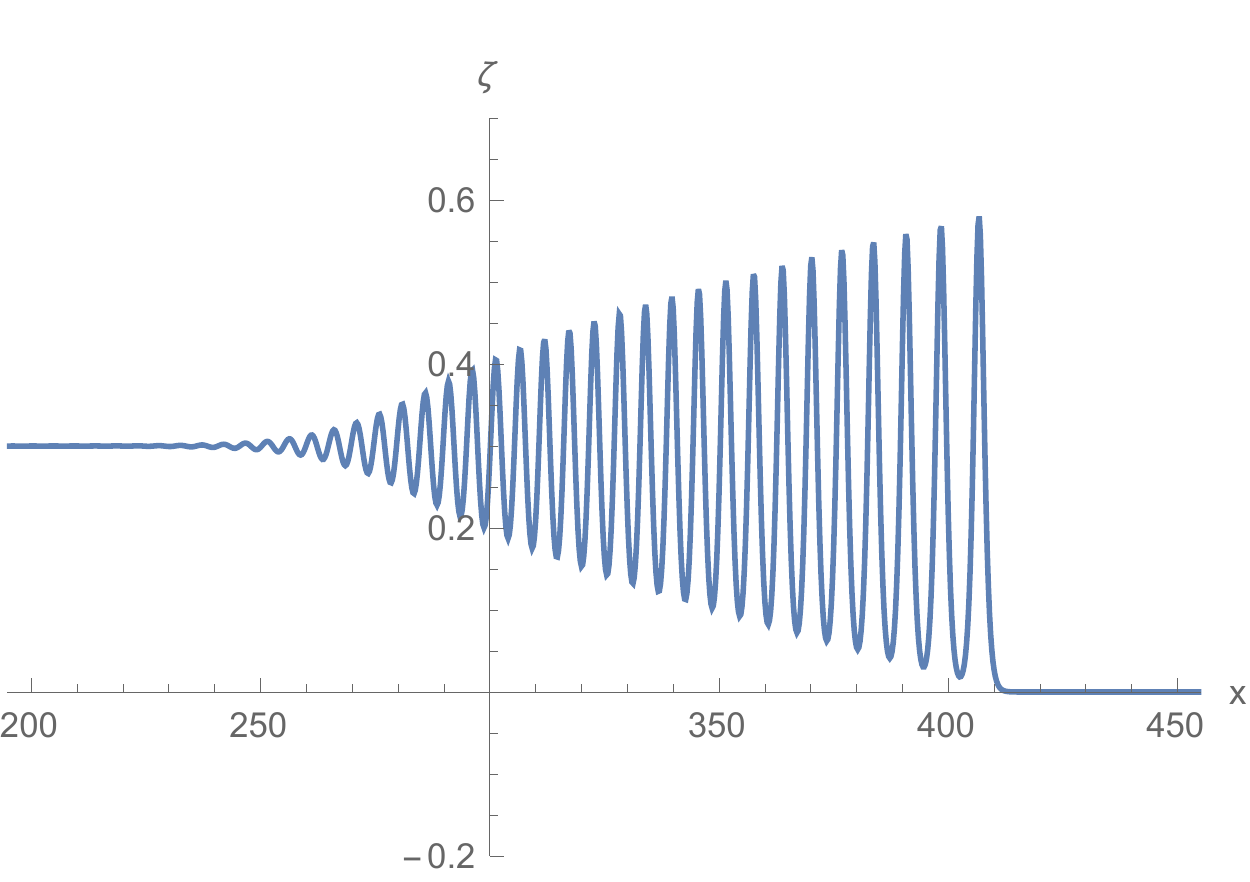}
		\includegraphics[width=0.49 \linewidth]{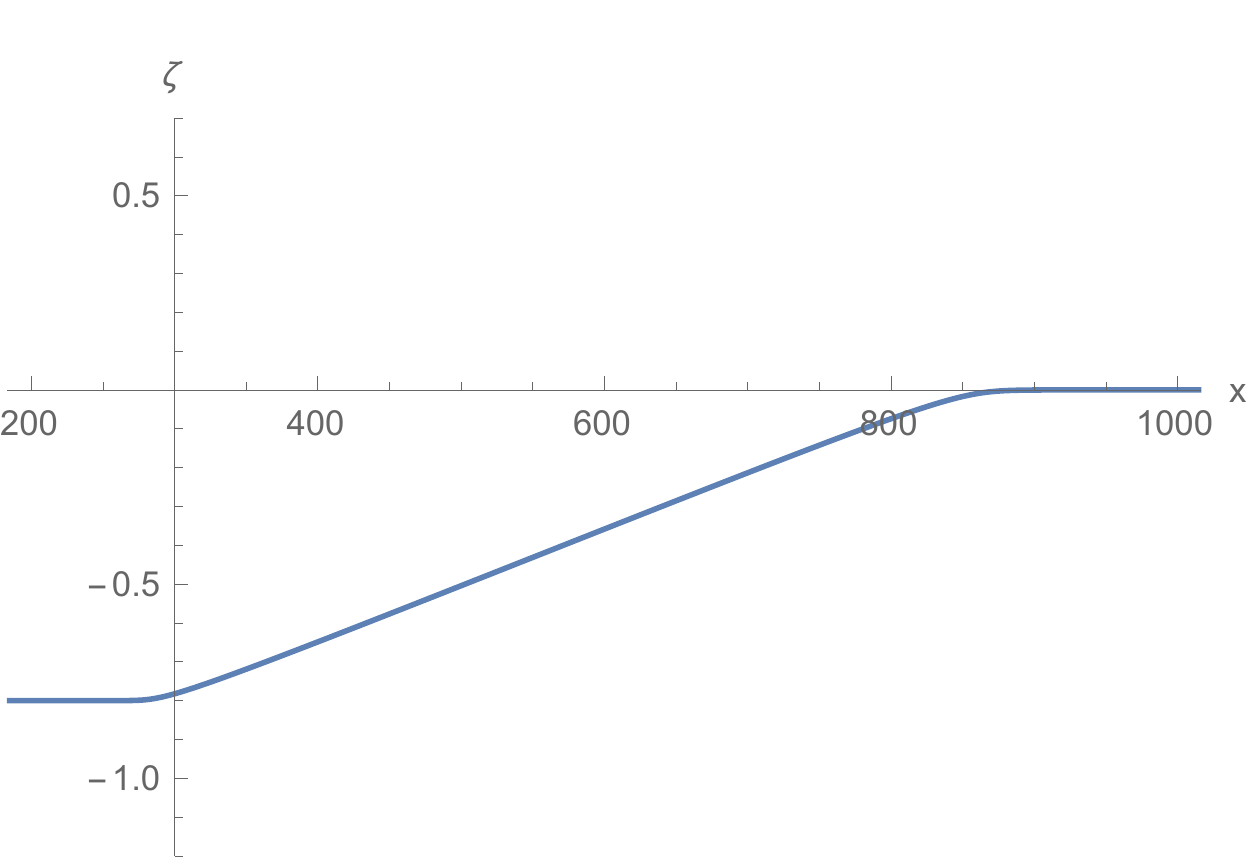}
		\caption{Here we show a DSW (left) and a SW (right) for the KdV equation. The SW is strictly scale-invariant, as along this solution $\zeta$ depends only on $x/t$ up to corrections not captured by the Whitham equations. For the DSW, the envelope and wavelength of the oscillations depend only on $x/t$.}\label{fig:solsKdV}
	\end{center}
\end{figure}

We now turn to the resolution of the Whitham modulation equations for \eq{eq:KdV}. Specifically, we look for self-similar solutions which interpolate between the homogeneous solution $\zeta = 0$ around $x=0$ and arbitrary asymptotic values $\zeta_\pm$ of $\zeta$ at $x \to \pm \infty$. Since the solution is independent on $\zeta_\pm$ around $x = 0$, we can consider independently the two regions $x<0$ and $x>0$.\footnote{Hence we avoid the additional complication of the GP equation, where the velocity of the solution around $x=0$ is determined by the asymptotic conditions.} On each side, we look for a DSW or simple wave with the two following properties:
\begin{itemize}
	\item It has correct values of $\zeta$ at its two edges, i.e., $\zeta = 0$ at its edge closest to the horizon $x=0$, and $\zeta = \zeta_\pm$ at its other edge;
	\item It moves away from $x=0$, i.e., both edges have a strictly positive velocity in the region $x > 0$ and a strictly negative velocity in the region $x<0$.  
\end{itemize}
Examples of SW and DSW are shown in \Fig{fig:solsKdV}.  
Using the results of above subsection, we find the following necessary and sufficient conditions to obtain 
\begin{itemize}
	\item SW 	for $x<0$: 
	\be 
	\zeta_- < 0 \wedge \sqrt{g h_-} + v_- < 0;
	\ee 
	\item SW	
	for $x>0$: 
	\be 
	\zeta_+ > 0 \wedge \sqrt{g h_+} + v_+ > 0;
	\ee
	\item DSW for $x<0$:
	\be 
	\zeta_- > 0 \wedge \sqrt{g h_-} + v_- + \sqrt{\frac{g}{h_-}} \zeta_- < 0;
	\ee
	\item DSW for $x>0$:
	\be 
	\zeta_+ < 0 \wedge \sqrt{g h_+} + v_+ + 3 \sqrt{\frac{g}{h_+}} \zeta_+ > 0.
	\ee
\end{itemize}
Hence, if the parameters $h_+$, $h_-$, $v_+$, and $v_-$ are such that the trivial homogeneous solution $\zeta = 0$ has a black-hole horizon, then there exists a solution which is identically equal to zero in a finite, time-dependent neighborhood $I_t$ of $x=0$ provided
\be \label{eq:fincondKdV}
\zeta_- < - \sqrt{\frac{h_-}{g}} \lp \sqrt{g h_-} + v_- \rp \wedge \zeta_+ > - \frac{1}{3} \sqrt{\frac{h_+}{g}} \lp \sqrt{g h_+} + v_+ \rp.
\ee
Moreover, $\lim_{t \to \infty} I_t = \mathbb{R}$. In this sense, the solution converges to the trivial one $\zeta = 0$. 
In conclusion, as in the case of the GP equation, when  \eq{eq:fincondKdV} is satisfied, the AH solution acts as an attractor. 

When \eq{eq:fincondKdV} is not satisfied, one dispersive shock wave computed using the Whitham equations has one edge moving towards the horizon. The wave is then scattered by the discontinuity of $h$ and $v$ at $x=0$. This process is not described by the modulation theory.\footnote{However, the resulting pattern should be describable using two-phase solutions of the Whitham equations~\cite{doublecnoidal,multiphase}.} Numerical simulations indicate that a stationary undulation is 
generically obtained at late times.

\subsection{Superluminal KdV equation}

A superluminal version of the KdV equation can be obtained by changing the sign of the last term in \eq{eq:LKdV}:
\be \label{eq:sKdV}
\pd_t \zeta + \pd_x \lp \lp \sqrt{g h(x)} + v(x) \rp \zeta + \frac{3}{4} \sqrt{\frac{g}{h(x)}} \zeta^2 - \pd_x \lp \frac{h(x)^2}{3} \sqrt{g h(x)} \pd_x \zeta \rp \rp = 0.
\ee
Like the GP equation, \eq{eq:sKdV} has a superluminal dispersion relation. However, like the KdV equation it describes only right-moving modes in the fluid frame $v=0$. Solutions for the superluminal KdV equation \eq{eq:sKdV} are in one-to-one correspondence with those of \eq{eq:KdV} through the transformation 
\be \label{eq:KdVtosKdV}
\begin{array}{rl}
	x & \hspace*{-0.1 cm} \to -x, \\
	\zeta & \hspace*{-0.1 cm} \to -\zeta, \\
	\sqrt{g h} + v & \hspace*{-0.1 cm} \to - \lp \sqrt{g h} + v \rp,
\end{array}
\ee
which preserves the black- or white-hole nature of the flow. The uniqueness result mentionned above for the KdV equation thus extends to the present case. The conditions of existence of the simple waves are unchanged. The conditions for the DSW become
\begin{itemize}
	\item DSW for $x<0$:
	\be 
	\zeta_- > 0 \wedge \sqrt{g h_-} + v_- + 3 \sqrt{\frac{g}{h_-}} \zeta_- < 0;
	\ee
	\item DSW for $x>0$:
	\be 
	\zeta_+ < 0 \wedge \sqrt{g h_+} + v_+ + \sqrt{\frac{g}{h_+}} \zeta_+ > 0.
	\ee
\end{itemize}
Hence, if the parameters $h_+$, $h_-$, $v_+$, and $v_-$ are such that the trivial homogeneous solution $\zeta = 0$ has a black-hole horizon, then there exists a solution which is identically equal to zero in a finite, time-dependent neighborhood $I_t$ of $x=0$ such that $\lim_{t \to \infty} I_t = \mathbb{R}$ provided
\be \label{eq:fincondsKdV}
\zeta_- < - \frac{1}{3} \sqrt{\frac{h_-}{g}} \lp \sqrt{g h_-} + v_- \rp \wedge \zeta_+ > - \sqrt{\frac{h_+}{g}} \lp \sqrt{g h_+} + v_+ \rp .
\ee
As in the previous case, this condition is always satisfied provided the asymptotic conditions are sufficiently close to those compatible with a homogeneous black hole flow (where $\zeta_+ = \zeta_- = 0$). 
\section{Linear no-hair theorem for KdV}
\label{App:linKdV}

In this appendix we consider the linear stability of a black-hole flow of the KdV equation (\ref{eq:KdV}). For definiteness and simplicity, we consider the trivial solution $\zeta = 0$. We follow a procedure similar to that of the Appendix C in~\cite{Michel:2015pra}, although we here use the exact scattering coefficients to match the solutions at $x=0$. The physical picture underlying the calculation is clear: an initially localized perturbation will generally split into two parts. One of them moves away from the horizon and is diluted due to dispersion. The other one is scattered on the horizon and stimulates the Hawking effect, that is, pairs of of modes carrying opposite energies are emitted on both sides of the horizon. At late times, providing the amplification from the scattering is not strong enough to compensate for the dilution, only the modes with very small group velocities remain, with an amplitude which decreases polynomially in time. 

To make this idea more precise, we first briefly discuss the structure of the modes of the linearized KdV equation. We then determine the mode content of a square perturbation and compute its late-time evolution using a saddle-point approximation. To carry out the explicit calculation, we consider a steplike black-hole flow. 

\subsection{Modes in a steplike black-hole flow}

We consider the setup of \eq{eq:ansatzvh}, with $\sqrt{g h_-} + v_- < 0$ and $\sqrt{g h_+} + v_+ > 0$. Linearising the KdV equation gives
\be  \label{eq:lKdV}
\pd_t \psi + \lp \sqrt{g h} + v \rp \pd_x \psi + \pd_x \lp \frac{h^2}{6} \sqrt{g h} \pd_x^2 \psi \rp = 0.
\ee 
The scalar product of two solutions of this equation is defined by
\be 
\lp \psi_1, \psi_2 \rp \equiv \frac{i}{2} \int_\mathbb{R} \lp \psi_1^* \pd_x \psi_2 - \psi_2 \pd_x \psi_1^* \rp dx.
\ee
A straightforward calculation shows that $\pd_t \lp \psi_1, \psi_2 \rp = 0$. 
In the following we shall loosely call $(\psi_1, \psi_1)$ the ``norm'' of the mode $\psi_1$. 

\begin{figure}
	\begin{center}
		\includegraphics[width = 0.5 \linewidth]{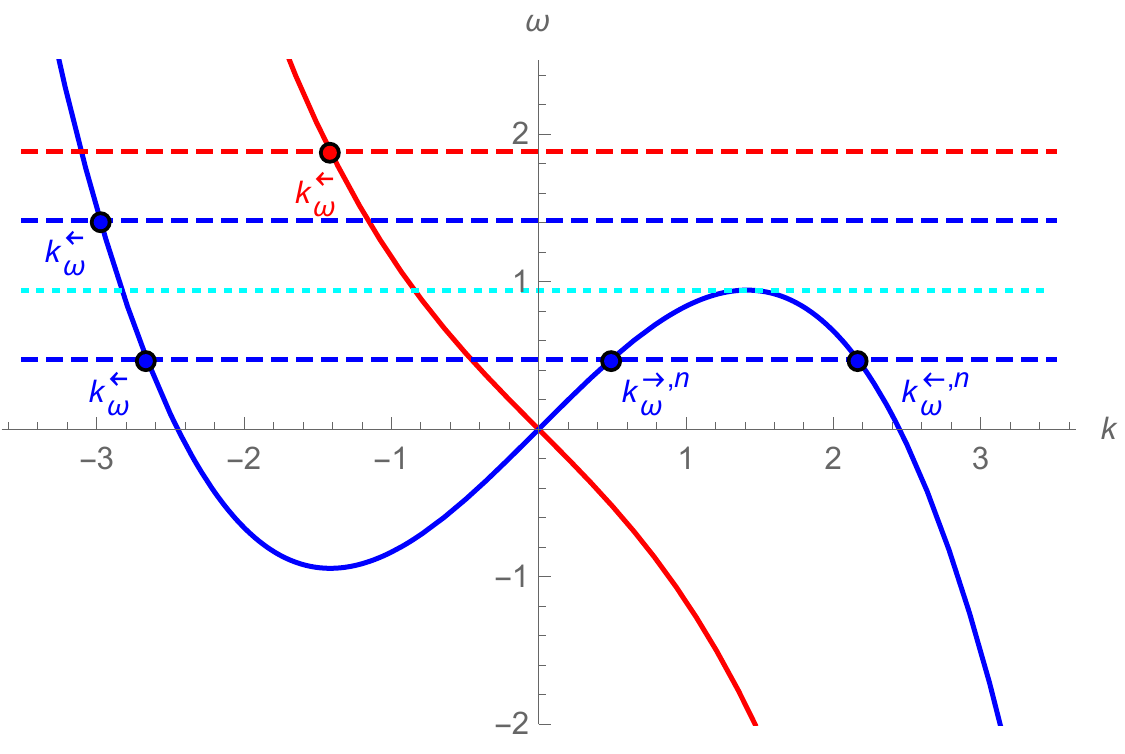}
	\end{center}
	\caption{Dispersion relation for the linearized KdV equation~(\ref{eq:lKdV}) in subcritical (blue) and supercritical (red) regions. Points give the real roots of the dispersion relation at fixed frequencies. The dotted, cyan line shows the value $\omc$ of $\om$ at which two roots in the subcritical region merge and become complex. Arrows indicate the direction of the group velocity, and a superscript ``$n$'' labels the modes with negative norms.} \label{fig:DRKdV}
\end{figure}

The dispersion relation relating the frequency $\om$ of a stationary mode to its wave vector $k$ in a homogeneous region is shown in \Fig{fig:DRKdV}. At fixed $\om \in \mathbb{R}$, we have only one real wave vector $k_\om^{\leftarrow}$ on the supercritical side. The corresponding mode has a negative group velocity and a positive norm for $\om > 0$. We also have two complex roots, giving exponentially increasing and decreasing modes for $x \to \infty$. We shall call the wave vector of the decreasing mode $k_\om^-$. In the subcritical region, the two complex wave vectors become real when $\om \in [-\omc, \omc]$, where
\be 
\omc = \frac{2 \sqrt{2}}{3} \sqrt{\frac{g}{h_+}} \lp  1 + \frac{v_+}{\sqrt{g h_+}} \rp^{3/2}.
\ee
In the following it will be convenient to work with globally defined incoming modes. For $\om \in [- \om_c, \om_c]$, there are two of them, which we call $\psi_{\om}^{\rm in}$ and $\psi_{\om}^{\rm in,n}$.  If $\om > 0$, $\psi_{\om}^{\rm in}$ has a positive norm while $\psi_{\om}^{\rm in,n}$ has a negative norm. They are given by
\be 
\psi_{\om}^{\rm in, (n)} =
\left\lbrace 
\begin{array}{ll}
	b_{\om,\leftarrow}^{(n)} e^{i k_{\om,L}^\leftarrow x} + b_{\om,-}^{(n)} e^{i k_{\om,L}^- x} & x<0 \\
	e^{i k_{\om,R}^{\leftarrow,(n)}} + b_{\om, \rightarrow}^{(n)} e^{i k_{\om,R}^{\rightarrow,n}x} & x>0
\end{array}
\right.,
\ee
where an index ``$L$'' or ``$R$'' indicates the region (left or right) where the wave vector is evaluated. The coefficients $b_{\leftarrow}^{(n)}$, $b_-^{(n)}$, and $b_{\rightarrow}^{(n)}$ are given by
\be \label{eq:epxplicitcoeff}
b_{\om,\rightarrow}^{(n)} = \frac{h_-^{5/2} \lp k_{\om,L}^- k_{\om,R}^{\leftarrow,(n)} + k_{\om,L}^{\leftarrow} k_{\om,R}^{\leftarrow,n} -k_{\om,L}^\leftarrow k_{\om,L}^- \rp - h_+^{5/2} \lp k_{\om,R}^{\leftarrow,(n)} \rp^2}{h_-^{5/2} \lp k_{\om,L}^- k_{\om,L}^{\leftarrow} - k_{\om,L}^\leftarrow k_{\om,R}^{\rightarrow,n} -k_{\om,R}^{\leftarrow,n} k_{\om,L}^- \rp + h_+^{5/2} \lp k_{\om,R}^{\rightarrow,n} \rp^2}, \nn
b_{\om,\leftarrow}^{(n)} = \frac{ \lp k_{\om,R}^{\leftarrow,(n)} - k_{\om,R}^{\rightarrow,n} \rp \lp h_+^{5/2} \lp k_{\om,R}^{\rightarrow,(n)} k_{\om,R}^{\leftarrow,(n)} + k_{\om,L}^- k_{\om,R}^{\rightarrow,n} -k_{\om,L}^- k_{\om,R}^{\leftarrow,(n)} \rp + h_-^{5/2} \lp k_{\om,L}^{-} \rp^2 \rp}{ \lp k_{\om,L}^- - k_{\om,L}^\leftarrow \rp \lp h_-^{5/2} \lp k_{\om,L}^- k_{\om,L}^{\leftarrow} - k_{\om,L}^\leftarrow k_{\om,R}^{\rightarrow,n} -k_{\om,R}^{\leftarrow,n} k_{\om,L}^- \rp + h_+^{5/2} \lp k_{\om,R}^{\rightarrow,n} \rp^2 \rp}, \nn
b_{\om,-}^{(n)} = \frac{ \lp k_{\om,R}^{\leftarrow,(n)} - k_{\om,R}^{\rightarrow,n} \rp \lp h_+^{5/2} \lp k_{\om,L}^\leftarrow k_{\om,R}^{\rightarrow,(n)} + k_{\om,L}^{\leftarrow} k_{\om,R}^{\leftarrow,n} -k_{\om,R}^{\leftarrow,n} k_{\om,R}^{\rightarrow,(n)} \rp + h_{\om,R}^{5/2} \lp k_{\om,L}^{\leftarrow} \rp^2 \rp}{ \lp k_{\om,L}^- - k_{\om,L}^\leftarrow \rp \lp h_-^{5/2} \lp k_{\om,L}^- k_{\om,L}^{\leftarrow} - k_{\om,L}^\leftarrow k_{\om,R}^{\rightarrow,n} -k_{\om,R}^{\leftarrow,n} k_{\om,L}^- \rp + h_+^{5/2} \lp k_{\om,R}^{\rightarrow,n} \rp^2 \rp}.
\ee
When $|\om| > \om_c$, only the mode $\psi_{\om}^{\rm in}$ remains, with $k_{\om,R}^{\rightarrow,n}$ replaced by the wave vector with a positive imaginary part. It can be shown that the only divergences of these coefficients occur for $\om \to 0$, where $b_{\om,\leftarrow}^{(n)}$ and $b_{\om,\rightarrow}^{(n)}$ diverge as $1/\om$ while $b_{\om,-}^{(n)}$ remains finite. Importantly, the diverging coefficients multiply exponentials with wave vectors which vanish linearly in $\om$ for $\om \to 0$. So, 
this divergence is regularized when expressing the results in terms of $\zeta = \pd_x \psi$. 

\subsection{Late-time evolution of a steplike perturbation}

Let us consider some initial condition $\psi(x,t=0) = \psi_0(x)$. One may expand it on the basis of incoming modes as
\be \label{eq:intmodes} 
\psi_0(x) = \int_{-\om_c}^{\om_c} \lp A_\om \psi_\om^{\rm in} (x,0) + A_\om^n \psi_\om^{\rm in,n} (x,0) \rp d \om + \int_{\mathbb{R} \backslash [-\om_c, \om_c]} A_\om \psi_\om^{\rm in}(x,0) d \om,
\ee
where
\be \label{eq:A}
A_\om^{(n)} = - \frac{\lp \psi_\om^{\rm in,(n)}, \psi_0 \rp}{2 \pi k_{\om,R}^{\leftarrow,(n)} \left\lvert \pd_\om k_{\om,R}^{\leftarrow,(n)} \right\rvert^{-1}}.
\ee
For definiteness, we work with a perturbation of the form
\be 
\psi_0 (x) = C \theta(x-x_-) \theta(x-x_+),
\ee
where $x_- < x_+$ are two real numbers. Notice that using a sum of such terms we can approximate any localized, smooth initial perturbation. Then,
\be \label{eq:explicitsc}
\lp \psi_\om^{\rm in,(n)}, \psi_0 \rp = i C \lp \psi_\om^{\rm in,(n)}(x_-,0) - \psi_\om^{\rm in,(n)}(x_+,0) \rp.
\ee
From then on, the analysis is long but straightforward. Since it does not present any particular difficulty, we shall only sketch it and give its main results. Using the linearity of \eq{eq:lKdV} we can assume without loss of generality that $x_-$ and $x_+$ have the same sign. Using Eqs.~(\ref{eq:epxplicitcoeff}, \ref{eq:A}, \ref{eq:explicitsc}), \eq{eq:intmodes} may be written as a sum of integrals of plane waves multiplied by prefactors which depend polynomially on $\om$. Writing everything explicitly, one checks that all divergences of the prefactors cancel each other when working with $\zeta$. At late times, the only contributions thus come from the saddle-points. These occur only when $\pd_\om k_\om = (x+...) / t$, where the three dots indicate the possible addition of $\pm x_+$ and/or $\pm x_-$. In the limit $t \to \infty$ at fixed $x$, only the waves with vanishing group velocities thus contribute, i.e., those of frequency $\om = \pm \om_c$. Using that the corresponding group velocity goes to zero as $\sqrt{\om_c - |\om|}$, we obtain after the Gaussian integration
\be \label{eq:zetat3/2}
\zeta(x,t) \mathop{\approx}_{t \to \infty} O (t^{-3/2}),
\ee
where the symbol $\approx$ is used to emphasize that logarithmic corrections not captured by the Gaussian integration may be present.

The same analysis can be done for a steplike perturbation where either $x_-$ or $x_+$ is sent to infinity. This is important as a localized perturbation on $\psi$ corresponds to a perturbation on $\zeta$ with a vanishing mean, and it is not a priori clear that the limits $t \to \infty$ and $x_\pm \to \pm \infty$ commute. To be complete, we must thus also consider one localized perturbation on $\zeta$ such that $\int \zeta dx \neq 0$. Choosing for instance $\psi_0(x) = C \theta(x)$, we obtain
\be 
\lp \psi_\om^{\rm in,(n)}, \psi_0 \rp = i C (1+b_{\om,\rightarrow}^{(n)}).
\ee 
The calculation then follows the same lines as the previous case. 
An additional divergency is present for $\om \to 0$, but does not contribute to the late-time solution as it multiplies an incoming wave with a non-vanishing group velocity. We thus find again \eq{eq:zetat3/2}.

\subsection{Case of \eq{eq:BdG}} 

As mentioned in the main text, the above calculation can be done for \eq{eq:BdG}.
The only point which does not directly extend to that case concerns the cancellation of divergences for $\om \to 0$. For the KdV equation, their cancellation when considering physical waves is due to two properties. First, all diverging terms multiply waves with a linearly vanishing wave vector. Second, the particular relationship between the solution $\phi$ of the linear equation and the physical observable introduces an additional factor $\om$. The first point carries on to the case of the BdG equation. The second one is replaced by the fact that the solution of \eq{eq:formlin} linearly vanishes when $\om \to 0$ for the corresponding waves.

\section{Black-hole solutions of the cubic-quintic GP equation}
\label{App:CQNLS} 

Both the GP equation \eqref{eq:GPE} and the KdV equation possess the specific property of being IST-integrable in a homogeneous background. Since this property played a crucial role in the derivation of DSW and SW solutions, it is of interest to consider the case of non-integrable equations. As an example, we consider the cubic-quintic GP (CQGP) equation 
\be \label{eq:CQGPE}
i \pd_t \psi(x,t) = -\frac{1}{2} \pd_x^2 \psi(x,t) +V(x) \psi(x,t) + g(x) \left\lvert \psi(x,t) \right\rvert^2 \psi(x,t) + \la \abs{\psi(x,t)}^4 \psi(x,t),
\ee
where $\la$ is a real parameter. To our knowledge, \eq{eq:CQGPE} is not integrable for $\la \neq 0$, even with uniform $V$ and $g$. It may thus be used to determine the effect of a small non-integrable deformation of \eq{eq:GPE}. 

To get the modified linearized equation, we consider a solution given by \eq{relpert}. Plugging it into \eq{eq:CQGPE} and extracting the first order in $\phi$ gives
\be 
i \lp \pd_t + v \pd_x \rp \phi = -\frac{1}{2} \pd_x^2 \phi - \frac{\pd_x \rho}{2\rho} \pd_x \phi + \lp g + 2 \la \rho \rp \rho \lp \phi + \phi^* \rp.
\ee
In a region where $\rho$ and $v$ are constant, we can again look for solutions of the form of \eq{eq:formlin}. The resulting system has solutions if and only if the dispersion relation:
\be 
\lp \om - v k \rp^2 = \lp g \rho + 2 \la \rho^2 \rp k^2 + \frac{k^4}{4}
\ee
is satisfied. The sound velocity, i.e., the velocity of long wavelength perturbations in the fluid frame, is now given by $c_S(\rho) = \sqrt{g \rho + 2 \la \rho^2}$. 

Let us briefly discuss the asymptotically homogeneous, stationary solutions of \eq{eq:CQGPE}. Rewriting $\psi$ as $\psi(x,t) = \sqrt{\rho(x,t)} e^{i \theta(x,t)}$, where $\rho$ and $\theta$ are real, \eq{eq:CQGPE} becomes
\be 
\left\lbrace 
\begin{array}{l}
	\pd_t \rho + \pd_x \lp \rho \pd_x \theta \rp = 0 \\
	\pd_t \theta - \frac{1}{2} \lp \frac{\pd_x^2 \sqrt{\rho}}{\sqrt{\rho}} - \lp \pd_x \theta \rp^2 \rp + V(x) + g(x) \rho + \la  \rho^2 = 0
\end{array}
\right. .
\ee
We consider a stationary solution where $\rho$ is independent on $t$ and $\pd_t \theta = \om$ is a constant. Then, $J \equiv \rho \pd_x \theta$ is also a constant and 
\be 
\pd_x^2 \sqrt{\rho} = \frac{J^2}{\rho^{3/2}} + 2 (V - \om) \sqrt{\rho} +2 g \rho^{3/2} + 2 \la \rho^{5/2}. 
\ee 
In a region where $V$ and $g$ are homogeneous, solutions with homogeneous densities exist iff 
$J^2 < J_c^2$, where $J_c$ is given by
\be 
J_c &=& \sqrt{-\lp 2 (V-\om) \rho_c^2 + 2 g \rho_c^3 + 2 \la \rho_c^4 \rp}, 
\nonumber \\
\rho_c &\equiv& \frac{\sqrt{9 g^2 + 32 (\om - V) \la} -3 g}{8 \la}.
\ee 
Then, as was the case for the standard GP equation, there exists two homogeneous solutions: a supersonic one with density $\rho_p$ and a subsonic one with density $\rho_b$. The latter can be seen as the limit of a stationary soliton solution when the center of the soliton is sent to infinity. 

We now focus on the steplike regime, with functions $g$ and $V$ given by \eq{eq:step}. If $g_+ < g_-$ and $V_+ > V_-$, there exist AH transonic solutions with density $\rho = \rho_0$, a conserved current satisfying $\sqrt{g_+ \rho_0 + 2 \la \rho_0^2} < (J/\rho_0) < \sqrt{g_- \rho_0 + 2 \la \rho_0^2}$, and a frequency 
\be 
\om = \frac{J^2}{2 \rho_0^2} + \frac{V_+ + V_-}{2} + \frac{g_+ + g_-}{2} \rho_0 + \la \rho_0^2.
\ee
The other possible black-hole solution with an asymptotically homogeneous density on both sides contains half a soliton in the subsonic region. Like in the case of the standard GP equation, one can show that it does not exist for $(J/\rho_0)$ inside the above interval.  

\begin{figure}[h]
	\begin{center}
		\includegraphics[width=0.49 \linewidth]{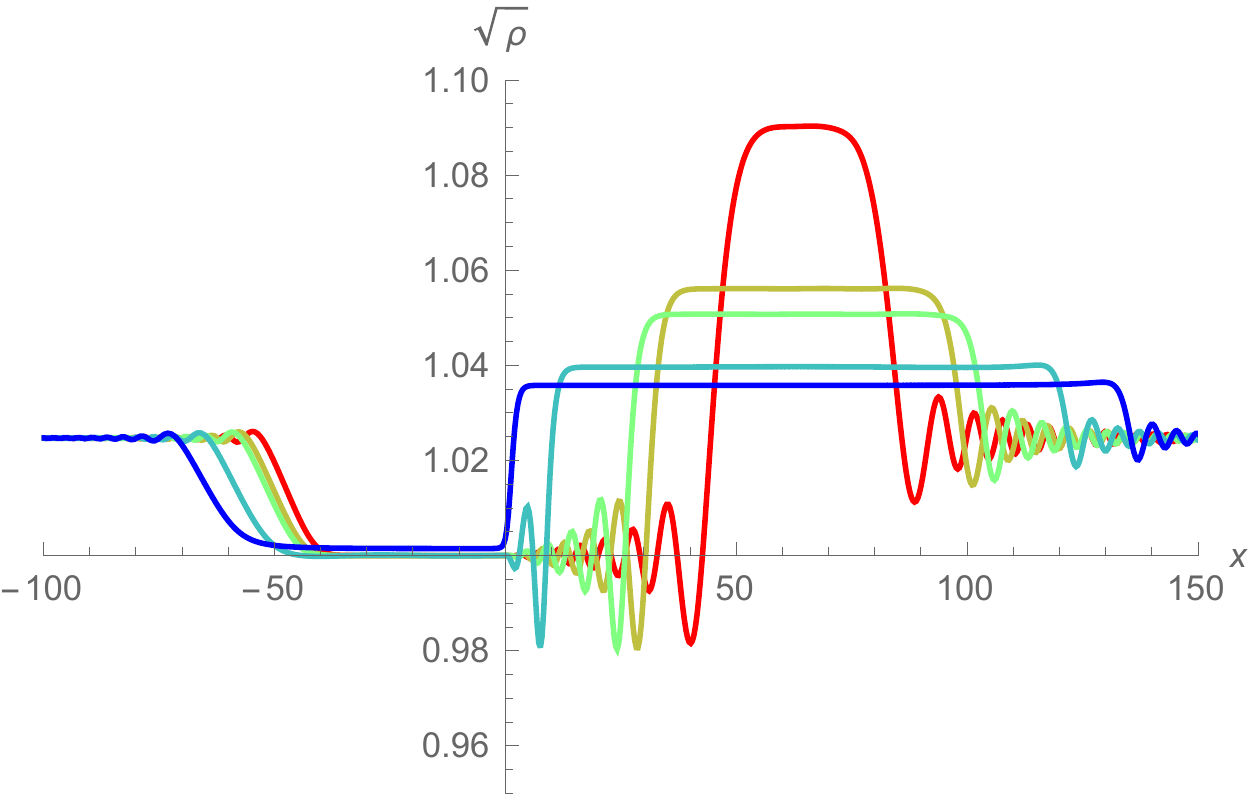} 
		\includegraphics[width=0.49 \linewidth]{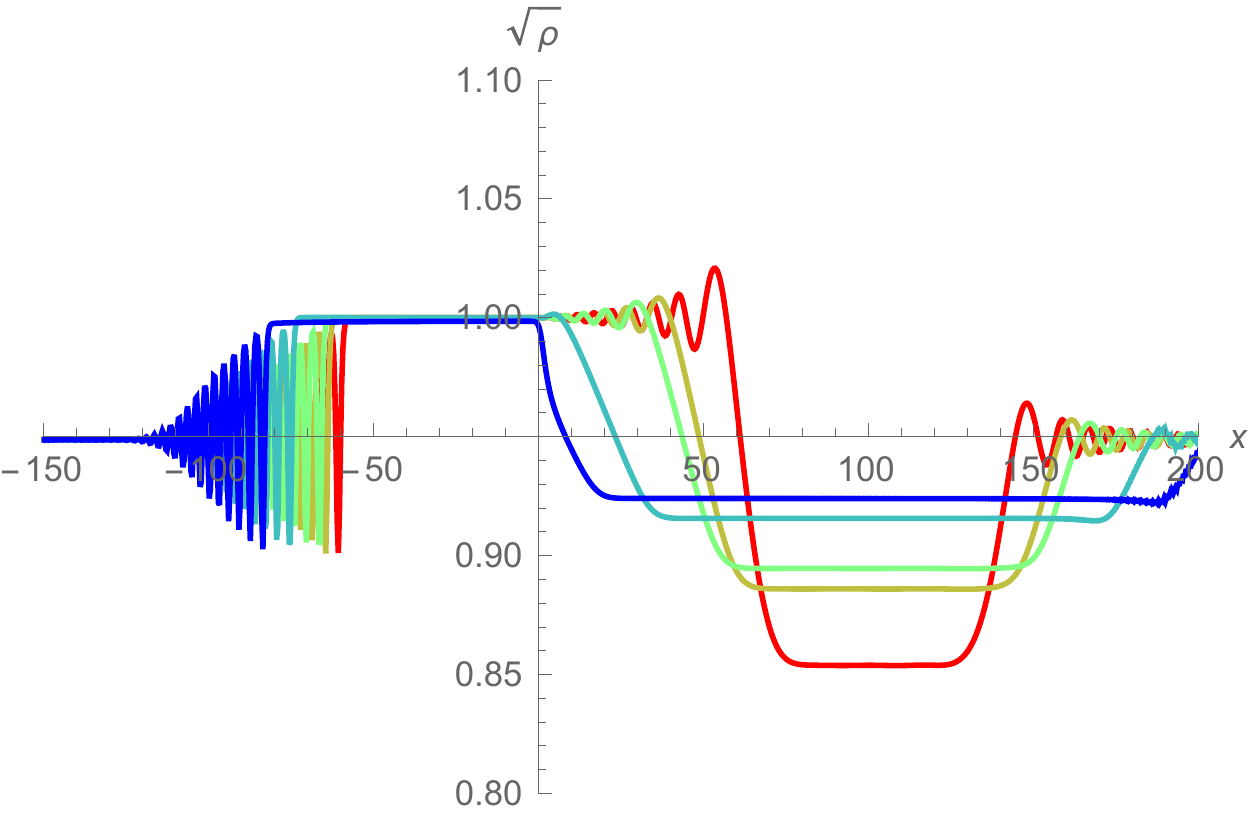}
	\end{center}
	\caption{Solutions obtained with the CQGP \eq{eq:CQGPE} in the steplike regime. The parameters $g_+$, $g_-$, $V_+$, $V_-$, and $J$ are the same as in the left panel of \fig{fig:solsNLSb}. The curves correspond to different values of $\la$: from red to blue, $-0.3$, $-0.1$, $0$, $0.5$, and $1$. The profiles are shown for  $t=40$. The initial configuration at $t=0$ is homogeneous with a density $\rho_i = 1.05$ (left) and $\rho_i = 0.9$ (right). One clearly sees that increasing $\lambda$ smoothly deforms the solutions without affecting their qualitative properties. In particular, the homogeneous flows still acts as an attractor.} \label{fig:CQGPE}
\end{figure}
In \fig{fig:CQGPE} we show results of numerical simulations for homogeneous initial conditions. For small values of $\la$, the solution still has three waves of the same type (SW or DSW) and with the same direction of propagation as for $\la = 0$. The main effects of $\la$ are to change the velocities of the waves and the value of the density between the two rightmost ones. A positive value of $\la$ seems to increase the velocity of the rightmost wave and reduce that of the two other ones, while bringing the density between the two rightmost waves closer to $\rho_0$. A small, negative value of $\la$ has the opposite effect. When increasing $\abs{\la}$, we observe a qualitative change in the solution. For $\la$ smaller than a critical value, close to $-0.4$ for the parameters of the figure, the solution diverges after a finite time. When $\la$ is larger than another critical value, close to $1$ in the figure, the velocity of the middle wave becomes negative and the solution develops a hairy black-hole flow.

\bibliography{../biblio/bibliopubli}

\end{document}